\documentclass[11pt]{article}

\usepackage{amsmath}
\usepackage{graphicx}
\usepackage{indentfirst}
\usepackage{amssymb}
\usepackage{cite}
\usepackage{color}
\usepackage{subfigure}
\usepackage{varwidth}
\usepackage{diagbox}

\begin{document}

\title{The observation image  of a soliton boson star illuminated by various accretions}

\date{}
\maketitle

\begin{center}
\author{Ke-Jian He,}$^{a}$\footnote{E-mail: kjhe94@.163com}
\author{Guo-Ping Li,}$^{b}$\footnote{E-mail: gpliphys@yeah.net}
\author{Chen-Yu Yang,}$^{a}$\footnote{E-mail: chenyuyang2024@163.com}
\author{Xiao-Xiong Zeng}$^{a}$\footnote{E-mail: xxzengphysics@163.com (Corresponding author)}
\\

\vskip 0.25in
$^{a}$\it{Department of Mechanics, Chongqing Jiaotong University, Chongqing 400000, People's Republic of China}\\
$^{b}$\it{School of Physics and Astronomy, China West Normal University, Nanchong 637000, People's Republic of China}\\
$^{c}$\it{College of Physics and Electronic Engineering, Chongqing Normal University, Chongqing 401331, People's Republic of China}\\
\end{center}

\vskip 0.6in
{\abstract
{ In this paper, we explore the observable signatures of solitonic boson stars by employing ray-tracing simulations, with celestial spheres and thin accretion disks serving as illumination sources. By numerically fitting the metric form, we solve the geodesic equation for photons under the influence of the soliton potential, enabling us to simulate the optical appearance of the soliton boson star in two distinct regimes. In the weak  coupling case (larger value of coupling parameter $\alpha$) with an initial scalar field $\psi_0$, the images on the screen predominantly show direct and lensed images, where $\psi_0$ and $\alpha$ modulate the image region size  while the observation inclination $\theta$ controls morphological asymmetry. In the case of strong coupling (small value of $\alpha$), the images on the screen show a nested sub-annulus within the Einstein ring in the celestial model, whereas thin disk accretion models reveal higher-order lensing images indicative that photons are capable of orbiting the equatorial plane of the boson star multiple times. We also analyze how the effective potential and redshift factor depend on the correlation parameter. At low inclination($\theta<30^{\circ})$, the redshift is the dominant effect, the image is characterized by a dim central cavity enclosed by a bright ring. At high inclination ($\theta>60^{\circ})$, the Doppler effect becomes more pronounced, resulting in a substantial brightness disparity between the left and right sides of the optical image. These findings offer robust theoretical underpinnings for differentiating solitonic boson stars from black holes via high-resolution astronomical observations.}}
\thispagestyle{empty}
\newpage
\setcounter{page}{1}\

\section{Introduction}
\label{sec:intro}
Since the formulation of general relativity (GR), Einstein's field equations have fundamentally reshaped our understanding of spacetime dynamics. In 1919, the solar eclipse measurements quantified light deflection, ushering in the era of experimental relativity\cite{crispino2019hundred}. Modern breakthroughs emerged in 2019 when the Event Horizon Telescope (EHT) captured the first direct image of  the supermassive black hole at the center of the galaxy Messier 87 (M87)\cite{EventHorizonTelescope:2019dse,EventHorizonTelescope:2019uob,EventHorizonTelescope:2019jan,
EventHorizonTelescope:2019ths,EventHorizonTelescope:2019pgp,EventHorizonTelescope:2019ggy}, revealing a characteristic bright ring formed by strongly lensed photons from accretion flows. Subsequently, observations of the Sagittarius A* (Sgr A*) black hole at the center of the Milky Way showed the same bright ring structure\cite{EventHorizonTelescope:2022wkp,EventHorizonTelescope:2022vjs,EventHorizonTelescope:2022wok,
EventHorizonTelescope:2022exc,EventHorizonTelescope:2022urf,EventHorizonTelescope:2022xqj}. Owing to the intense gravitational influence exerted by the compact object, the trajectory of light is significantly deflected in the vicinity of the object, a phenomenon known as gravitational lensing \cite{wambsganss1998gravitational}. The underlying mechanism involves relativistic synchrotron emission from magnetized plasma in the compact object's vicinity, where spacetime curvature modifies photon trajectories via multi-order lensing. The imaging results of the EHT not only validate the correctness of general relativity at strong fields, but also provide the possibility for the accurate measurement of ultra-compact objects\cite{EventHorizonTelescope:2020qrl}.

The critical curve constitutes a cornerstone in analyzing optical signatures of ultra-compact objects. In the spacetime of a Kerr black hole, unstable photon orbits form a toroidal structure known as the photon shell. This corresponds to the critical point of the radial potential, where light rays asymptotically approach but do not cross\cite{hou2022multi,cunha2018shadows}. In static spacetimes, the photon shell degenerates into a single circular critical curve corresponding to the extremum of the effective potential, which is referred to as the photon sphere. A significant body of research has explored this theoretical framework and its potential empirical implications \cite{cunningham1972optical,
cardoso2019moving,cardoso2021light}. Research has demonstrated that for any compact object situated within its photon sphere, light rays are capable of orbiting multiple times, thereby forming a distinct photon ring embedded within the primary bright ring of emitted radiation. These photon rings vanish at the outer edge of the central brightness depression, i.e., the shadow. The dimensions, intensity of the shadow, as well as the positioning and luminosity of the photon rings, are determined by the synergistic effects of the background spacetime geometry, optical characteristics, and radiative properties of the accretion disk surrounding the object\cite{vincent2022images}. Separating the contributions of the background geometry and the astrophysical properties of the disk in shadow images remains a critical challenge\cite{lara2021separating,wielgus2021photon}, and whether the Kerr solution can effectively describe every compact object remains a subject of debate\cite{vincent2021geometric}.

During the imaging of black holes, the radiative accretion flow surrounding them plays a critical role. Moreover, the realistic image results from the complex interactions between the strong gravitational lensing effect of the black hole and the electromagnetic properties of the plasma within the accretion flow. This phenomenon necessitates extensive numerical simulations based on general relativistic magnetohydrodynamics (GRMHD)\cite{{EventHorizonTelescope:2019pcy}}. In particular, the Sgr A* data were compared with observations at other wavelengths, including 86 GHz, 136 THz, and X-ray bands, as well as with theoretical models derived from GRMHD simulations and general relativistic radiative transfer (GRRT) calculations\cite{EventHorizonTelescope:2022urf}. Time-averaged GRMHD simulations have provided strong evidence that the photon ring persists as a stable relativistic feature\cite{EventHorizonTelescope:2019ths}. Meanwhile, short-term variations in the accretion flow may lead to transient fluctuations in both brightness and morphological structure\cite{Dexter:2020hqv}. Using the GRRT and GPU-accelerated GRMHD code, Chatterjee et al. generated synthetic radio images of (highly) tilted disk/jet models and compared them with EHT observations, suggesting the possibility that M87 hosts a tilted disk/jet system\cite{Chatterjee:2020eqc}. Recently, Moriyama et al. systematically investigated the dependence of the light curve amplitude in non-Kerr GRMHD simulations on deviations from Schwarzschild spacetime, under the constraints imposed by weak gravitational fields and the observed size of the Sgr A* shadow\cite{Moriyama:2025isl}. The results show that the dynamical properties of the accretion flow exhibit a systematic dependence on the degree of deviation. Interestingly, based on GRRT calculations of magnetized accretion flows around naked singularities and black holes, Dihingia et al. found that the accreting matter near the singularity is repelled by the effective potential energy, thereby preventing it from reaching the naked singularity\cite{Dihingia:2024cch}. Meanwhile, they also observed that naked singularities produce light arcs rather than the halos typically seen around black holes, and the fluid dynamics and radiative characteristics of the surrounding matter exhibit significant differences compared to those around black holes. Currently, extensive research has been conducted on the simulation of black hole images using  GRMHD  simulations and GRRT calculations, yielding several significant results\cite{McKinney:2012vh,Moscibrodzka:2015pda,Hou:2023bep,Mao:2025ibt}.

It is widely recognized that research on black hole imaging based on GRMHD simulations will significantly advance the understanding of magnetohydrodynamics in the vicinity of black holes and contribute to the development of black hole imaging techniques. However, significant challenges are encountered in solving the equations of GRMHD and in applying numerical simulation techniques.
Therefore, it is essential to develop simplified accretion models to investigate the optical observational features of black holes. Theoretically, simplified accretion models are generally sufficient for capturing the essential features of black hole images and have thus been widely investigated, including the spherical accretion  and  thin disk accretion model\cite{Narayan:2019imo,Zeng:2020dco,Heydari-Fard:2023ent,Gralla:2019xty,Zeng:2020vsj,Peng:2020wun,He:2022yse,Li:2021ypw,Zeng:2021mok,Li:2021riw,Gao:2023mjb,He:2021htq,Wang:2022yvi
,Hou:2022eev,Yang:2024nin,He:2024amh,He:2024qka}. In addition, some important results have been obtained on black hole shadows and photon rings using wave optics in the holographic framework\cite{Zeng:2024ptv,Zeng:2023zlf,Liu:2022cev,He:2024mal,Li024mdd:2,He:2024bll}. However, an emerging frontier in compact object research involves not black hole entities that theoretically populate the universe, as evidenced by alternative spacetime solutions\cite{cardoso2019testing,barack2019black}. In terms of optical images (or shadows), these objects may exhibit differences in the structure of their photon spheres, such as possessing multiple photon rings or lacking photon rings entirely \cite{wielgus2020reflection,tsukamoto2021gravitational,olmo2022new,guerrero2022light,tsukamoto2022retrolensing}, providing a theoretical basis for distinguishing various compact objects through optical images. A notable example arises in asymptotically flat exotic compact objects: topological triviality of their spacetime enforces an even number of photon rings despite their event horizon absence \cite{cunha2017light}. This study focuses on boson stars, which are ultra-compact objects composed of bosons bound by self-gravity, with their mass and size ranging from atomic to astrophysical scales, depending on the boson mass. Since the pioneering work of Kaup \cite{kaup1968klein} and Ruffini and Bonnazol \cite{ruffini1969systems}, boson stars have attracted significant attention, and their stability and dynamical properties have been extensively studied \cite{liebling2023dynamical,di2020dynamical}. Subsequent numerical relativity studies have established that both static and rotating boson star configurations with multi-solar-mass profiles remain observationally viable alternatives to black holes. Intriguingly, the supermassive object Sgr A* has been proposed as a potential boson star candidate, motivating detailed comparisons between their optical signatures\cite{vincent2016imaging}. Boson stars, along with their optical images under various potentials and gravitational modifications, have been extensively studied \cite{maso2021boson,rosa2023imaging,rosa2024accretion}, including the specific case of mini boson stars \cite{zeng2025opticalimagesminiboson}. Due to the resemblance between black hole shadows and boson star optical images, several studies have proposed that boson stars can replicate the shadow of a Schwarzschild black hole under a truncated accretion disk \cite{herdeiro2021imitation} or at reduced observer inclination angles \cite{rosa2022shadows}, thereby presenting boson stars as a plausible alternative to black hole models.

A pivotal distinction between boson stars and black holes stems from the self-interaction potential \cite{collodel2022solitonic}, which enables bound state formation without relying solely on gravitational confinement, that is, a phenomenon analogous to Q-ball solutions in scalar field theories \cite{coleman1985q}. In shift-symmetric models, the potential generally incorporates quadratic, quartic, and sextic terms, as expressed by $V = c_2\psi^2 + c_4\psi^4 + c_6\psi^6$, where the coefficients satisfy $c_2, c_6 > 0$ and $c_4 < 0$. We specialize to an interesting special case of the sextic potential, commonly referred to as the solitonic potential, which is expressed as  $V = u^2 \psi^2 (1 - \frac{\psi^2}{\alpha^2})^2$ \cite{friedberg1976class}, where the interaction term is governed by the parameter $\alpha$. Crucially, this potential admits a false vacuum state at $\psi = \alpha$ with $V (\alpha)=0$. The parameter $\alpha$ controls solution stiffness: large $\alpha$ regimes the system recover a mini boson star, while decreasing $\alpha$  enhances nonlinearities in the field equations, significantly complicating numerical solutions.
In this work, two background light source models, celestial sphere and thin disk, are used to study the optical image features of solitonic boson stars. We sequentially investigate the effects of the initial scalar field $\psi_0$, weak coupling, and strong coupling regimes, which correspond to smaller and larger coupling parameters $\alpha$, respectively. In the first two cases, a stable photon ring cannot be formed because there is no closed photon orbit (critical curve). Observed images display a single bright annulus from direct disk emission, with central intensity suppression due to gravitational lensing. However, under conditions of strong coupling, the optical images of solitonic boson stars exhibit remarkable resemblance to black hole.

The structure of this paper is organized as follows: Section \ref{sec2} derives the equations of motion by utilizing the solitonic potential. Section \ref{sec3} outlines the numerical methods employed for spherical light sources, whereas Section \ref{sec4} details the methods applied to thin accretion disks. Section \ref{sec5} presents the numerical results and offers an in-depth analysis of the optical images and redshift factors of solitonic boson stars under various conditions. Lastly, a concise conclusion and discussion are provided.

\section{Solutions of solitonic boson star}\label{sec2}
In consideration of the subsequent action involving the minimal coupling of a complex scalar field with a gravitational field, that is
\begin{equation}
S=\int d^4x \sqrt{-g} [\frac{R}{2 \kappa}-\nabla_b \overline{\Psi}\nabla^b\Psi-V(|\Psi|^2)].
\end{equation}
Here, $g$ denotes the determinant of the metric tensor, $R$ represents the scalar curvature, $\kappa = 8\pi$ is the gravitational coupling constant, $\overline{\Psi}$ is the complex conjugate of the scalar field $\Psi$, and $V$ is the scalar potential.
We are primarily focused on the solitonic potential, which is
\begin{equation}
V(|\Psi^2|)= u^2 \psi^2 (1-\frac{\psi^2}{\alpha^2})^2, \label{vphi}
\end{equation}
where $u$ is the scalar field mass and $\alpha$ is a free parameter controlling the self-interaction. The self-interaction potential employed here represents a conventional choice, as it is capable of providing configurations that can exist even in flat spacetime\cite{cahn1985production}.
The equation of motion given by the corresponding action with respect to the variation of the metric and the field is expressed as
\begin{equation}
R_{bc}-\frac{1}{2}Rg_{bc}=\kappa T_{bc},\label{gbc}
\end{equation}
\begin{equation}
g^{bc}\nabla_b\nabla_c\Psi=\Psi \frac{dV}{d|\Psi|^2},\label{gphi}
\end{equation}
And, the energy-momentum tensor is
\begin{equation}
T_{bc}=\nabla_b\overline{\Psi}\nabla_c\Psi+\nabla_c\overline{\Psi}\nabla_b\Psi-g_{bc}\label{tab}
(\nabla_d\overline{\Psi}\nabla^d\Psi+V).
\end{equation}
It is reasonable to assume that the solution bears resemblance to the generalized spherically symmetric metric in Schwarzschild-like coordinates, which can be expressed as
\begin{equation}
ds^{2}=-f(r)dt^{2}+g(r)^{-1}dr^{2}+r^{2}(d\theta^2+\sin^2\theta d\phi^2),\label{metric1}
\end{equation}
and
\begin{equation}
\Psi(r,t)=\psi(r) e^{i \omega t} \label{psi}.
\end{equation}
By substituting Eqs. (\ref{vphi}), (\ref{metric1}), and (\ref{psi}) into Eqs. (\ref{gbc}) and (\ref{gphi}), one can get
\begin{equation}
\frac{d f}{dr}= \frac{f(1-g)}{gr}+\kappa r (\frac{\omega ^2 \psi^2}{g}+f(\frac{d \psi }{dr})^2-\frac{f}{g }u^2 \psi^2 (1-\frac{\psi^2}{\alpha^2})^2 ),
\end{equation}

\begin{equation}
\frac{d g}{dr}=\frac{1-g}{r}-g \kappa  r\frac{ d\psi}{dr}^2 -\frac{\kappa  r \left(f \mu ^2 \psi ^2 \left(1-\frac{\psi ^2}{\alpha ^2}\right)^2+\psi ^2 \omega ^2\right)}{f},
\end{equation}

\begin{equation}
\frac{d^2 \psi}{dr^2}=-\frac{1}{2}(  \frac{df}{f}+\frac{dg}{g}+\frac{4}{r})\frac{d\psi}{dr}-\frac{\psi  \left(\mu ^2 \left(-\frac{3 \psi ^4}{\alpha ^4}+\frac{4 \psi ^2}{\alpha ^2}-1\right)+\frac{\omega ^2}{f}\right)}{g}.
\end{equation}
It is worth noting that the aforementioned equations of motion remain invariant under the following scale transformation, which are
\begin{equation}
 r\rightarrow ur,~~f\rightarrow u^2 f,~~g\rightarrow r,~~V\rightarrow u^{-2}V.
\end{equation}
In order not to lose generality, the value of $u$ is taken as $u = 1$ in subsequent calculations. To solve the aforementioned equation of motion, it is necessary to determine the boundary conditions for the functions $f(r)$, $g(r)$ and $\psi(r)$.  At infinity, the asymptotic behavior of this spacetime is expected to resemble that of the Schwarzschild spacetime, that is
\begin{equation}
 f(r\rightarrow \infty)\sim f_{ \infty} (1-\frac{2M}{r}),
\end{equation}
\begin{equation}
 g(r\rightarrow \infty)\sim   (1-\frac{2M}{r}),
\end{equation}
\begin{equation}
\psi(r\rightarrow \infty)\sim   0.
\end{equation}
Since we are interested in localized solutions of the scalar field $\Psi$ that preserve asymptotic flatness, the wave function $\psi$ therefore decays exponentially as $r \to \infty$. On the other hand,  at $r = 0$, the solutions should be convergent. By expanding $f(r)$, $g(r)$, and $\psi(r)$ into a power series of $r$ at $r = 0$, one can obtain that
\begin{equation}
 f(r\rightarrow 0)\sim f_0,  ~~g(r\rightarrow 0)\sim 1,~~  \psi(r\rightarrow 0)\sim \psi_0.
\end{equation}
In reality, $f_{ \infty}$ and $f_{ 0}$ are not independent of each other; once $f_{ \infty}$  is specified, $f_{ 0}$ becomes determined accordingly.
Since the equations of motion are independent of the time $t$, it is possible to reparameterize $t$, $f$ and $\omega$ with any constant without altering their functional relationships. Specifically, this can be expressed as
\begin{equation}
  t\rightarrow a t,    ~~f\rightarrow a^{-2} f, ~~\omega \rightarrow a^{-1} \omega.
\end{equation}
In this case, one can set $f_0=1$. Given that boson stars exhibit similarities to ideal fluid stars, their solutions can also be characterized in terms of their total mass $M$ and radius $R$. Both of these quantities can be defined as
\begin{equation}
 m(r)=\frac{r}{2}(1-g(r)).
\end{equation}
The mass is defined as $M=m(r\rightarrow \infty)$. Generally, we define the radius $R$ of the boson star as the value of $r$ at which the mass function satisfies $M(r) = 0.98M$. Therefore, under the aforementioned boundary conditions, the equation of motion can be solved by assigning fixed values to $\alpha$ and $\psi_0$. It should be noted that in defining the radius $ R $ of a boson star, we do not strictly require the numerical solution to exactly coincide with the external Schwarzschild geometry at $ r = R $; in other words, $ r = R $ is not treated as the matching surface. Instead, as $r$ increases, the spacetime naturally and progressively transitions toward the Schwarzschild spacetime through the overall numerical evolution. This approach reflects the continuous attenuation of the boson star's matter field and energy density in the outer regions, rendering the area far from the center nearly vacuum, with the spacetime metric approaching the Schwarzschild form. Since the mass of a boson star is generally highly concentrated, the radius values calculated using different definitions do not exhibit substantial discrepancies.

\section{Celestial sphere model and the selection of the observer}\label{sec3}
In this section, we shall examine the observed images of a solitonic boson star illuminated by the celestial light source. The first step is to determine the behavior of photons around the boson star, which is governed by the Euler-Lagrange equation, which is
\begin{equation}
\frac{d }{d\gamma}(\frac{\partial\mathcal{L} }{\partial \dot{x}^{\alpha } } )=\frac{\partial\mathcal{L} }{\partial x^{\alpha } }.
\end{equation}
Here, the parameter $\gamma$ represents the affine parameter, $\dot{x}^{\alpha}$ denotes the four-velocity of the photon. In addition, the term of $\mathcal{L}$ represents the Lagrangian, which can be expressed as
\begin{eqnarray}
\mathcal{L}=\frac{1}{2} g_{\alpha \beta }\dot{x}^{\alpha}\dot{x}^{\beta}
=\frac{1}{2}(-f(r)\dot{t}^{2}+\frac{1}{g(r)} \dot{r}^{2}+r^{2}\label{lequa}(\dot{\theta}^{2}+\sin^{2}\theta\dot{\phi }^{2})).
\end{eqnarray}
In which, $g_{\alpha \beta }\dot{x}^{\alpha}\dot{x}^{\beta}=-m^2_p$. Moreover, $m_p$ denotes the mass of a particle, and in the case of a photon, $m_p = 0$.
The metric function does not explicitly depend on time $t$ or the azimuthal angle $\phi$, thereby giving rise to two conserved quantities, which can written as
\begin{equation}
E=\frac{\partial \mathcal{L} }{\partial \dot{t} } =f(r)\frac{d t}{d \gamma },
\label{eque}
\end{equation}
\begin{equation}
L=-\frac{\partial \mathcal{L} }{\partial \dot{\phi } } =r^{2}\frac{d \phi }{d \gamma }.
\end{equation} \label{equl}
With the assistance of Eq.(\ref{lequa})- Eq.(21), it is possible to derive the four-velocity components corresponding to the time, azimuthal angle, and radial direction, which are
\begin{equation}
   \frac{d t}{d \gamma} =\frac{1}{b f(r)},
   \label{equ-7}
\end{equation}
\begin{equation}
    \frac{d \phi }{d  \gamma } =\pm \frac{1}{r^{2}},
    \label{equ-8}
\end{equation}
\begin{equation}
    \frac{d r}{d  \gamma } =\sqrt{\frac{1}{b^{2}}\frac{g(r)}{f(r)}-\frac{1}{r^{2}} g(r)}.
    \label{equ-9}
\end{equation}
Here,  the term of $b$ is called the impact parameter, which is formally  defined as
\begin{equation}
    b=\frac{\left |  L\right | }{E}.
\end{equation}
Given that the spacetime of a boson star exhibits spherical symmetry and is static, it is feasible to derive the corresponding effective potential, that is
\begin{equation}
\dot{r}^2=V_{eff}, \quad  V_{eff}=-m^2_p-\left[ \frac{1}{b^{2}}\frac{g(r)}{f(r)}-\frac{1}{r^{2}} g(r)\right]. \label{veff}
\end{equation}
According to the extremum condition of the effective potential, one can determine whether the soliton boson star has  photon ring, and the position of the photon ring is determined by the first derivative of the effective potential energy, i.e., $V_{eff}'(r) = 0$. Although the geodesic equation describes the motion of photons in curved spacetime, specifying initial conditions necessitates the definition of a physical observer frame. We adopt the zero angular momentum observer (ZAMO), a fiducial observer in stationary spacetime whose four-velocity aligns with the timelike Killing vector. Specifically, the ZAMO is positioned at coordinates $(t_o, r_o, \theta_o, \phi_o)$, and a locally orthogonal normalized frame may exist in the neighborhood of the observer, which are
\begin{align}
	&\hat{e}_{(t)} =\left(\sqrt{-\frac{1}{g_{tt}}}, 0, 0, 0\right),\quad \hat{e}_{(r)} =\left(-\sqrt{\frac{1}{g_{rr}}}, 0, 0, 0\right),\label{eq:tetrad1}\\
	&\hat{e}_{(\theta)} =\left(0, \sqrt{\frac{1}{g_{\theta \theta}}}, 0, 0\right),\quad \hat{e}_{(\varphi)} =\left(-\sqrt{\frac{1}{g_{\varphi \varphi}}}, 0, 0, 0\right). \label{eq:tetrad2}
\end{align}
In the observer's reference frame, the photon's four-momentum is expressed as $ p_{(\mu)} = p_\nu \hat{e}^\nu_{(\mu)} $, where $ p_{(\mu)} $ represents the four-momentum in the ZAMO frame, while $ p_\nu $ corresponds to that in the Boyer-Lindquist coordinate system. Then, the tangent vector of the null geodesic can be written  as
\begin{equation}
\dot{s}_o=\mid \overrightarrow{OP}\mid  (-\hat{e}_{(t)}+\cos \Theta \hat{e}_{(r)}+\sin \Theta \cos \Xi \hat{e}_{(\theta)}+\sin \Theta \sin \Xi \hat{e}_{(\theta)}),\label{ccort}
\end{equation}
in which $\mid \overrightarrow{OP}\mid$  represents  the tangent vector
of the null geodesic at the point $O$ in the three-dimensional
subspace, and $\Theta $ denotes the angle between  $\mid \overrightarrow{OP}\mid$   and $\hat{e}_{(r)}$, $\Xi$ denotes the angle between  $\mid \overrightarrow{OP}\mid$   and $\hat{e}_{(\theta)}$. The celestial coordinates are the coordinates described by  $(\Theta ,\Xi)$.
On the other hand, for each light ray $s(\alpha)$ with coordinate representation  $t(\alpha)$, $r(\alpha)$, $\theta(\alpha)$ and  $\phi(\alpha)$, the general form of tangent vector is
\begin{equation}
\dot{s}=\dot{t}\partial_t+\dot{r}\partial_r+\dot{\theta}\partial_{\theta}+\dot{\phi}\partial_{\phi},\label{zcort}
\end{equation}
By comparing Eq.(\ref{ccort}) and Eq.(\ref{zcort}), one can conclude that the celestial coordinates are uniquely determined once the four-momentum of a photon is specified. Conversely, if the celestial coordinates are known, the four-momenta can be determined through coordinate transformation. Thus, by incorporating the observer's position, the initial values for the photon motion equation can be directly obtained. In order to acquire the image of the solitonic boson star, it is essential to map the celestial coordinates $(\Theta , \Xi)$ to the corresponding points $(x, y)$ on the imaging plane systematically. In accordance with the work \cite{Hu:2020usx}, the celestial coordinates can be expressed as
\begin{align} \label{TM3}
\cos \Theta=\frac{p^{(r)}}{p^{(t)}}, \quad \tan \Xi= \frac{p^{(\phi)}}{p^{(\theta)}}.
\end{align}
On the image plane $(x, y)$, the conversion between Cartesian coordinates and spherical celestial coordinates can be formulated as
\begin{align}\label{TM4}
x = -2\tan\frac{\Theta}{2}\sin \Xi,\quad y = -2\tan\frac{\Theta}{2}\cos \Xi.
\end{align}
Additionally, the image plane is divided into $\mathcal{N} \times \mathcal{N}$ equally sized square regions. For any position with pixel coordinates $(i, j)$, the corresponding celestial coordinates can be expressed as
\begin{equation}
\tan \Xi =\frac{2j -(\mathcal{N}+1)}{2i -(\mathcal{N}+1)},
\end{equation}
and
\begin{equation}
\tan \Theta =\frac{1}{\mathcal{N}} \tan\frac{1}{2} \beta_{\mathrm{fov}}\sqrt{(i-\frac{\mathcal{N}+1}{2})^2+(j-\frac{\mathcal{N}+1}{2})^2}.
\end{equation}
Here, $\beta_{\mathrm{fov}}$ is the angle of field of view. Once the pixel resolution and field of view angle are determined, the celestial sphere coordinates and initial conditions are established. Consequently, based on the photon motion equation, the behavior of photons within the soliton boson star spacetime can be accurately characterized.

\section{The configuration of the thin accretion disk model}\label{sec4}
To obtain the optical observational appearance of a solitonic boson star, it is essential to model the background light source, i.e., the accretion disk. Astronomical observations indicate that Sgr A* and M87 are low-luminosity Active Galactic Nucleus (AGNs), typically associated with radiatively inefficient accretion flows (RIAFs). These flows are expected to be geometrically thick and optically thin due to their low accretion rates and advection-dominated dynamics. However, for theoretical simplicity and to facilitate direct comparison with prior studies, the geometrically thin accretion disk model can also be taken into account. This approximation is justified by the following considerations:
\begin{itemize}
    \item The thin disk model provides a tractable framework for analyzing qualitative trends of observational features (e.g., photon rings and shadows) without requiring full radiative transfer simulations of RIAFs.
    \item As demonstrated in Ref. \cite{Gralla:2019xty} by Wald et al., even simplified thin disk models are capable of capturing key observational signatures, such as the decomposition of a shadow image into direct emission, lensed ring, and photon ring components. These features have been extensively utilized in studies involving various compact objects and modified gravity theories.
\end{itemize}
In this context, we consider an axisymmetric, geometrically and optically thin accretion disk located in the equatorial plane of the solitonic boson star, where the plasma within the disk moves in circular orbits along timelike trajectories. In addition, the inner edge of the accretion disk is set at the innermost stable circular orbit (ISCO), whose position is determined by the effective potential, that is
\begin{align} \label{ISCO}
\begin{cases} V_{eff}(r)|_{r=r_{isco}}=0,  \\
      \partial_r V_{eff}(r)|_{r=r_{isco}}=0, \\
      \partial^2_r V_{eff}(r)|_{r=r_{isco}}=0.
    \end{cases}
\end{align}
Due to the particle within the accretion disk moving along timelike circular orbits, their four-velocity is expressed as
\begin{align}\label{Fvelocity}
 u^\mu=\left(-\sqrt{\frac{2 g_{tt}^2}{r \partial_r g_{tt}-2 g_{tt}}}, 0, 0, \sqrt{\frac{r^3 \partial_r g_{tt}}{2 g_{tt}- r \partial_r g_{tt}}}\right).
\end{align}

\begin{figure}[h]
\centering
\includegraphics[width=.4\textwidth]{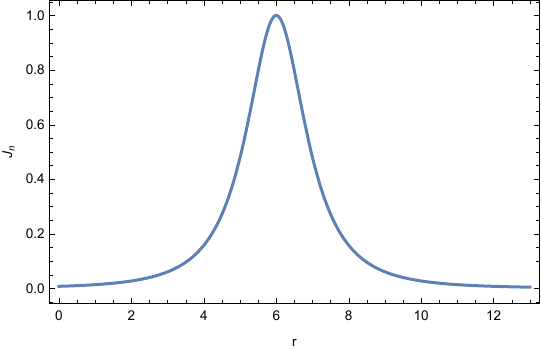}
\caption{The intensity distribution curve of emissivity $J_n$.\label{figjn}}
\end{figure}

On the other hand, the observed intensity is modified by relativistic effects including Doppler boosting, gravitational redshift, and absorption-emission coupling at the disk surface. As in \cite{Li:2024ctu}, the refraction effect is neglected, and the variation in light intensity can be obtained by
\begin{equation}
\frac{d}{d \gamma}(\frac{Q_{\nu}}{\nu^3})=\frac{J_{\nu}-k_{\nu} Q_{\nu}}{\nu^2}\label{lintensity},
\end{equation}
in which $\gamma$ is the affine parameter of null geodesics as mentioned above, $Q_{\nu}$, $J_{\nu}$, and $k_{\nu}$ are the specific intensity, emissivity and absorption coefficient at the frequency $\nu$,  respectively. When light propagates in a vacuum, both $J_{\nu}$ and $k_{\nu}$ become zero, and consequently, $\frac{Q_{\nu}}{\nu^3}$ is conserved along the geodesics. In the thin disk approximation, it is necessary to consider only the instantaneous emission and absorption occurring on the equatorial plane. Hence, $J_{\nu}$ and $k_{\nu}$ are negligible except on the equatorial plane. Under these conditions, the total light intensity observed by the observer can be expressed as
\begin{equation}
Q_{o}=\sum\limits_{n}^{n_{\max}} f_n g_n^3J_n \label{Q0},
\end{equation}
where $n$ is the number of times that the light passes through the equatorial plane, $f_n$ is  the absorption of the accretion disk, $J_n$ is  the emission of the accretion disk, and $g_n$ is  the  redshift factor. To determine the intensity $Q_{o}$, it is necessary to first calculate $f_n$, $J_n$, and $g_n$. As emphasized in \cite{Yang:2024nin}, since $f_n$ primarily affects the strength of the narrow photon ring and has a limited influence on the overall image, it is set to 1. In previous studies, there have been numerous choices regarding the specific form of radiation $J_n$. To compare with astronomical observations, such as the images of M87 and Sgr A*, the emissivity is frequently modeled as a second-order polynomial in log-space. Here, we employ the functional form derived from Johnson's SU distribution\cite{Gralla:2020srx}, that is
\begin{equation}
J_n=\frac{e^{ -\frac{1}{2} \left(\epsilon +\sinh ^{-1}\left(\frac{r-\zeta }{\eta }\right)\right)^2}}{\sqrt{(r-\zeta )^2+\eta ^2}}.
\end{equation} \label{jf}
In above equation, the term of $\epsilon$, $\zeta$ and $\eta$ are  parameters controlling the shape of the emission profile, which are called as the rate of increase,  radial translation, and the dilation of the profile, respectively. These three parameters ($\epsilon$, $\zeta$, $\eta$)  collectively characterize the thin accretion disk model, and their selection was originally designed to align with the results of GRMHD simulations of Kerr black holes. In principle, we can adjust these parameters   to select  proper intensity profiles for the models under study. In this work, we take $\epsilon=0$, $\zeta=6M$, and $\eta=M$, and present in Figure \ref{figjn} the variation trend of the radiative $J_n$ with the radial coordinate $r$.
Through detailed analysis and calculation, the results indicate that the ISCO of the boson stars discussed in this paper are either extremely small or effectively non-existent. In other words, the inner edge of the emission model does not need to be strictly defined in terms of its position. Therefore, we took the distribution of the accretion disk to be in the form of Figure 1, with the brightest position at $6M$, and then it rapidly decays towards both ends. Hence,   its inner edge can be considered to be at a position close to $r\rightarrow0$, but when $r < 2M$, its luminosity is very small and can be ignored.

Another quantity that requires calculation is the redshift factor $g_n$. Since the plasma within the accretion disk propagates along the timelike geodesic, its angular velocity $\Omega_{n}$ satisfies
\begin{equation}\label{angular}
\Omega_{n}=\frac{u^{\phi}}{u^t } \mid_{r=r_n}.
\end{equation}
In this case, the redshift factor can be expressed as
\begin{equation}
g_{n}=-\frac{1}{ \mathcal{A} (1-b \Omega_{n})},
\end{equation}
in which
\begin{equation}
\mathcal{A} =\sqrt{\frac{1}{g_{tt}+g_{\phi\phi}\Omega_{n}^2}} |_{r=r_n}.
\end{equation}
By utilizing this equation, we not only substitute it into Eq. (\ref{Q0}) to examine the light intensity, but also conduct a further analysis of the redshift phenomenon occurring in the vicinity of the soliton boson star.

\section{Optical observation images of soliton boson stars}\label{sec5}
With the aforementioned preparations,  one can obtain the  solution of a boson star through numerical methods, with systematic investigation of their optical signatures under different emission scenarios. In principle, the solution of the boson star depends on the initial  scalar field  $\psi_0$ and the coupling parameter $\alpha$. Hence, we shall concentrate on investigating the impact of variations in relevant parameters as well as the observation inclination angle on the observational characteristics of the boson star.

\subsection{Influence of the initial  scalar field  $\psi_0$  on optical images}
\begin{figure}[h]
\centering
\includegraphics[width=.4\textwidth]{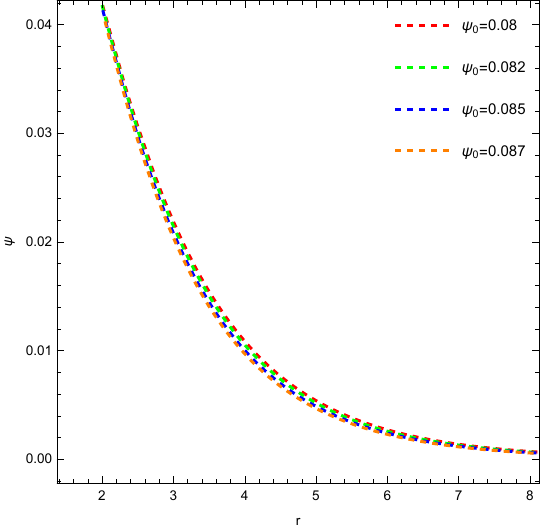}
\caption{Variation of the scalar field $\psi$ as a function of the radial distance $r$ for different initial scalar field values $\psi_0$, with the coupling parameter $\alpha=0.9$.\label{fig1}}
\end{figure}

\begin{figure}[htbp]
\centering
\includegraphics[width=.34\textwidth]{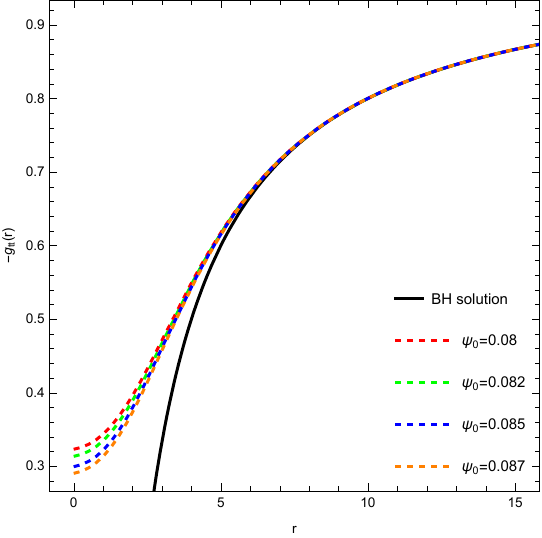}
\qquad
\includegraphics[width=.34\textwidth]{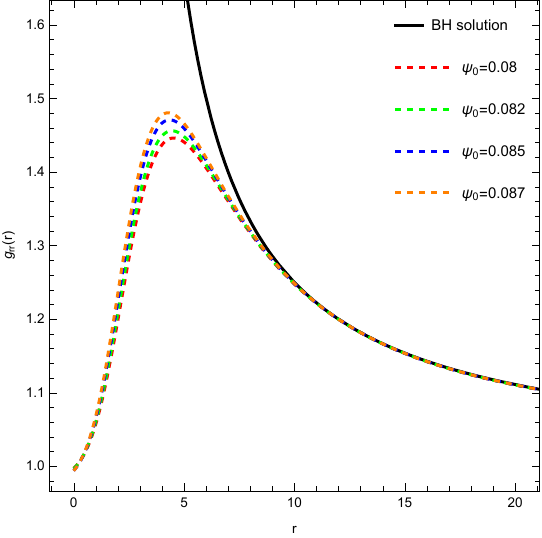}
\caption{Comparison between the numerical metrics and the Schwarzschild black hole metric components for different initial scalar field values $\psi_0$, with the coupling parameter $\alpha=0.9$.}
\label{fig2}
\end{figure}

\begin{figure}[h]
\centering
\includegraphics[width=.34\textwidth]{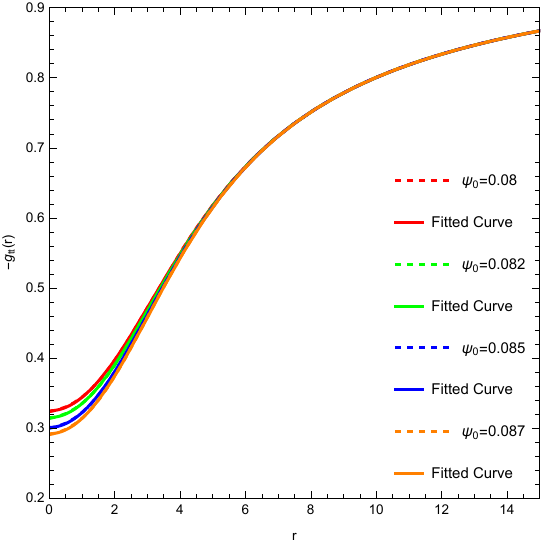}
\qquad
\includegraphics[width=.34\textwidth]{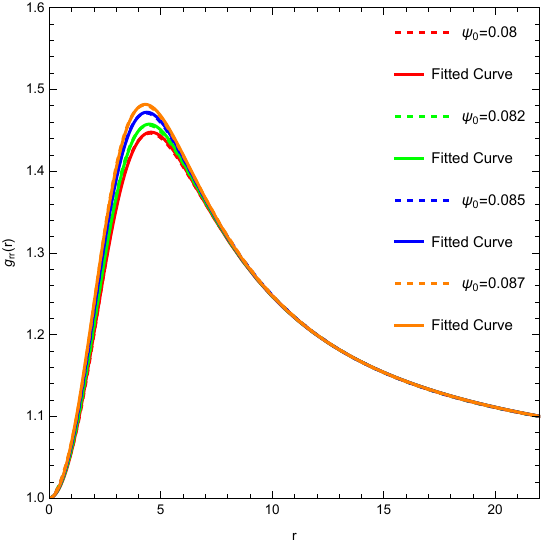}
\caption{Comparison between the numerical metrics and the fitting functions for different initial scalar field values $\psi_0$, with the coupling parameter $\alpha=0.9$. \label{fig3}}
\end{figure}

\begin{figure}[h]
	\centering
	\includegraphics[width=.4\textwidth]{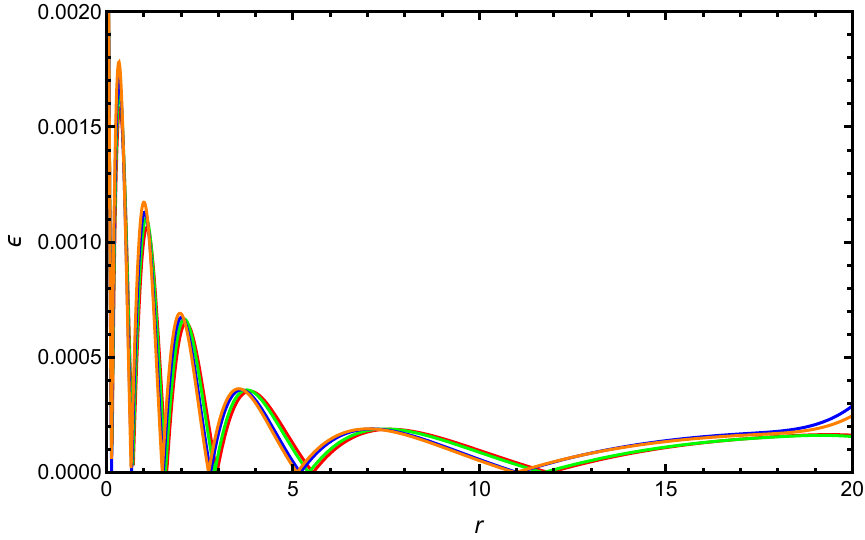}
	\qquad
	\includegraphics[width=.4\textwidth]{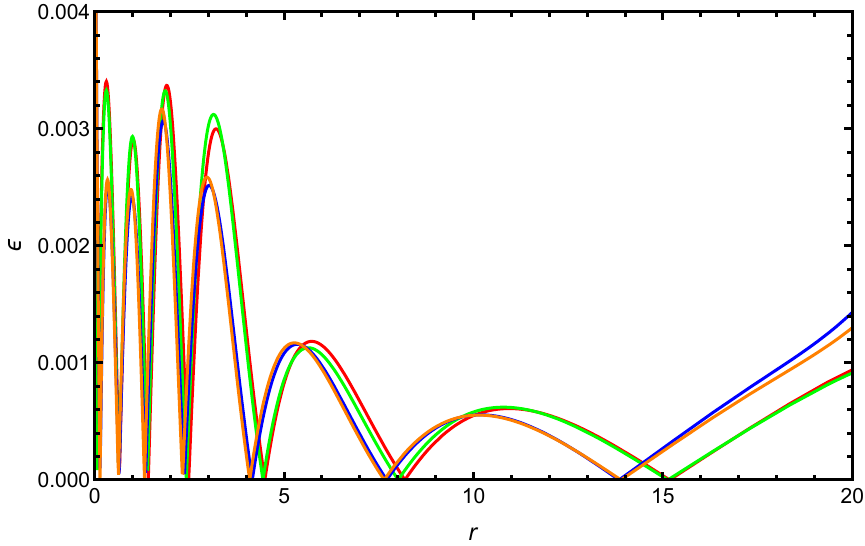}
	\caption{ The relative error between the numerical metric and the fitting metric, where the left panel corresponds to the metric component $-g_{tt}$, and the right panel corresponds to the metric component $g_{rr}$. The red, green, blue, and orange curves represent the cases with $\psi_0 = 0.08, 0.082, 0.085,$ and $0.087$, respectively.\label{fig32}}
\end{figure}

In this part, we will explore how the scalar field $\psi_0$ influences the structure and observational features of boson stars. In order to obtain the fitting function of the metric, it is necessary to solve the numerical solutions of the scalar field and the metric. The numeric results for different scalar field $\psi_{0}$ is shown in Figure \ref{fig1}, where  the coupling parameter is taken as $\alpha=0.9$. One can find that the scalar field exist only in a narrow range and it decline rapidly to zero as $r$ increases for all the boson star models.
The metric of the boson stars is presented in Figure \ref{fig2}. For comparison with the simplest spherically symmetric black hole, the metric of a Schwarzschild black hole is depicted by a solid black line. From the left subfigure, one can observe that boson stars, as anticipated, since the boson star has no event horizon, the numerical result of the metric will not be zero. This characteristic distinguishes them from Schwarzschild black holes. However, the asymptotic behavior of boson stars is precisely identical to that of Schwarzschild black holes. As $r$ increases, the metrics approach to one. In particular, when $r$ is approximately $10M$, boson stars exhibit identical characteristics to Schwarzschild black holes. The right subfigure in Figure \ref{fig2} illustrates the $g_{rr}$ components of both boson stars and Schwarzschild black holes, the resulte shows that  they are not divergent as they possess no horizons. Notably, the metric components $g_{rr}$ converge to the same value under various initial conditions $\psi_0$, which is an intriguing finding.
It should be noted that the existence of numerical infinity makes direct numerical calculation impractical. Therefore, specific functions can be employed to approximate the metric components. It has been found that the following functions provide an excellent fit, which are
\begin{equation}
g_{tt}=-\exp\left[p_7 \left(\exp \left(-\frac {1+p_1 r+p_2 r^2}{p_3+p_4 r+p_5r^2+p_6r^3}\right)-1\right)\right],\label{fit1}
\end{equation}
and
\begin{equation}
g_{rr}=\exp\left[q_7 \left(\exp \left(-\frac {1+q_1 r+q_2 r^2}{q_3+q_4 r+q_5r^2+q_6r^3}\right)-1\right)\right].\label{fit2}
\end{equation}
To visually illustrate the alignment between the fitting metric and the numerical metric, we present the fitted results of the  metric components $-g_{tt}$ and $g_{rr}$, see Figure\ref{fig3}. It can be clearly observed that, for both $-g_{tt}$ and $g_{rr}$, the numerical results exhibit a high degree of consistency with the fitted results.
Nevertheless, we can still quantitatively analyze the differences between the numerical metric and the fitting metric. The relative errors between them can be calculated by
\begin{equation}
	\epsilon = \left| \frac{g_\mathrm{num} - F_\mathrm{fit}}{g_\mathrm{num}} \right|,
\end{equation}
where $g_\mathrm{num}$ denotes the numerical metric (either $-g_{tt}$ or $g_{rr}$), and $F_\mathrm{fit}$ represents the fitted metric function. The variation trend of the relative error $\epsilon$  between the numerical metric and the fitting metric with respect to the radial coordinate $r$ is presented in Figure~\ref{fig32}. It can be observed that as $r$ increases, the relative error between the numerical metric and the fitted one shows a damped oscillatory form. The maximum relative error for metric component $-g_{tt}$ is approximately $ \epsilon \sim 0.0018$, whereas the maximum relative error for metric component $g_{rr}$ is slightly higher, reaching $\epsilon \sim 0.0034$. Therefore, the maximum deviation values of the fitting results for both $-g_{tt}$ and  $g_{rr}$ are within the acceptable range, which ensures the robustness of the fitting degree metric. When the parameter $\psi_0$ takes on different values and the coupling parameter $\alpha$ is held constant at $\alpha = 0.9$, Table 1 and Table 2 summarize the resulting parameter values in the fitted metric components as well as the corresponding boson star masses. The results indicate that the mass of the boson star decreases with an increase in the initial value $\psi_0$.

\begin{figure}[!h]
\centering 
\subfigure[$\psi_0=0.08,\theta=45^{\circ}$]{\includegraphics[scale=0.3]{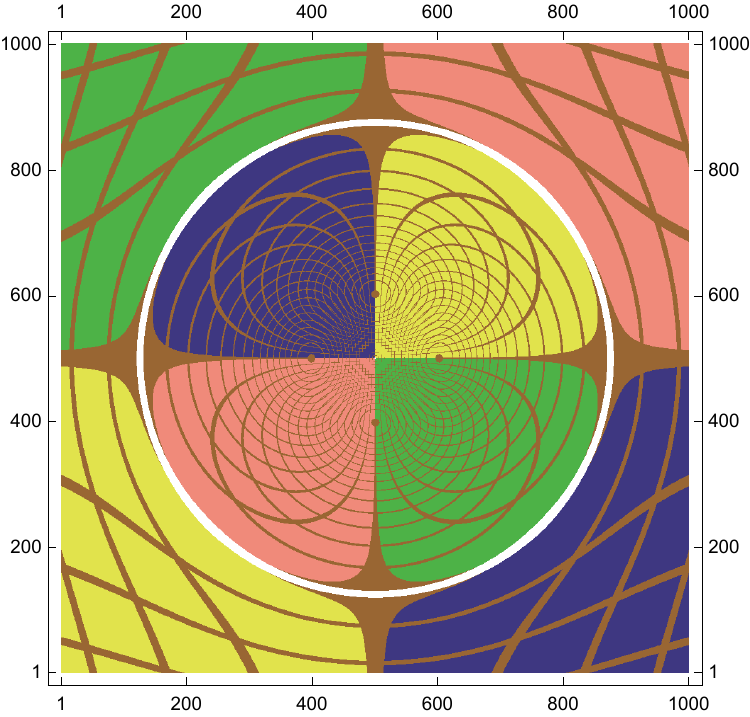}}
\subfigure[$\psi_0=0.082,\theta=45^{\circ}$]{\includegraphics[scale=0.3]{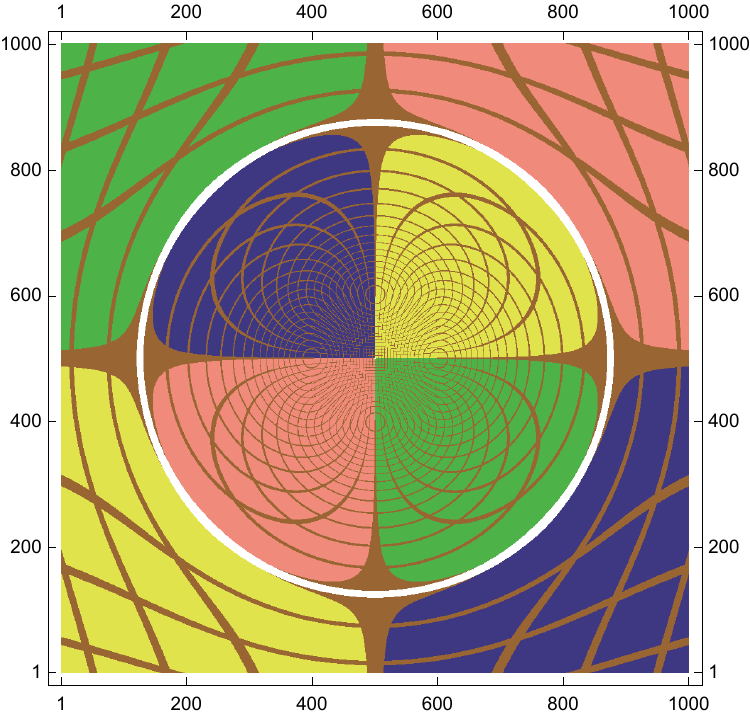}}
\subfigure[$\psi_0=0.085,\theta=45^{\circ}$]{\includegraphics[scale=0.3]{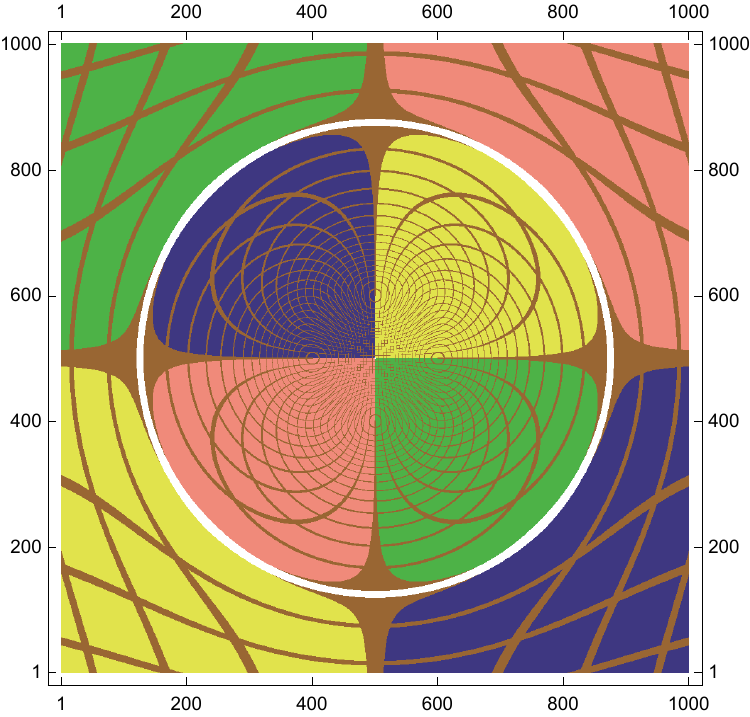}}
\subfigure[$\psi_0=0.087,\theta=45^{\circ}$]{\includegraphics[scale=0.3]{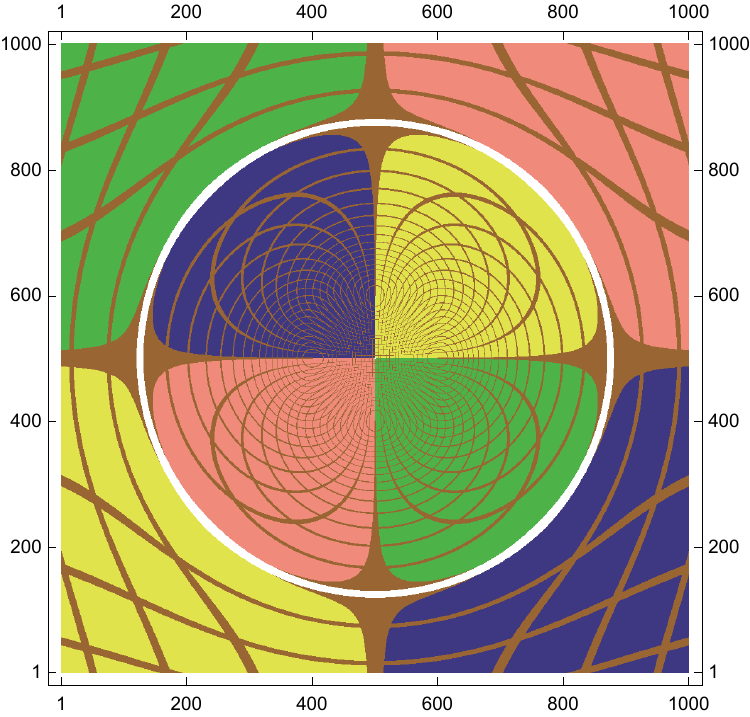}}
\caption{\label{fig5}  Optical images of boson stars illuminated by a  celestial sphere light source for different initial scalar field values $\psi_0$. Here, the observer inclination angles $\theta=45^{\circ}$, and the coupling parameter $\alpha=0.9$.}
\end{figure}

\begin{table}[h]
\begin{center}
{\footnotesize{\bf Table 1.}
When $\psi_0$ takes on different values, the corresponding values of $p_i$ (where $i = 1, \dots, 7$) in the fitting metric component $-g_{tt}$ and the associated boson star mass $m(r)$.} \\
\vspace{2mm}
\begin{tabular}{c|c|c|c|c|c|c|c|c|c}
\hline
Type & $\psi_0$ & $m(r)$ & $p_1$ & $p_2$ & $p_3$ & $p_4$ & $p_5$ & $p_6$ & $p_7$ \\ \hline
SBS1 & 0.08 & 0.603& 0.024& 0.113& -0.367& -0.013& -0.051& -0.007& -0.079 \\ \hline
SBS2 & 0.082 & 0.6 &0.026& 0.119& -0.367& -0.014& -0.054& -0.008& -0.082\\ \hline
SBS3 & 0.085 & 0.595&0.027& 0.127& -0.365& -0.014& -0.057& -0.009& -0.083\\ \hline
SBS4 & 0.087 &0.591 &0.029& 0.134& -0.365& -0.015& -0.06& -0.01& -0.086\\ \hline
\end{tabular}
\end{center}
\end{table}
\begin{table}[h]
\begin{center}
{\footnotesize{\bf Table 2.} When $\psi_0$ takes on different values, the corresponding values of $q_i$ (where $i = 1, \dots, 7$) in the fitting metric component $g_{rr}$ and the associated boson star mass $m(r)$.} \\
\vspace{2mm}
\begin{tabular}{c|c|c|c|c|c|c|c|c|c}
\hline
Type & $\psi_0$ & $m(r)$ & $q_1$ & $q_2$ & $q_3$ & $q_4$ & $q_5$ & $q_6$ & $q_7$ \\ \hline
SBS1 & 0.08 & 0.603& -17.515& -2.315& 6.652& 1.03& 1.545& 0.032& 0.02 \\ \hline
SBS2 & 0.082 &0.6  & -16.16& -2.199& 6.127& 0.935& 1.486& 0.033& 0.022\\ \hline
SBS3 & 0.085 &0.595&-10.702& -1.356& 4.323& 0.553& 1.075& 0.026& 0.029\\ \hline
SBS4 & 0.087 & 0.591&-10.945& -1.454& 4.333& 0.572& 1.132& 0.029& 0.029\\ \hline
\end{tabular}
\end{center}
\end{table}
Through the application of appropriate fitting metric, one can simulate the observed image of the soliton boson star that is enveloped by the light source of the celestial sphere, which are shown in Figure \ref{fig5}. In Figure \ref{fig5}, the celestial sphere is divided into four distinct quadrants, each differentiated by a unique color to enhance visual clarity. The grid consisting of both longitude and latitude lines is represented by brown lines, spaced at intervals of $10^{\circ}$. Clearly, these images illustrate the spacetime distortion caused by boson stars as well as the phenomenon of gravitational lensing. Interestingly, the central regions of these images still exhibit the presence of light rays, which contrasts with black holes, where the central regions are entirely dark. Boson stars lack an event horizon, implying that light can propagate through their interior and subsequently be detected by an external observer after exiting the star. As in the case of black holes, the Einstein ring is still visible, as shown by the white circle. Under the same observation inclination angle, as the initial scalar field $\psi_0$ increases, the overall characteristics of the image remain largely unchanged. However, certain differences in the distribution of brown patterns can still be observed in the central regions of the images, as shown in Figures \ref{fig5}(a) and (d). In other words, the distribution of light paths in spacetime is affected by changes in the scalar field. Consequently, these variations suggest that fluctuations in the scalar field can modify the structure of spacetime geometry, thereby influencing the gravitational lensing effect.

\begin{figure}[!h]
\centering 
\subfigure[$\psi_0=0.08,\theta=0^{\circ}$]{\includegraphics[scale=0.35]{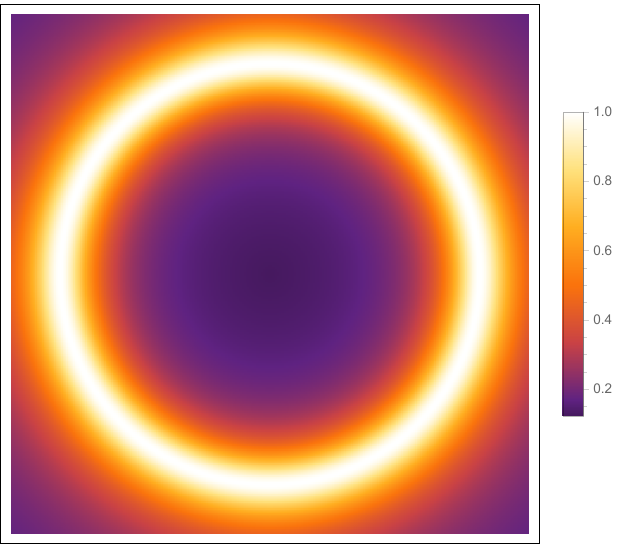}}
\subfigure[$\psi_0=0.08,\theta=30^{\circ}$]{\includegraphics[scale=0.35]{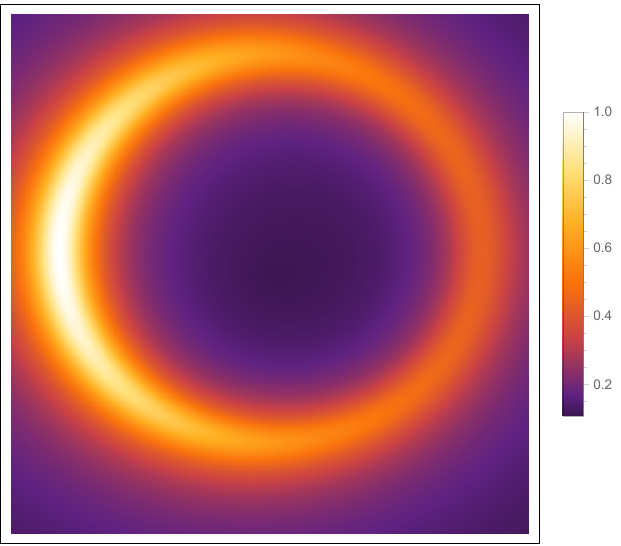}}
\subfigure[$\psi_0=0.08,\theta=60^{\circ}$]{\includegraphics[scale=0.35]{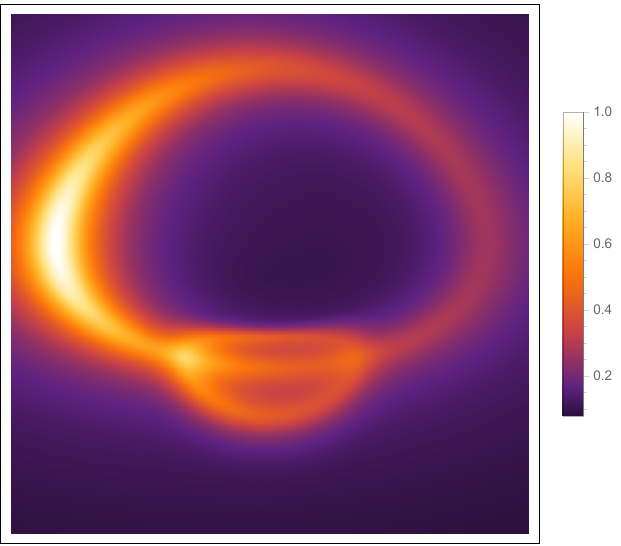}}
\subfigure[$\psi_0=0.08,\theta=75^{\circ}$]{\includegraphics[scale=0.35]{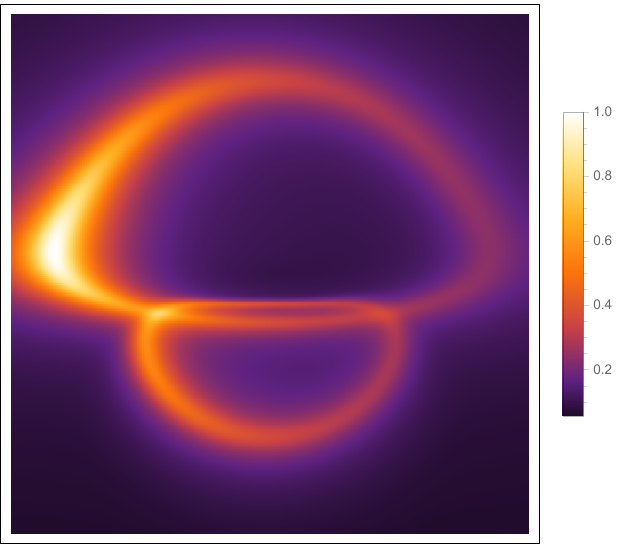}}
\subfigure[$\psi_0=0.082,\theta=0^{\circ}$]{\includegraphics[scale=0.35]{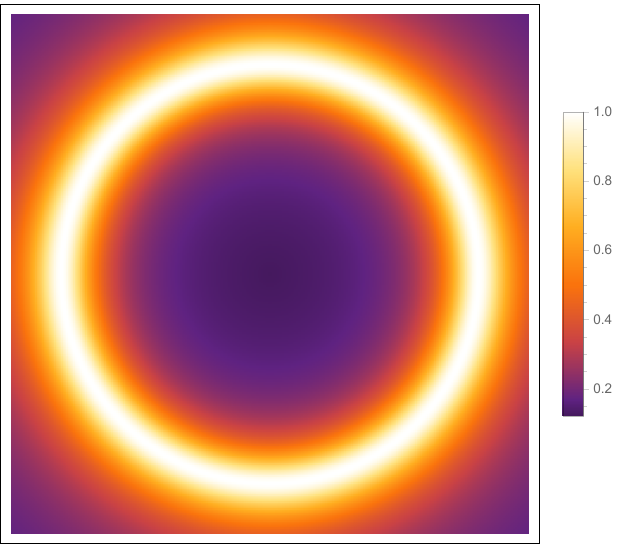}}
\subfigure[$\psi_0=0.082,\theta=30^{\circ}$]{\includegraphics[scale=0.35]{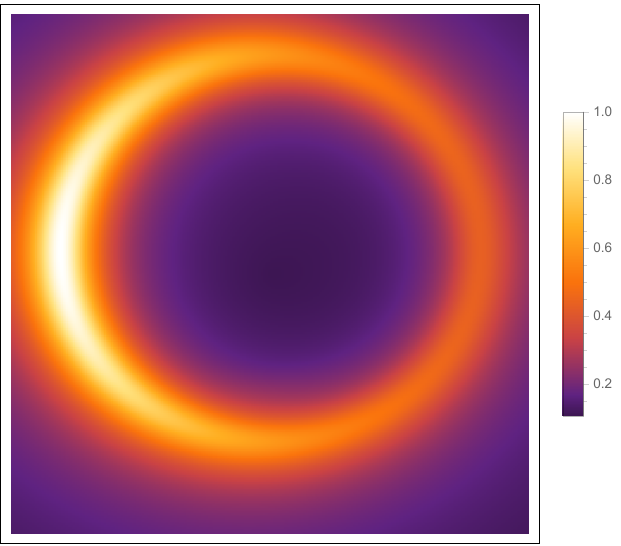}}
\subfigure[$\psi_0=0.082,\theta=60^{\circ}$]{\includegraphics[scale=0.35]{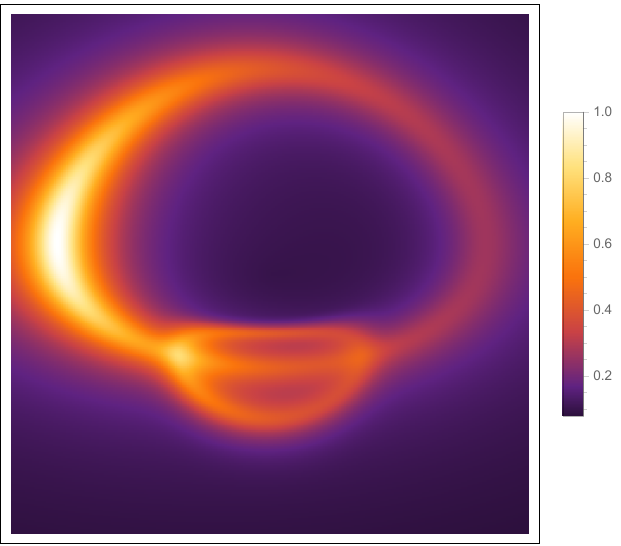}}
\subfigure[$\psi_0=0.082,\theta=75^{\circ}$]{\includegraphics[scale=0.35]{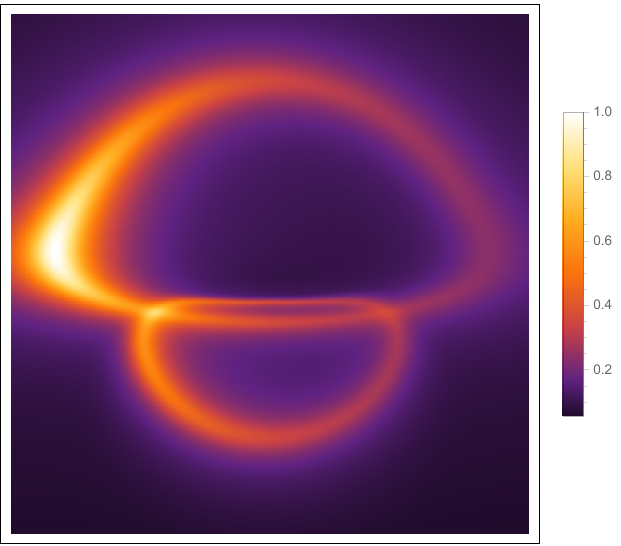}}
\subfigure[$\psi_0=0.085,\theta=0^{\circ}$]{\includegraphics[scale=0.35]{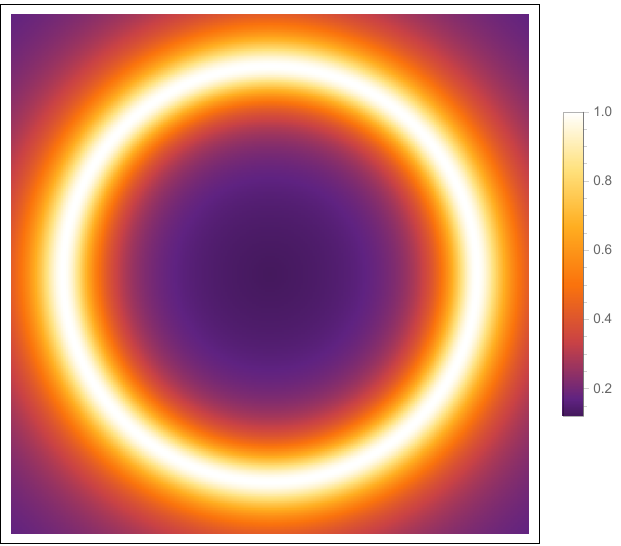}}
\subfigure[$\psi_0=0.085,\theta=30^{\circ}$]{\includegraphics[scale=0.35]{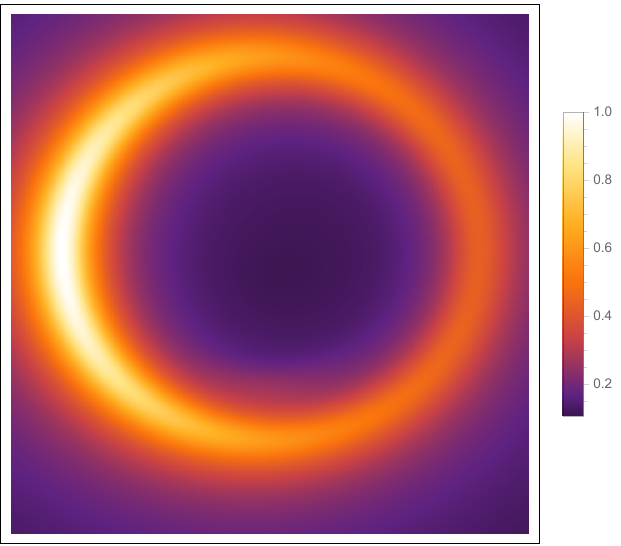}}
\subfigure[$\psi_0=0.085,\theta=60^{\circ}$]{\includegraphics[scale=0.35]{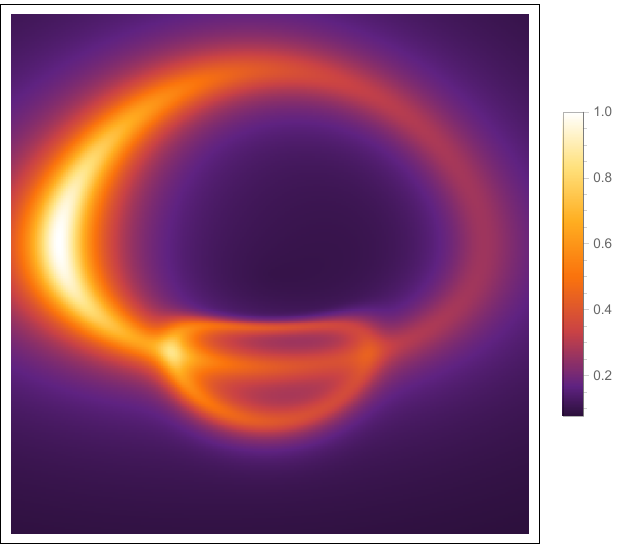}}
\subfigure[$\psi_0=0.085,\theta=75^{\circ}$]{\includegraphics[scale=0.35]{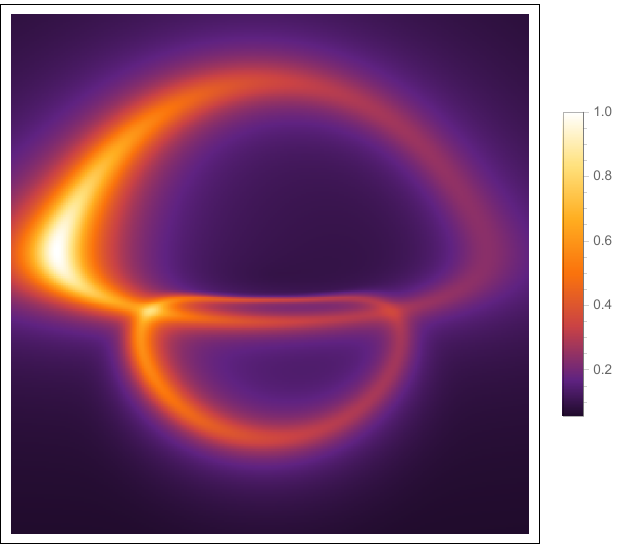}}
\subfigure[$\psi_0=0.087,\theta=0^{\circ}$]{\includegraphics[scale=0.35]{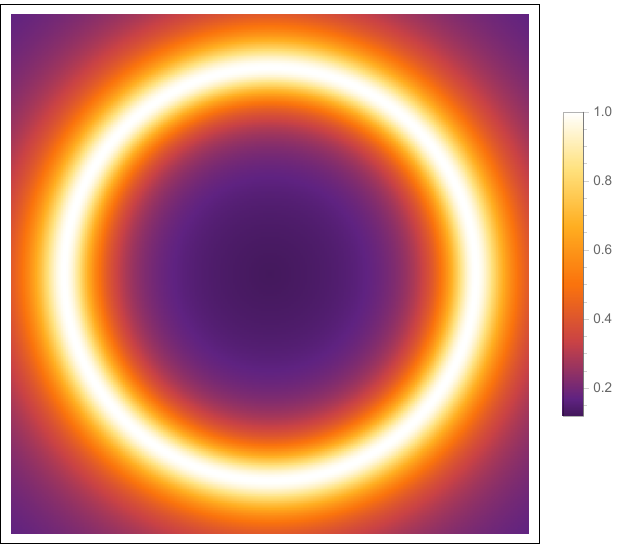}}
\subfigure[$\psi_0=0.087,\theta=30^{\circ}$]{\includegraphics[scale=0.35]{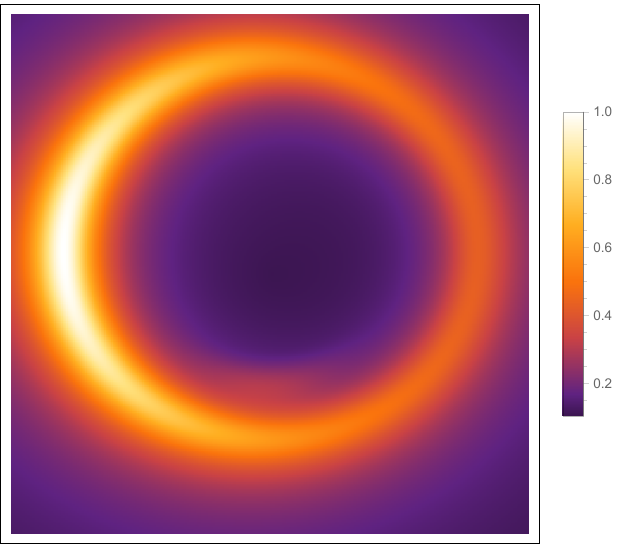}}
\subfigure[$\psi_0=0.087,\theta=60^{\circ}$]{\includegraphics[scale=0.35]{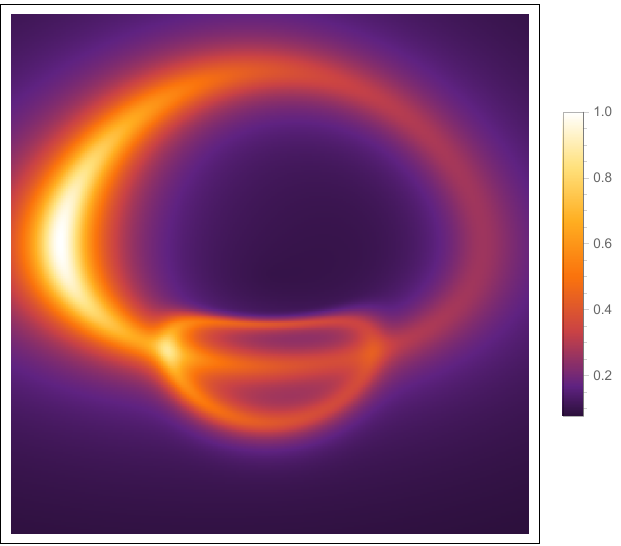}}
\subfigure[$\psi_0=0.087,\theta=75^{\circ}$]{\includegraphics[scale=0.35]{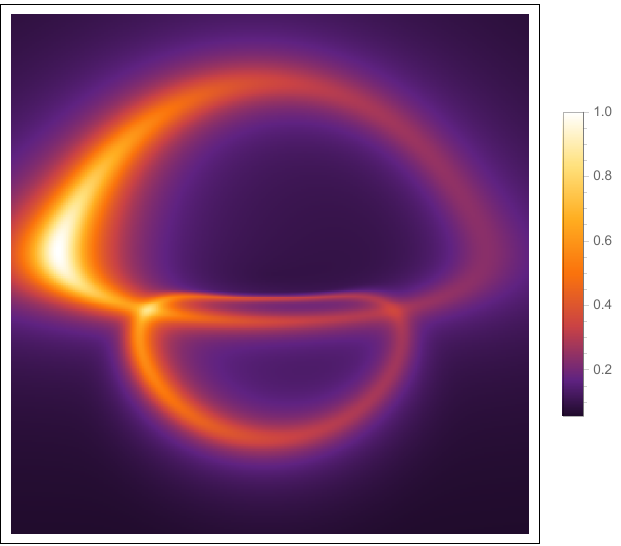}}
\caption{\label{fig6}  Optical images of boson stars illuminated by a thin accretion disk for different initial scalar field values $\psi_0$, with the angle of field of view $\beta_{\mathrm{fov}}=10^{\circ}$, the coupling parameter $\alpha=0.9$, and the observer at $r_{\mathrm{obs}}=200$. Each row from top to bottom corresponds to $\psi_0=0.08, 0.082, 0.085, 0.087$, while each column from left to right corresponds to observer inclination angles $\theta=0^{\circ}, 30^{\circ}, 60^{\circ}, 75^{\circ}$.}
\end{figure}
In Figure \ref{fig6}, we present the optical images of boson stars illuminated by a thin accretion disk as the light source. The angle of the field of view is set to $\beta_{\mathrm{fov}}=10^{\circ}$, the coupling parameter is determined as $\alpha=0.9$, and the observer distance is established at $r_{\mathrm{bos}}=200$. It is important to note that in the optical images of compact stars, light rays passing through the accretion disk a different number of times may generate distinct images. This occurs because the intensity of the light accumulates with each passage through the accretion disk.
Specifically, the primary image  (direct image) corresponds to the situation where $n = 1$, the secondary image (lens image) is linked to $n = 2$, while three or more intersections indicate higher-order images (photon ring).
In previous literature, the concept of the photon ring has been defined in different ways. For instance, according to Refs.~\cite{Gralla:2019drh,Hou:2022gge}, the photon ring encompasses all images situated outside the direct image. In contrast, Refs.~\cite{Gralla:2019xty,He:2024amh} use the term to specifically refer to those images located beyond both the primary and secondary images. In our study, we follow the definition provided in the latter references.

In the first column of Figure \ref{fig6}, the observer inclination is $\theta=0^{\circ}$, it can be observed that
the optical image of the boson star appears as a symmetric circular ring, corresponding to the direct image, with no lensed image or photon ring present. For the second column, the observer inclination is $\theta=30^{\circ}$, the bright ring structure still exists in the image, but it has undergone a slight deformation. As the value of $\psi_0$ increases, a clear upward trend is observed in the intensity on the left side of the bright ring in the image. When $\theta$ increases to $60^{\circ}$ (the third column), a noticeable lensed image emerges, and as $\psi_0$ increases, the direct image remains almost unchanged while the lensed image grows. In this case, both the direct and lensed images undergo significant deformation into a hat-shaped structure. When $\theta$ further increases to $75^{\circ}$ (the fourth column), the lensed image size increases significantly, and as $\psi_0$ continues to grow, the lensed image enlarges further. Although the initial scalar field $\psi_0$ does not significantly affect the size or shape of the direct image, it does exert a measurable influence on the observed intensity of that image, as demonstrated in Table 3. Table 3 presents the proportion of the direct image's ($n = 1$) observed intensity relative to the total observed intensity under varying initial scalar field conditions $\psi_0$. It can be observed that the intensity of direct imaging decreases as the value of $\psi_0$ increases.
At lower observation angles ($\theta=17^\circ, 30^\circ$), the observed flux is predominantly contributed by the direct image, especially at $\theta=17^\circ$, where the direct image accounts for more than 95$\%$ of the total observed intensity.
However, an increase in the observation inclination angle leads to a reduction in the proportion of the observed intensity of the direct image. For instance, when $\theta=60^\circ$, the proportion of the observed intensity of the direct image decreases by approximately 10$\%$ compared to the case when $\theta=30^\circ$.
\begin{table}[h]
\begin{center}
	{\footnotesize{\bf Table 3.} The  proportion of the observed intensity of the direct image ($n = 1$) to the total observed intensity under different initial scalar field $\psi_0$. Here, the value of coupling parameter is $\alpha=0.9$. } \\
	\vspace{3mm}
	\begin{tabular}{c|c|c|c|c}
		\hline
		\diagbox[width=2.5em, height=2.5em]{$\psi_0$}{$\theta$}& $17^\circ$ & $30^\circ$ & $60^\circ$ & $75^\circ$ \\ \hline
		0.080 & 98.7\% & 88.8\% & 72.2\% & 70.1\% \\ \hline
		0.082 & 98.3\% & 87.9\% & 71.9\% & 68.9\% \\ \hline
		0.085 & 97.6\% & 86.9\% & 71.6\% & 68.5\% \\ \hline
		0.087 & 96.6\% & 86.3\% & 71.4\% & 67.6\% \\ \hline
	\end{tabular}
\end{center}
\end{table}

\begin{figure}[!h]
\centering
\includegraphics[width=0.4\textwidth]{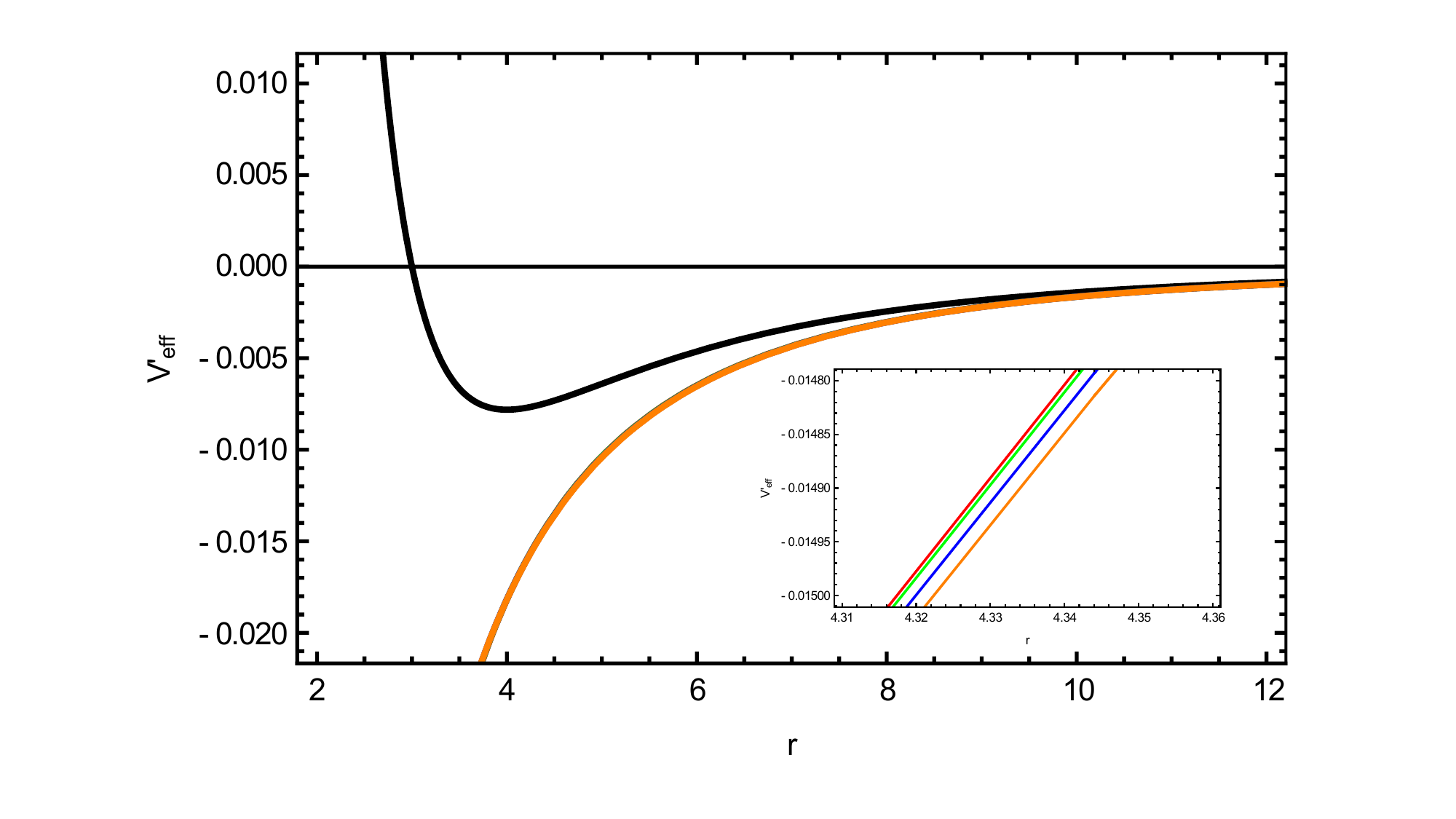}
\caption{The relationship between the first derivative of  effective potential $V'_{eff}$ and the radial coordinate $r$ when the parameter $\psi_0$ takes different values. The red, green, blue, and orange curves correspond to $\psi_0 = 0.08, 0.082, 0.085,$ and $0.087$, respectively. The black curve corresponds to the Schwarzschild black hole.} \label{fig4}
\end{figure}

\begin{figure}[!h]
	\centering
	\subfigure[$\theta=17^\circ$]{\includegraphics[scale=0.45]{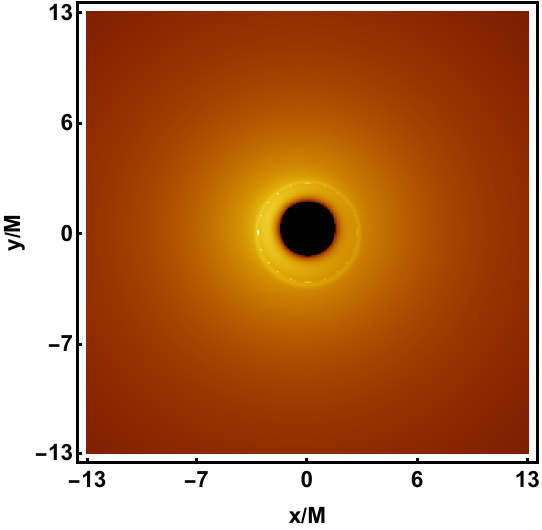}}
	\subfigure[$\theta=17^\circ$]{\includegraphics[scale=0.45]{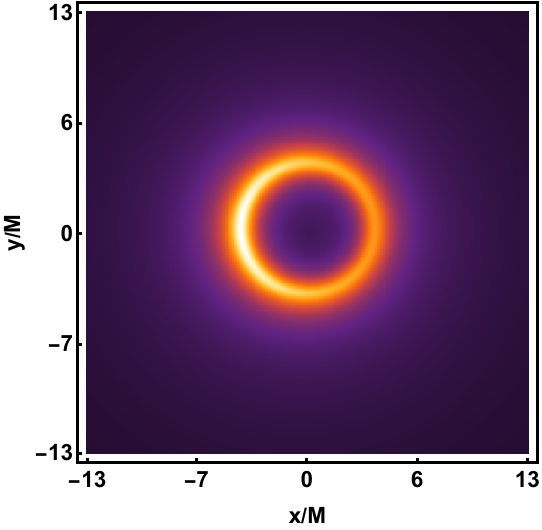}}
	\subfigure[$\theta=75^\circ$]{\includegraphics[scale=0.45]{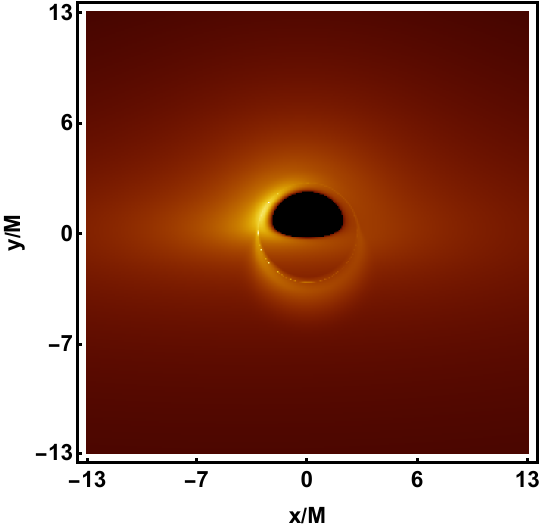}}
	\subfigure[$\theta=75^\circ$]{\includegraphics[scale=0.45]{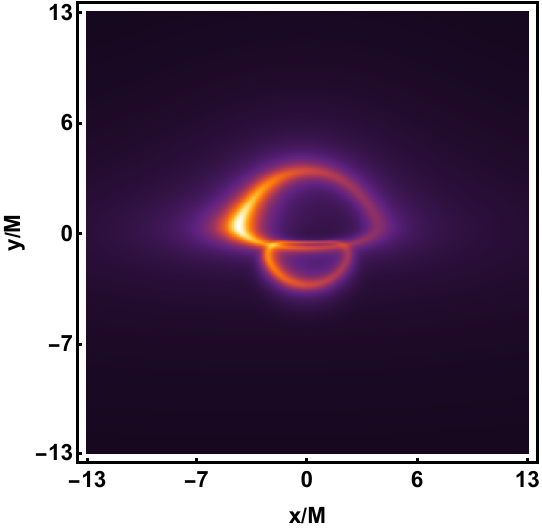}}
\caption{A comparison between the observed images of a boson star and a Schwarzschild black hole, both possessing the same field of view angle $\beta_{\mathrm{fov}}$ and an identical mass of $M = 0.603$. Panels (a) and (c) depict the Schwarzschild black hole, whereas panels (b) and (d) illustrate the boson star with an initial scalar field value of $\psi_0 = 0.08$. }
	\label{fig88}
\end{figure}

Interestingly, no photon ring structure was observed in any of the images, and a measurable level of light intensity was detected in the central region of the images. To further determine whether the photon ring exists in boson star spacetime, accordingto Eq. (\ref{veff}), we plot the first derivative $V_{eff}'(r)$ of the effective potential  in Figure \ref{fig4}. Among them, the black curve corresponds to the Schwarzschild black hole. The results indicate that, under varying initial scalar field $\psi_0$, the effective potential $ V'_{{eff}}(r) $ of the boson star exhibits a monotonic increase with respect to the radial coordinate $r$. As $r\rightarrow\infty$, the effective potential derivative $V'_{{eff}}(r)\rightarrow 0$ without crossing the horizontal axis. This implies that the equation $V'_{{eff}}(r)= 0$ does not admit any real solutions, meaning that no photon ring exists in the the boson star spcaetime. In contrast to the boson star, the first derivative of the effective potential for the Schwarzschild black hole vanishes at $r = 3M$, which corresponds to the location of the photon ring associated with the Schwarzschild black hole. When the mass of the boson star is equal to that of the Schwarzschild black hole, Figure~\ref{fig88} displays the corresponding optical observational images captured at  the same field of view $\beta_{\mathrm{fov}}$. Regardless of the observation angle, the Schwarzschild black hole consistently exhibits a photon ring, which is characterized as a notably bright yet extremely narrow axisymmetric circular feature. In contrast, such a structure is absent in boson stars. In addition, when the mass of the central compact object is the same, the diameter of the bright ring observed in the image of a boson star is consistently larger than that of the Schwarzschild black hole. It can be achieved by comparing Figure ~\ref{fig88}(a) with (b), or alternatively, Figure~\ref{fig88}(c) with (d).

\begin{figure}[!h]
\centering 
\subfigure[$\psi_0=0.08,\theta=0^{\circ}$]{\includegraphics[scale=0.35]{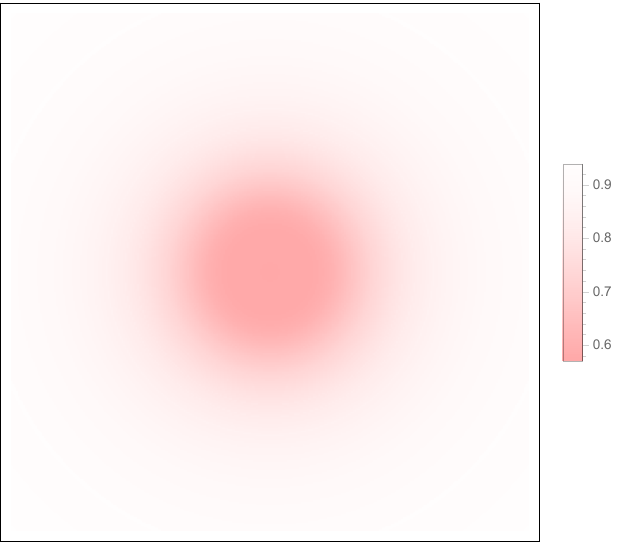}}
\subfigure[$\psi_0=0.08,\theta=30^{\circ}$]{\includegraphics[scale=0.35]{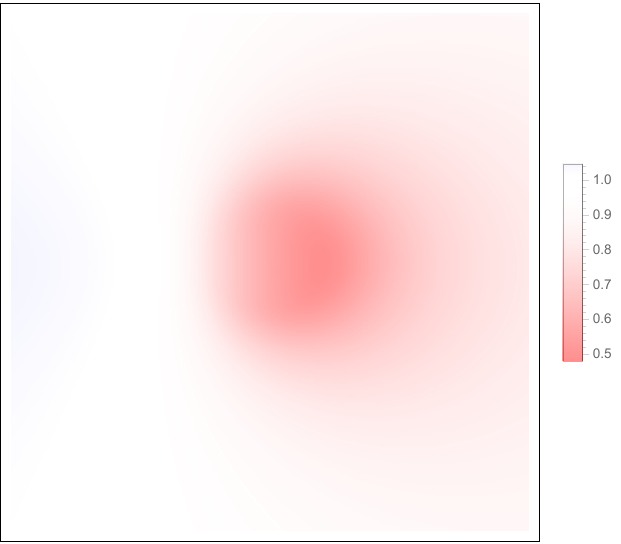}}
\subfigure[$\psi_0=0.08,\theta=60^{\circ}$]{\includegraphics[scale=0.35]{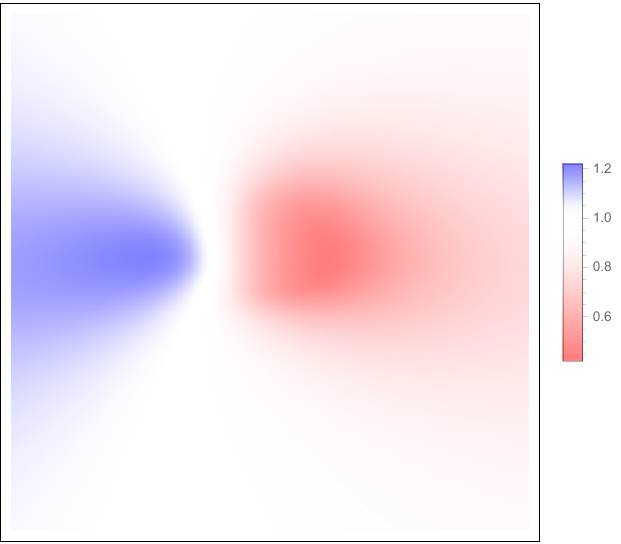}}
\subfigure[$\psi_0=0.08,\theta=75^{\circ}$]{\includegraphics[scale=0.35]{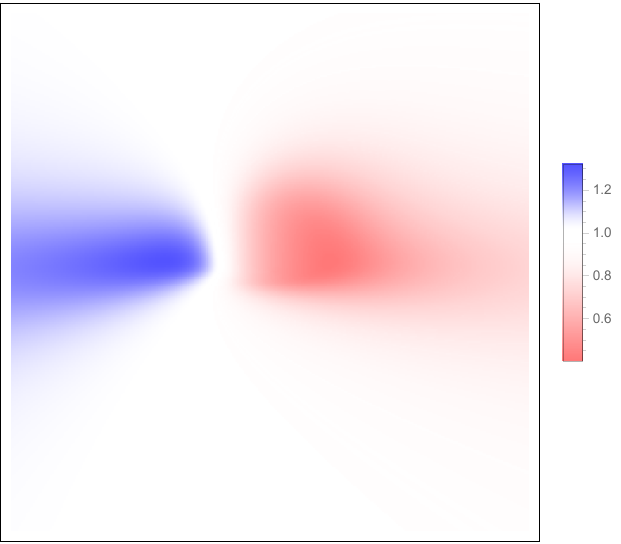}}
\subfigure[$\psi_0=0.082,\theta=0^{\circ}$]{\includegraphics[scale=0.35]{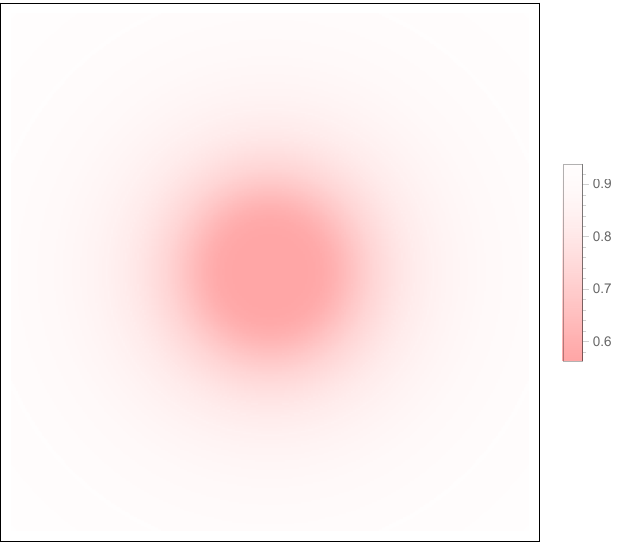}}
\subfigure[$\psi_0=0.082,\theta=30^{\circ}$]{\includegraphics[scale=0.35]{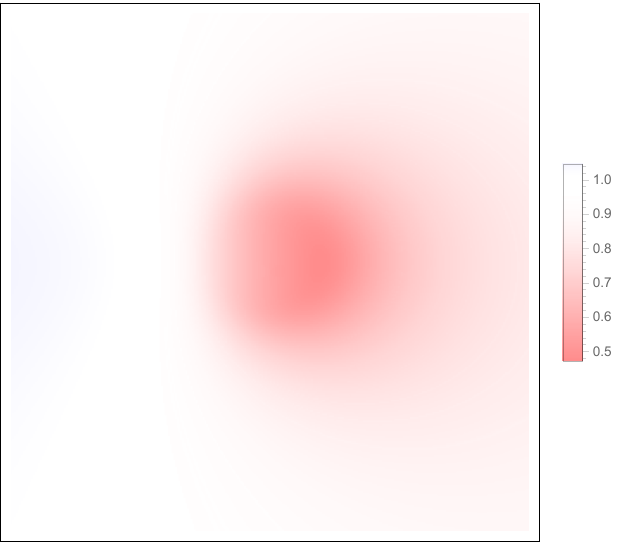}}
\subfigure[$\psi_0=0.082,\theta=60^{\circ}$]{\includegraphics[scale=0.35]{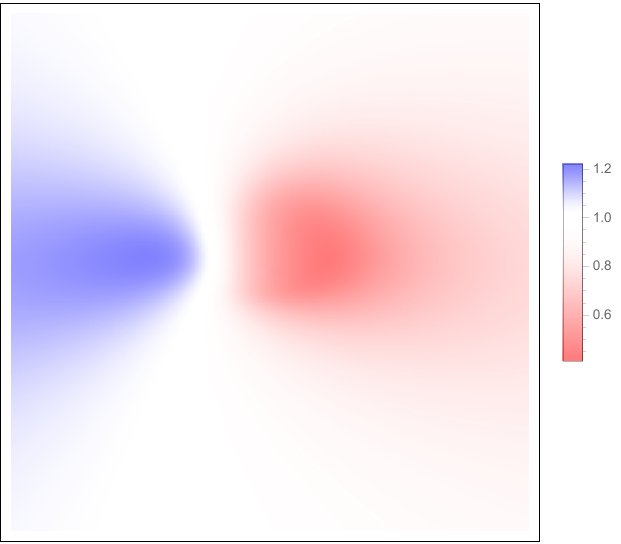}}
\subfigure[$\psi_0=0.082,\theta=75^{\circ}$]{\includegraphics[scale=0.35]{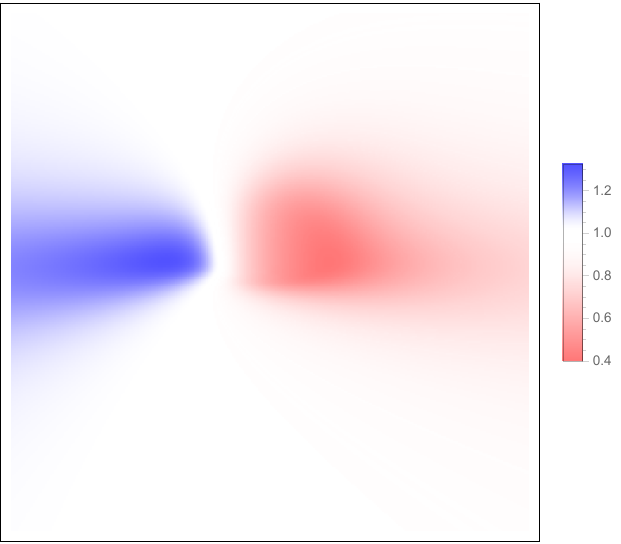}}
\subfigure[$\psi_0=0.085,\theta=0^{\circ}$]{\includegraphics[scale=0.35]{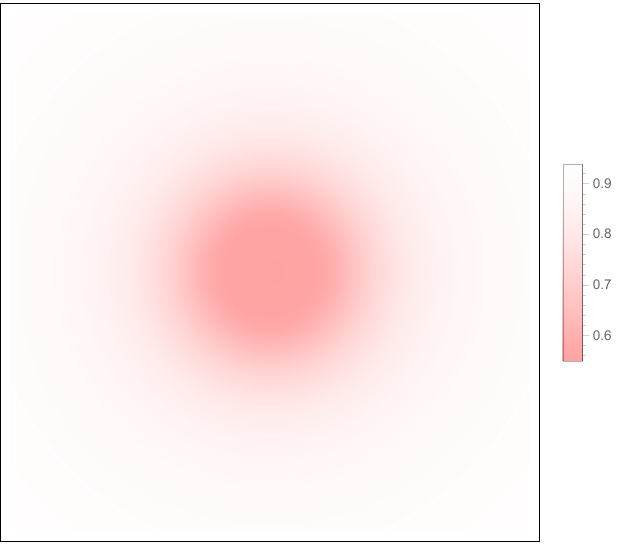}}
\subfigure[$\psi_0=0.085,\theta=30^{\circ}$]{\includegraphics[scale=0.35]{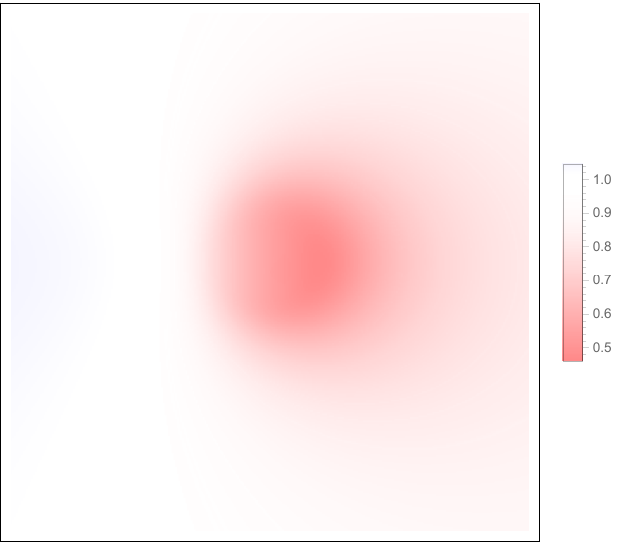}}
\subfigure[$\psi_0=0.085,\theta=60^{\circ}$]{\includegraphics[scale=0.35]{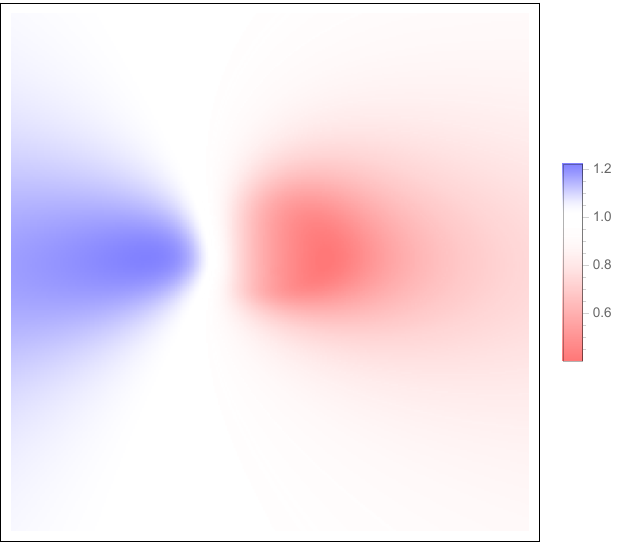}}
\subfigure[$\psi_0=0.085,\theta=75^{\circ}$]{\includegraphics[scale=0.35]{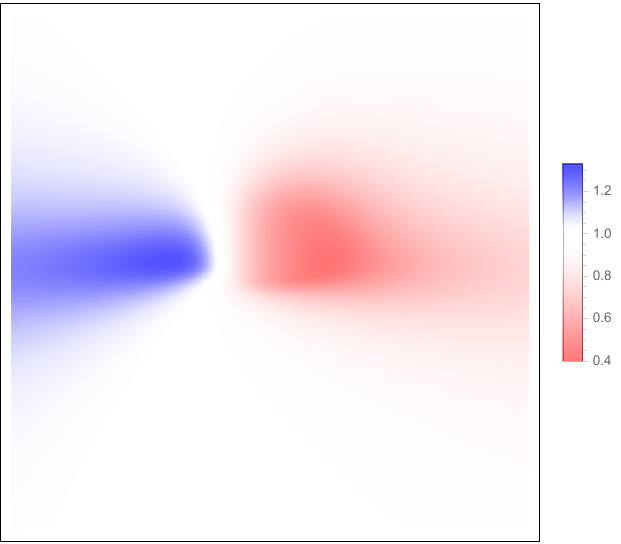}}
\subfigure[$\psi_0=0.087,\theta=0^{\circ}$]{\includegraphics[scale=0.35]{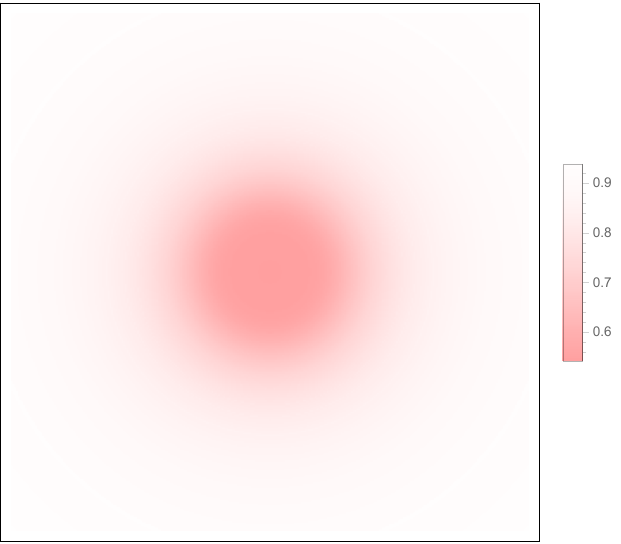}}
\subfigure[$\psi_0=0.087,\theta=30^{\circ}$]{\includegraphics[scale=0.35]{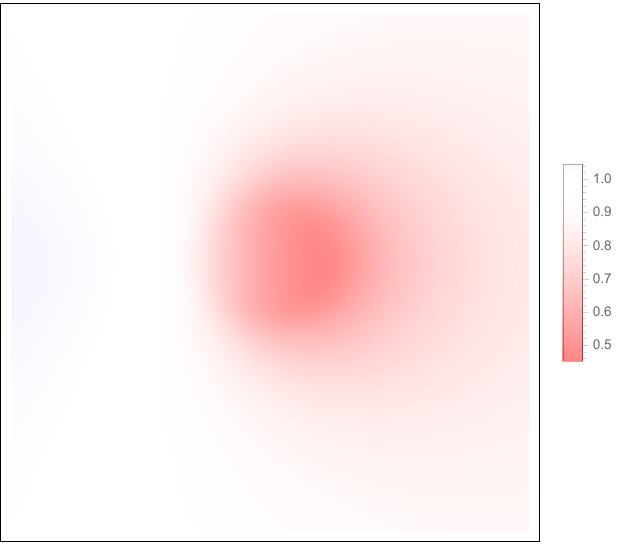}}
\subfigure[$\psi_0=0.087,\theta=60^{\circ}$]{\includegraphics[scale=0.35]{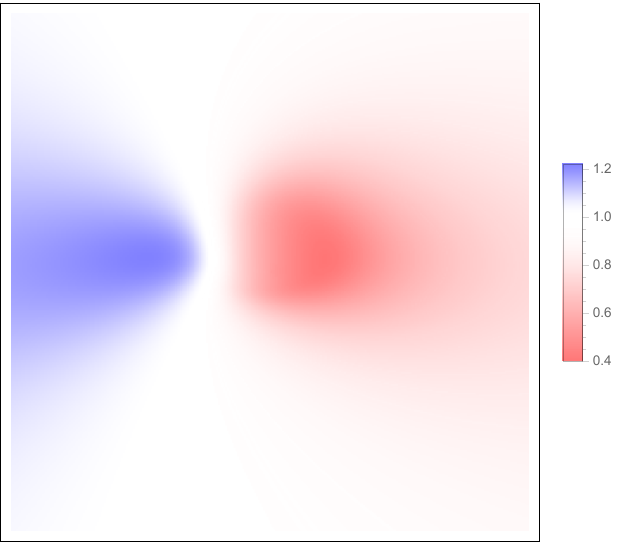}}
\subfigure[$\psi_0=0.087,\theta=75^{\circ}$]{\includegraphics[scale=0.35]{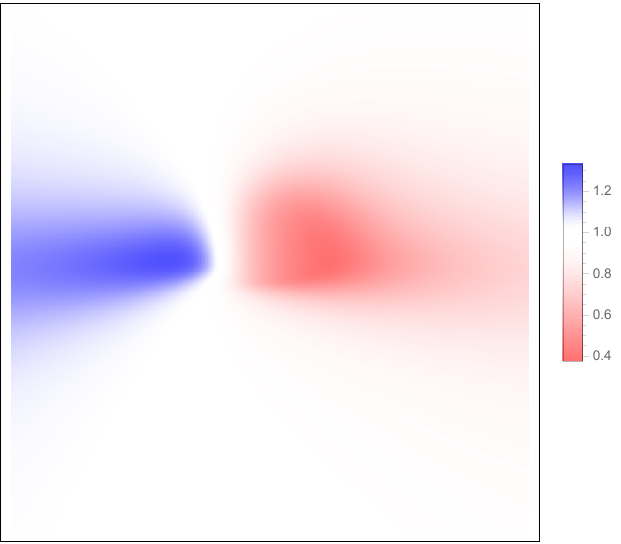}}
\caption{\label{fig7} The redshift factor distribution corresponding to the direct image under different initial scalar field $\psi_0$, with the coupling parameter $\alpha=0.9$. Red and blue colors represent redshift and blueshift, respectively.}
\end{figure}

The redshift distribution corresponding to the optical observation appearance in Figure \ref{fig6} is shown in Figure \ref{fig7}. In these images, deeper colors indicate stronger redshifts or blueshifts, adhering to a linear relationship. It can be observed from the images that when the observer inclination approaches $\theta \to 0^{\circ}$ (the first column), only redshift is present, with no blueshift, and the redshift exhibits a symmetric distribution. This is because the observer's line of sight is perpendicular to the equatorial plane and coincides with the symmetry axis of the accretion disk and the boson star. Hence, the radial velocity of each point on the accretion disk relative to the observer approaches zero, with only the tangential velocity contributing significantly. Consequently, the observed redshift is predominantly attributed to gravitational redshift. The initial scalar field $\psi_0$ does not significantly affect the degree of redshift. When $\theta=30^{\circ}$ (the second column), a slight blueshift appears on the left side of the image, causing the overall redshift to shift towards the right. When $\theta$ is larger (the third and fourth columns), a distinct blueshift appears on the left side of the image, with a sharp protrusion at the right endpoint of the blueshift region. As $\theta$ increases, the blueshift region becomes more concentrated. By comparing the images in each column, it can be seen that when $\theta$ is small, gravitational redshift dominates; however, as $\theta$ increases, the radial velocity component of the accreting material grows, leading to an increasing contribution from Doppler effect.

\subsection{Observational characteristics of boson stars in the weak coupling regime}
\begin{figure}[htbp]
	\centering
	\includegraphics[width=.4\textwidth]{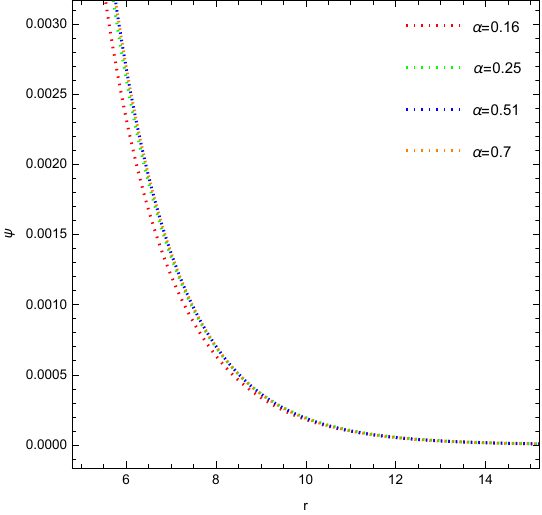}
	\caption{Variation of the scalar field $\psi$ as a function of the radial coordinate $r$ under weak coupling conditions, with the initial scalar field value set to $\psi_0 = 0.08$.\label{fig8}}
\end{figure}

\begin{figure}[h]
	\centering
	\includegraphics[width=.34\textwidth]{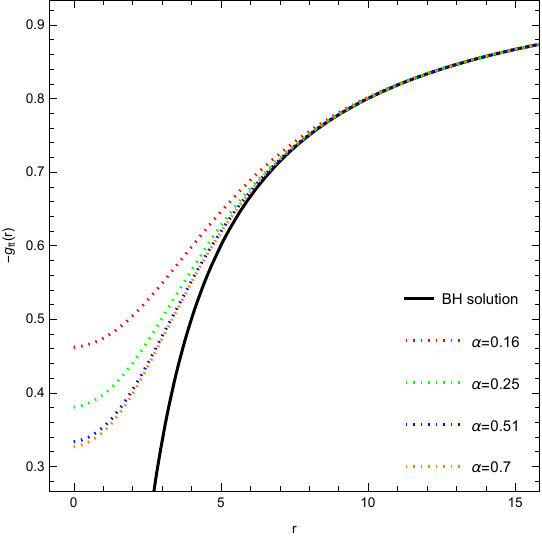}
	\qquad
	\includegraphics[width=.34\textwidth]{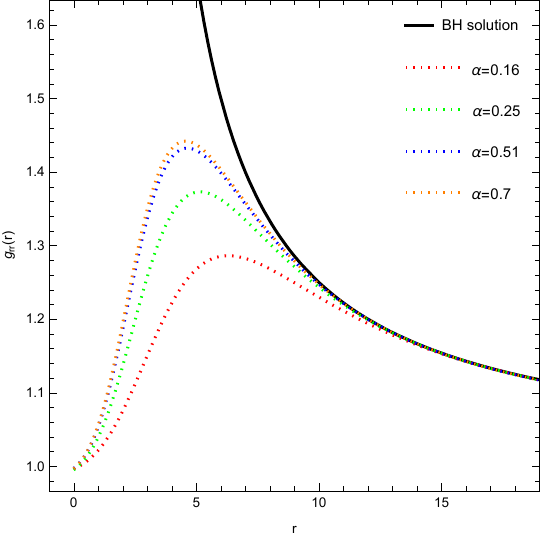}
	\caption{Comparison between the numerical metrics and the Schwarzschild black hole metric components under weak coupling, with the initial scalar field $\psi_0=0.08$.}
	\label{fig9}

\end{figure}

By fixing the initial scalar field value at $\psi_0 = 0.08$, we proceed to examine the evolution of the optical observational characteristics of the boson star under conditions of a larger coupling parameter, which corresponds to the weak coupling regime. The numerical results of $\psi$ for different values of $\alpha$ are shown in Figure \ref{fig8}. It can be observed that even when coupling parameter increases from $\alpha=0.16$ to $\alpha=0.7$, the overall variation pattern of  $\psi$ remains consistent, exhibiting a continuous decreasing trend with respect to $r$ and ultimately approaching zero $\psi\rightarrow0$. The numerical results of the metric components $-g_{tt}$ and $g_{rr}$ are presented in Figure \ref{fig9}, where the Schwarzschild metric components are plotted as black solid lines for comparison.
For a fixed $r$, as $\alpha$ increases, $-g_{tt}$ decreases while $g_{rr}$ increases. Unlike the Schwarzschild metric components, the metric components of the boson star do not diverge as $r \to 0$, whereas the Schwarzschild metric components tend to infinity. This distinction reflects the presence of an event horizon in the Schwarzschild black hole, whereas the boson star lacks an event horizon. On the other hand, as $r \to \infty$, the metric components $-g_{tt}$ and $g_{rr}$ of the boson star gradually approach those corresponding to the Schwarzschild metric, thereby ensuring the asymptotic flatness of the spacetime geometry associated with the boson star at spatial infinity.

\begin{figure}[h]
	\centering
	\includegraphics[width=.34\textwidth]{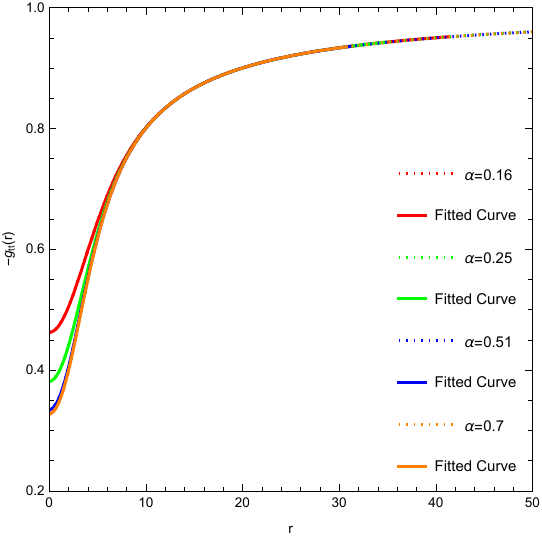}
	\qquad
	\includegraphics[width=.34\textwidth]{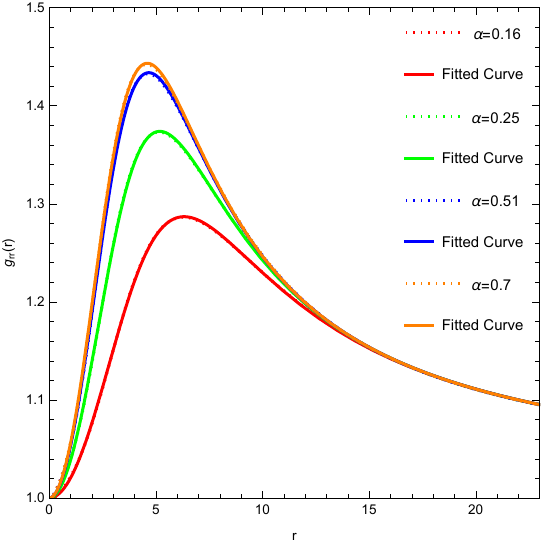}
	\caption{Comparison between the numerical metrics and the fitting functions under weak coupling, with the initial scalar field $\psi_0=0.08$. The dashed lines represent the numerical metrics, while the solid lines represent the fitting functions. \label{fig10}}
\end{figure}

\begin{figure}[htbp]
	\centering
	\includegraphics[width=.4\textwidth]{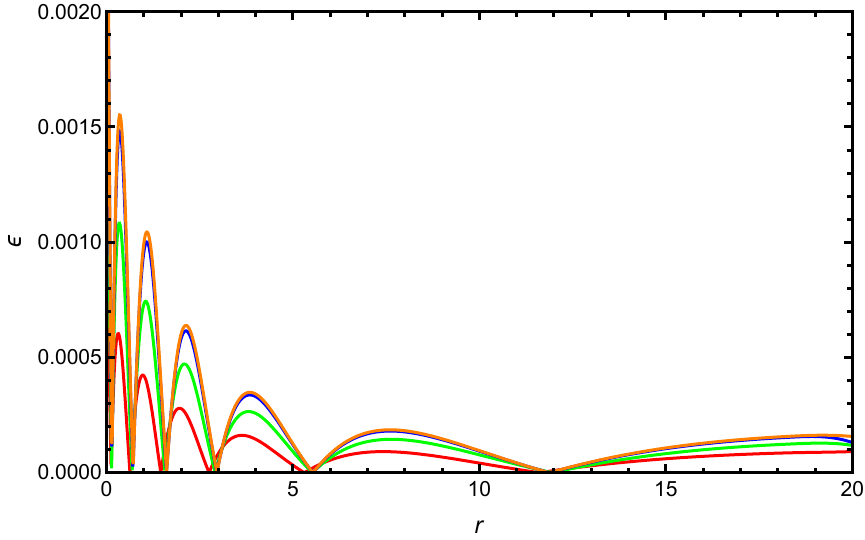}
	\qquad
	\includegraphics[width=.4\textwidth]{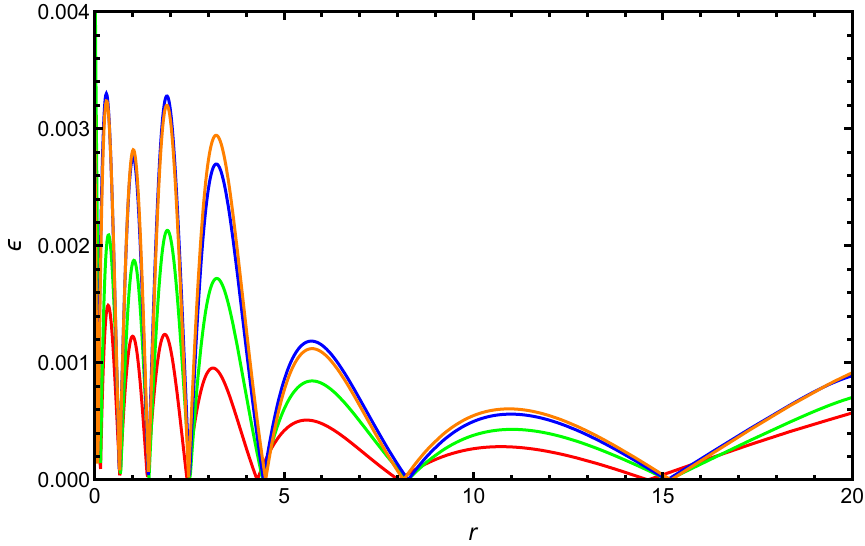}
	\caption{ The relative error between the numerical metric and the fitting metric, where the left panel corresponds to the metric component $-g_{tt}$, and the right panel corresponds to the metric component $g_{rr}$. The red, green, blue, and orange curves correspond to $\alpha = 0.16, 0.25, 0.51,$ and $0.7$, respectively, respectively.} \label{fig10_2}
	
\end{figure}

Similarly, we continue to employ the formulations presented in Eqs.(\ref{fit1}) and (\ref{fit2}) to fit the numerical metric under the weak coupling condition. The fitting results are presented in Figure~\ref{fig10}, where the dashed lines correspond to the numerical metric components, and the solid lines represent the fitted functions. When coupling parameter $\alpha$ takes different values, the numerical metric component $-g_{tt}$ decreases with increasing $\alpha$ , whereas the value of $g_{rr}$ increases correspondingly. This change aligns with the effect induced by the initial scalar field $\psi_0$. However, as the value of $r$ increases, all of them gradually converge to the metric components of the Schwarzschild black hole. As $r \to \infty$, the fitting functions approach unity asymptotically, thereby fulfilling the condition of asymptotic flatness.  Figure  \ref{fig10_2} illustrates the relative error $\epsilon$ between the numerical metric  component and the fitted metric component as the value of $\alpha$ varies. As the value of parameter $\alpha$ increases, the relative error associated with the fitting result of the metric component $-g_{tt}$  exhibits a slight upward trend, as well as metric component $g_{rr}$. When $\alpha = 0.7$, the fitting error of $-g_{tt}$ is $\epsilon \sim 0.0016 $, while that of $g_{rr}$ is  $\epsilon \sim 0.0034 $. Since the relative deviation of the metric components' fitting results is consistently lower than both of these values when $\alpha < 0.7$, the fitting results fall within an acceptable range. The parameter estimates of the fitting functions are presented in Table 4 and Table 5. Additionally, the coupling parameter $\alpha$ is found to be inversely related to the boson star mass $m(r)$.


\begin{table}[h]
\begin{center}
	{\footnotesize{\bf Table 4.}
	Parameter estimates of $p_i$ $(i = 1, \dots, 7)$ for the metric component $g_{tt}$ under weak coupling conditions, with the initial scalar field value $\psi_0 = 0.08$. Here, $m(r)$ represents the mass of the boson star.} \\
	\vspace{2mm}
	\begin{tabular}{c|c|c|c|c|c|c|c|c|c}
		\hline
		Type & $\alpha$ & $m(r)$ & $p_1$ & $p_2$ & $p_3$ & $p_4$ & $p_5$ & $p_6$ & $p_7$ \\ \hline
		SBS1 & 0.16 & 0.433 & 0.106& 0.124& -0.381& -0.043& -0.063& -0.009& -0.061 \\ \hline
		SBS2 & 0.25 &0.534  &0.056& 0.114& -0.373& -0.025& -0.054& -0.008& -0.071\\ \hline
		SBS3 & 0.51 & 0.591 &0.03& 0.113& -0.368& -0.015& -0.052& -0.007& -0.078\\ \hline
		SBS4 & 0.7 & 0.599  &0.026& 0.113& -0.368& -0.014& -0.051& -0.007& -0.079\\ \hline
	\end{tabular}
\end{center}
\end{table}

\begin{figure}[!h]
\centering 
\subfigure[$\alpha=0.16,\theta=45^{\circ}$]{\includegraphics[scale=0.3]{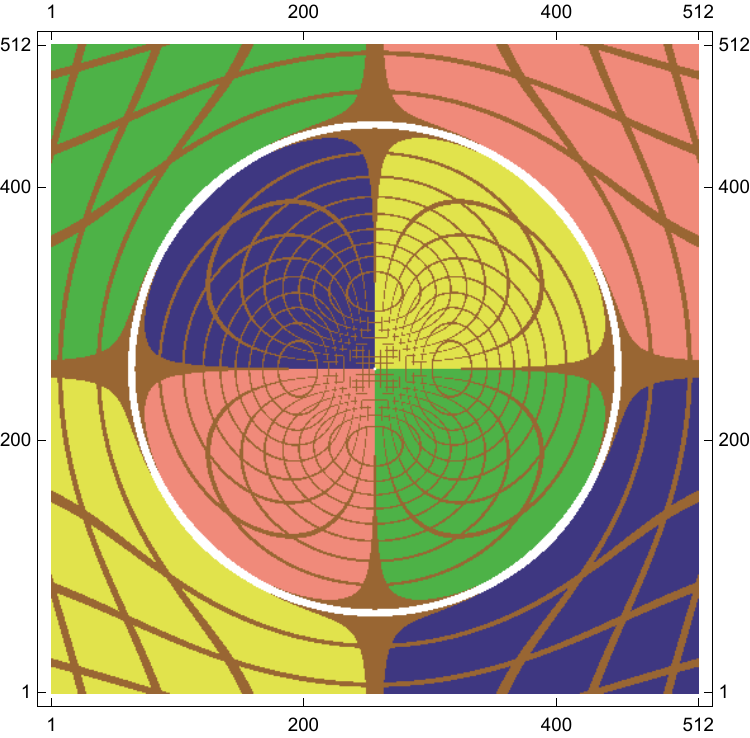}}
\subfigure[$\alpha=0.25,\theta=45^{\circ}$]{\includegraphics[scale=0.3]{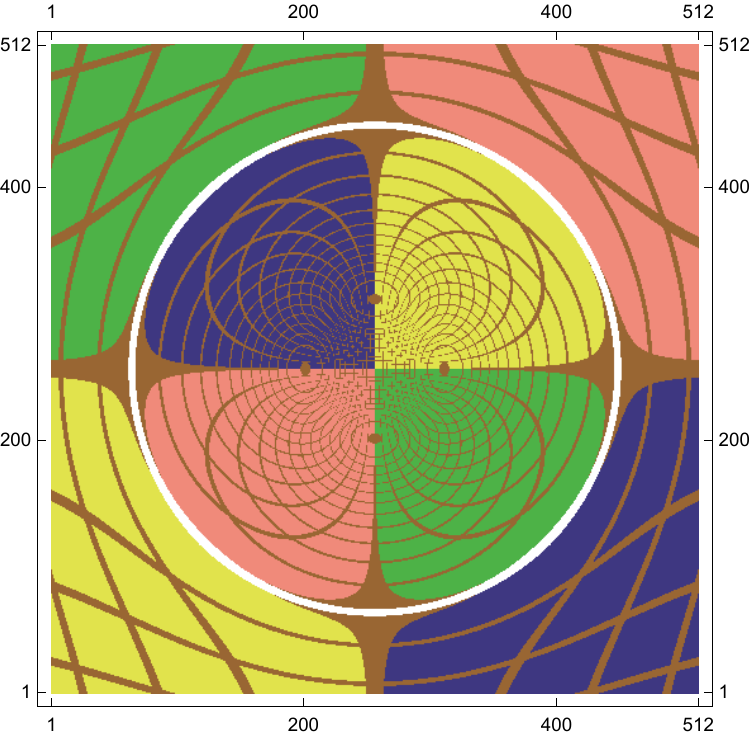}}
\subfigure[$\alpha=0.51,\theta=45^{\circ}$]{\includegraphics[scale=0.3]{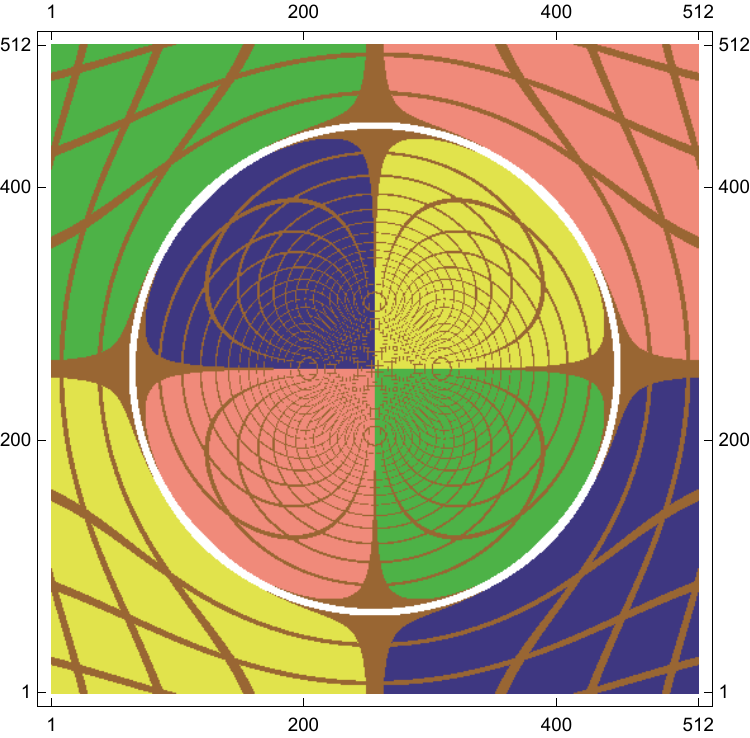}}
\subfigure[$\alpha=0.7,\theta=45^{\circ}$]{\includegraphics[scale=0.3]{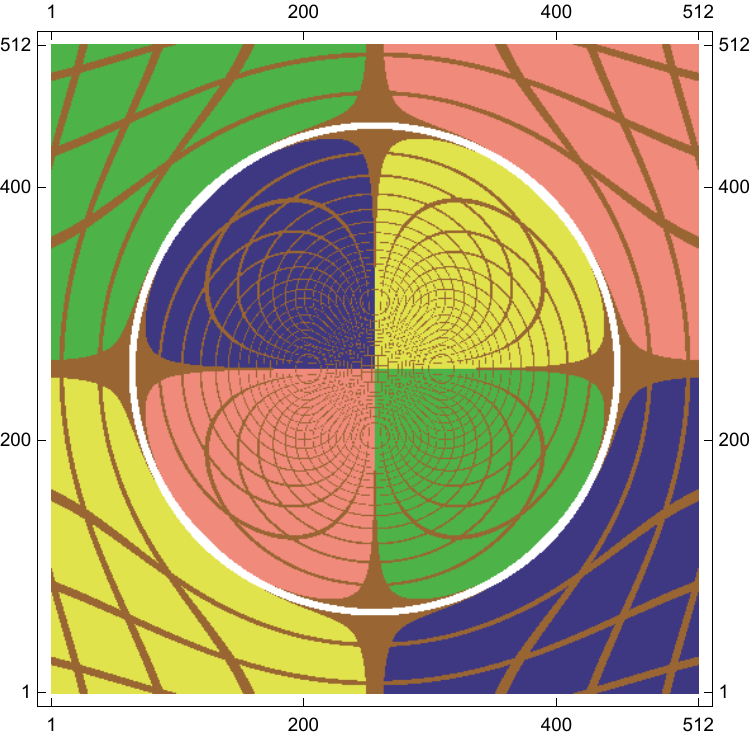}}
\caption{\label{fig12} Optical images of boson stars illuminated by a spherical light source under weak coupling, with the initial scalar field $\psi_0=0.08$, the angle of field of view $\beta_{\mathrm{fov}}=45^{\circ}$, and the observer at $r_{\mathrm{obs}}=50m(r)$, where $m(r)$ represents the mass of the boson star.  }
\end{figure}

\begin{table}[h]
\begin{center}
	{\footnotesize{\bf Table 5.}
		Parameter estimates of $q_i$ $(i = 1, \dots, 7)$ for the metric component $g_{rr}$ under weak coupling conditions, with the initial scalar field value $\psi_0 = 0.08$. Here, $m(r)$ represents the mass of the boson star.} \\
	\vspace{2mm}
	\begin{tabular}{c|c|c|c|c|c|c|c|c|c}
		\hline
		Type & $\alpha$ & $m(r)$ & $q_1$ & $q_2$ & $q_3$ & $q_4$ & $q_5$ & $q_6$ & $q_7$ \\ \hline
		SBS1 & 0.16 & 0.433 & -10.343& -1.522&4.217& 0.604& 1.028& 0.026& 0.017 \\ \hline
		SBS2 & 0.25 & 0.534 &-10.251& -1.35& 4.407& 0.546& 0.999& 0.025& 0.023\\ \hline
		SBS3 & 0.51 & 0.591 &-17.42& -2.308& 6.593& 1.043& 1.531& 0.031& 0.019\\ \hline
		SBS4 & 0.7 &0.599 &-15.783& -2.075& 6.122& 0.908& 1.414& 0.03& 0.021\\ \hline
	\end{tabular}
\end{center}
\end{table}
The optical observation image of the soliton boson star wrapped by celestial light sources under different values of $\alpha$ (the weak coupling regime), as illustrated in Figure \ref{fig12}.
Here, the initial scalar field is set to $\psi_0 = 0.08$, the observer's inclination angle is specified as $\beta_{\mathrm{fov}} = 45^{\circ}$, and the observer distance is defined as $r_{\mathrm{obs}} = 50 \, m(r)$, where $m(r)$ represents the mass of the boson star. It can be observed from these images that boson stars lack an event horizon, allowing light to penetrate their interiors, which results in the appearance of pixels within the central region of the image. Since the boson star is non-rotating, the color background within the Einstein ring (the white circular ring) exhibits a symmetric and undistorted distribution, indicating the absence of a frame-dragging effect. However, as the value of $\alpha$ increases, the distribution and morphology of the brown patterns within the Einstein ring undergo noticeable changes. This indicates that variations in $\alpha$ influence the structure of spacetime geometry, thereby altering the gravitational lensing effect.

\begin{figure}[!h]
\centering 
\subfigure[$\alpha=0.16,\theta=0^{\circ}$]{\includegraphics[scale=0.35]{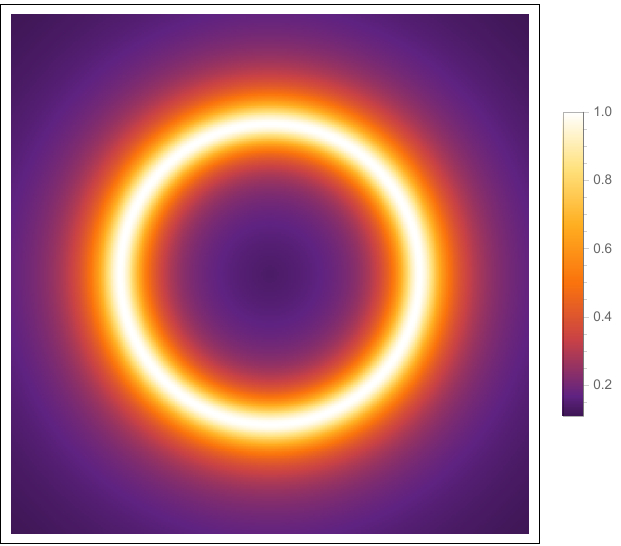}}
\subfigure[$\alpha=0.16,\theta=30^{\circ}$]{\includegraphics[scale=0.35]{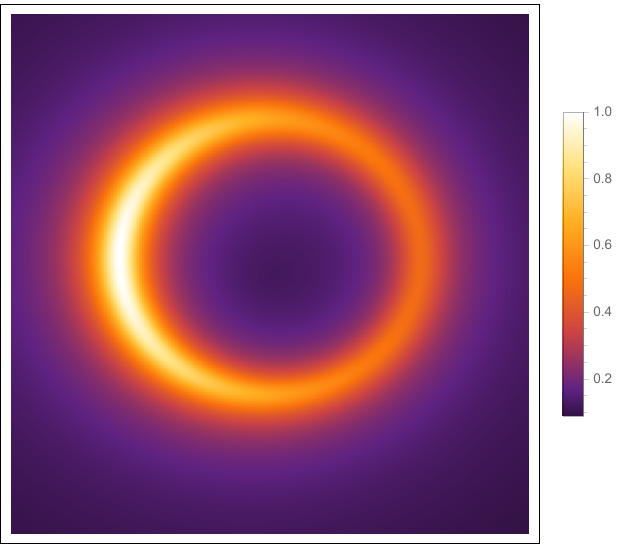}}
\subfigure[$\alpha=0.16,\theta=60^{\circ}$]{\includegraphics[scale=0.35]{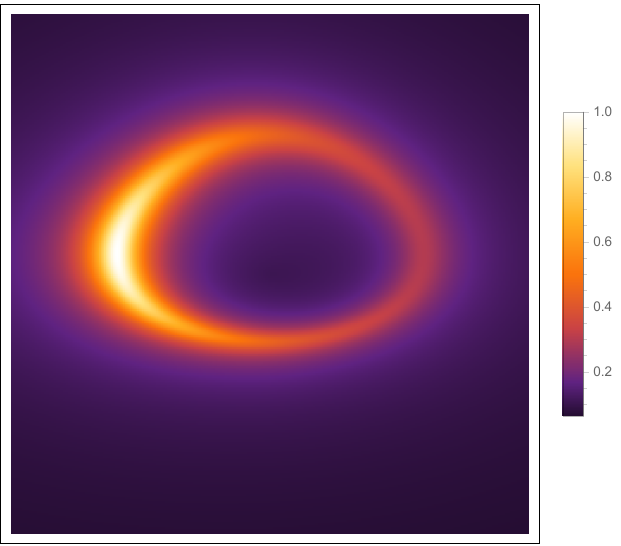}}
\subfigure[$\alpha=0.16,\theta=75^{\circ}$]{\includegraphics[scale=0.35]{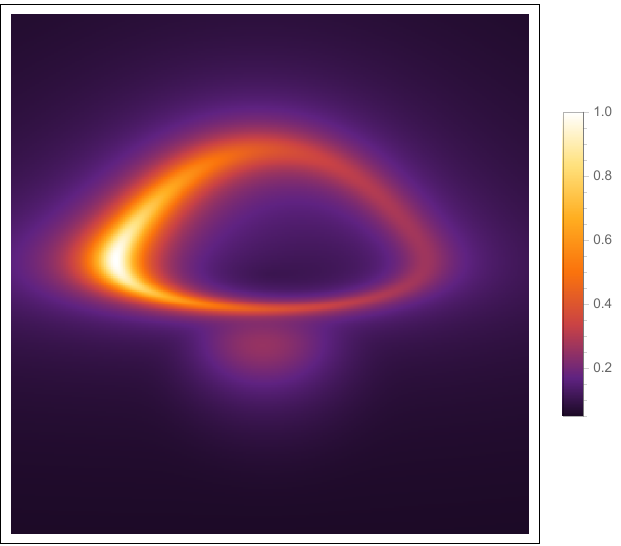}}
\subfigure[$\alpha=0.25,\theta=0^{\circ}$]{\includegraphics[scale=0.35]{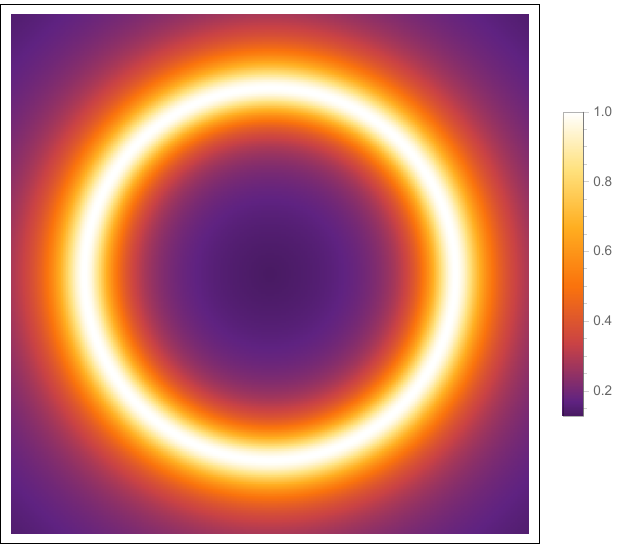}}
\subfigure[$\alpha=0.25,\theta=30^{\circ}$]{\includegraphics[scale=0.35]{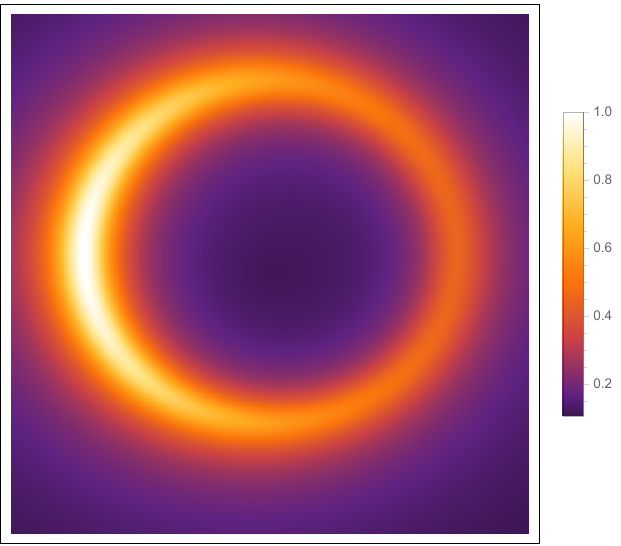}}
\subfigure[$\alpha=0.25,\theta=60^{\circ}$]{\includegraphics[scale=0.35]{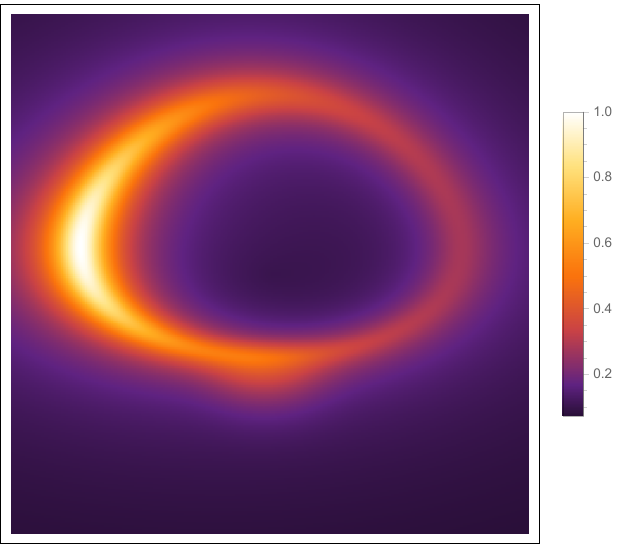}}
\subfigure[$\alpha=0.25,\theta=75^{\circ}$]{\includegraphics[scale=0.35]{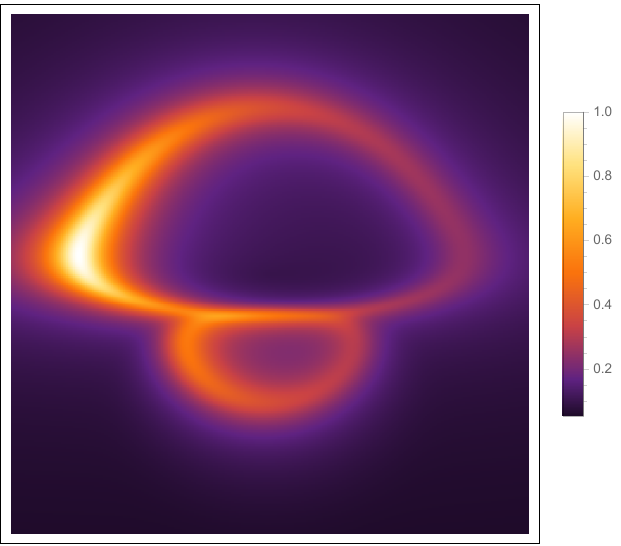}}
\subfigure[$\alpha=0.51,\theta=0^{\circ}$]{\includegraphics[scale=0.35]{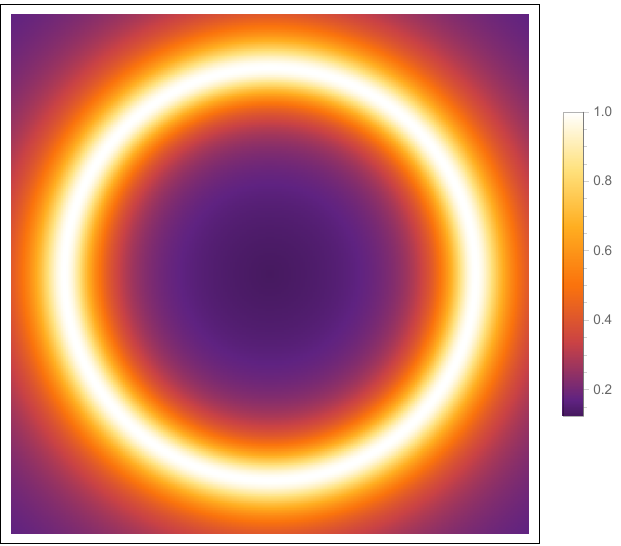}}
\subfigure[$\alpha=0.51,\theta=30^{\circ}$]{\includegraphics[scale=0.35]{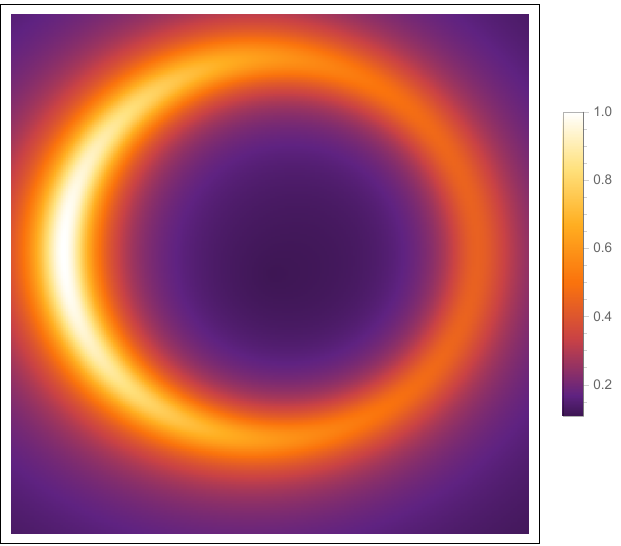}}
\subfigure[$\alpha=0.51,\theta=60^{\circ}$]{\includegraphics[scale=0.35]{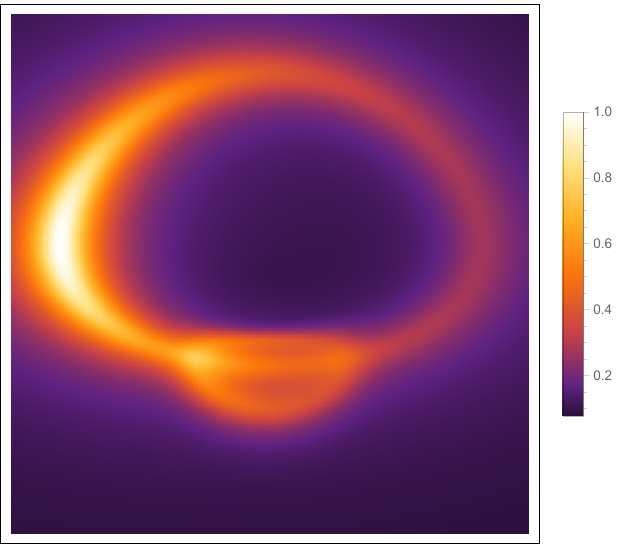}}
\subfigure[$\alpha=0.51,\theta=75^{\circ}$]{\includegraphics[scale=0.35]{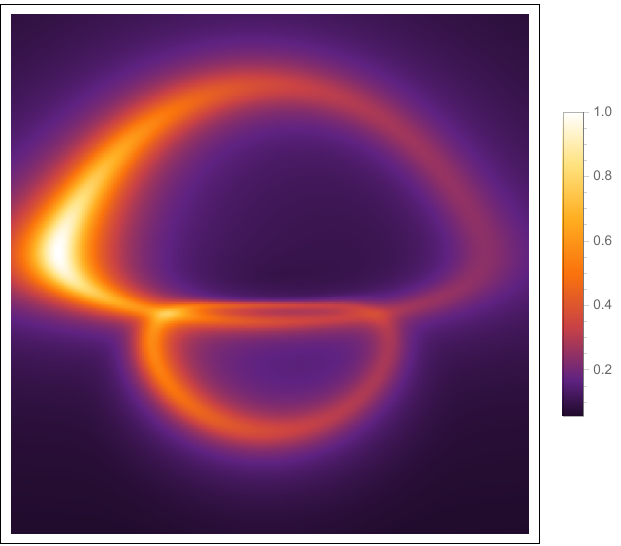}}
\subfigure[$\alpha=0.7,\theta=0^{\circ}$]{\includegraphics[scale=0.35]{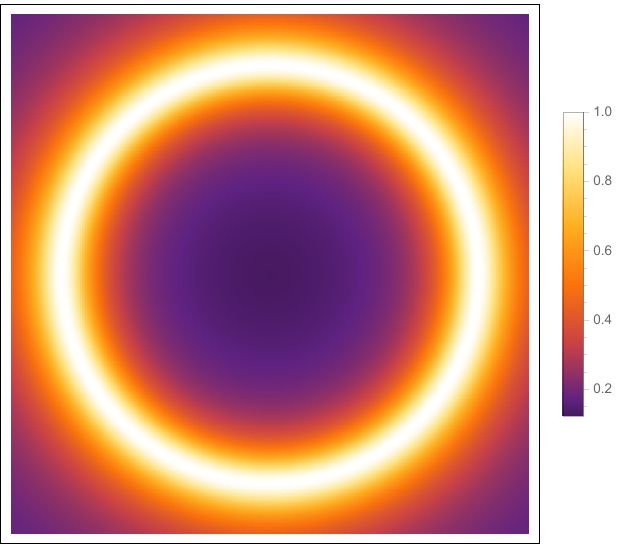}}
\subfigure[$\alpha=0.7,\theta=30^{\circ}$]{\includegraphics[scale=0.35]{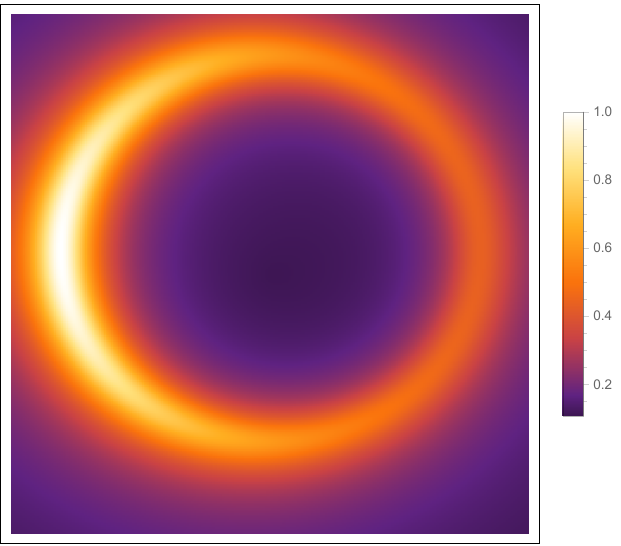}}
\subfigure[$\alpha=0.7,\theta=60^{\circ}$]{\includegraphics[scale=0.35]{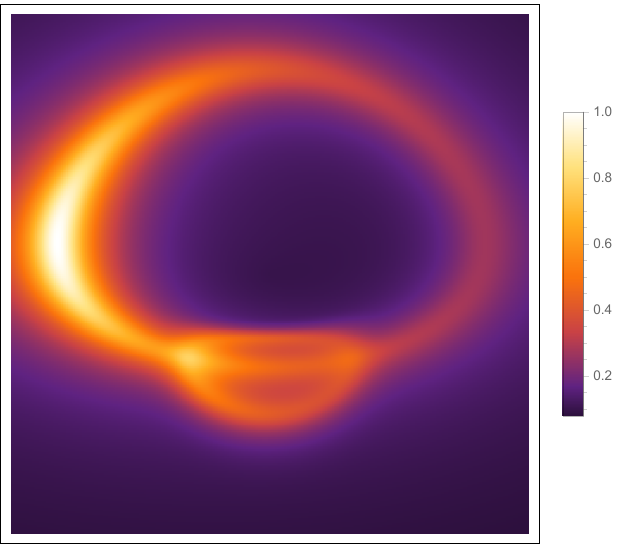}}
\subfigure[$\alpha=0.7,\theta=75^{\circ}$]{\includegraphics[scale=0.35]{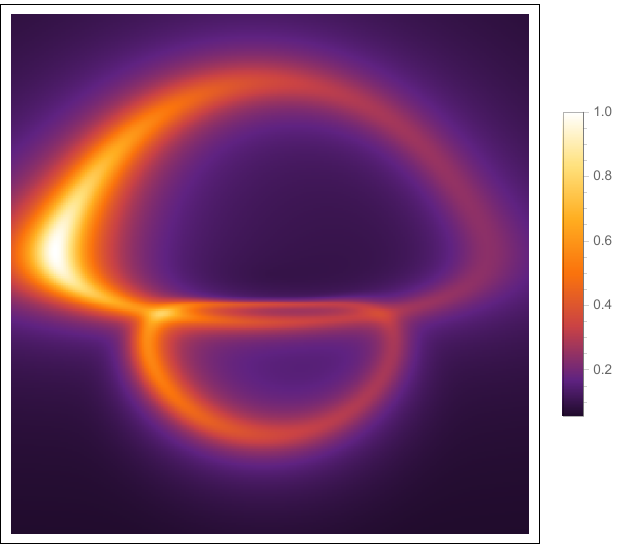}}
\caption{\label{fig13}Optical images of boson stars illuminated by a thin accretion disk under weak coupling, with the angle of field of view $\beta_{\mathrm{fov}}=3^{\circ}$, the initial scalar field $\psi_0=0.08$, and the observer at $r_{\mathrm{obs}}=200$. Each row from top to bottom corresponds to coupling parameters $\alpha=0.16, 0.25, 0.51, 0.7$, while each column from left to right corresponds to observer inclination angles $\theta=0^{\circ}, 30^{\circ}, 60^{\circ}, 75^{\circ}$.}
\end{figure}

Figure \ref{fig13} presents the optical images of boson stars under a thin accretion disk, with varying values of $\alpha$. The observer is positioned at $r_{\mathrm{obs}} = 200$, and the scalar field is initially set with $\psi_0 = 0.08$.
When $\theta = 0^{\circ}$ (the first column), the image displays a bright, axisymmetric circular ring, with a certain level of light intensity also present within the interior region of the ring. The bright ring in the image is the direct image of the accretion disk. Meanwhile, as the coupling parameter $\alpha$ increases, the diameter of the direct image expands notably.
When $\theta=30^{\circ}$ (the second column), the bright ring slightly deforms into an elliptical shape, and the enhanced Doppler effect causes the left side of the bright ring to be brighter than the right.
When $\theta = 60^{\circ}$ (the third column), the deformation of the bright rings becomes more obvious.
Although the lens image can already be observed in the image, this is for larger $\alpha$ values ($\alpha = 0.51, 0.7$). For a smaller value of $\alpha$ ($\alpha = 0.16$), only the direct image is present in the image. When $\theta = 75^{\circ}$ (the fourth column), even with a relatively small $\alpha$ value ($\alpha = 0.16$), the lensed image remains clearly observable, and its size increases with increasing $\alpha$.
It indicates that the coupling parameter $\alpha$ primarily affects the size of the direct image, while the observer inclination $\theta$ mainly influences its shape. Larger values of $\alpha$ and $\theta$ lead to the emergence of the lensed image, while the Doppler effect results in an asymmetric brightness distribution.

\begin{figure}[htbp]
	\centering
	\includegraphics[width=0.4\textwidth]{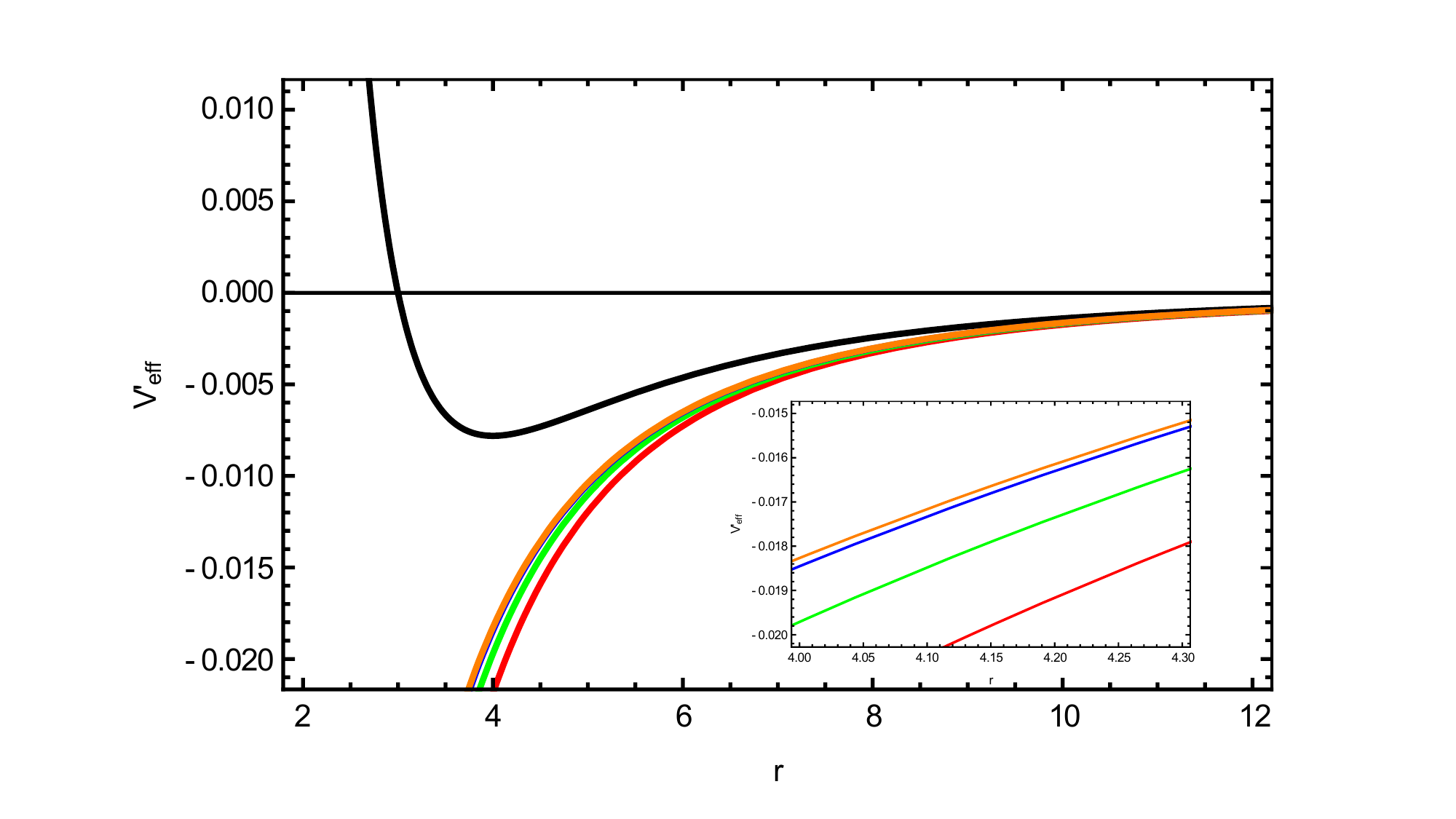}
	\caption{The relationship between the first derivative of  effective potential $V'_{eff}$ and the radial coordinate $r$ when the parameter $\alpha$ takes different values in  weak coupling case.
The red, green, blue, and orange curves correspond to $\alpha = 0.16, 0.25, 0.51,$ and $0.7$, respectively. The black curve corresponds to the Schwarzschild black hole. \label{fig11}}
\end{figure}

Corresponding to Figure \ref{fig13}, we present the proportion of the observed intensity of direct images ($n = 1$) relative to the total observed flux in Table 6. The results indicate that when the observation inclination angle is relatively low, the majority of the observed flux in the image still originates from the direct image.
Particularly, when both $\alpha$ and the observation inclination angle $\theta$ are relatively small ($\alpha=0.16, \theta=17^\circ$), the proportion of observed light intensity originating from the direct image accounts for as much as 99.8$\%$. However, as the observation inclination angle increases, the proportion of the directly observed image intensity gradually diminishes, with a more pronounced decrease observed at higher $\alpha$ values.
\begin{table}[h]
\begin{center}
	{\footnotesize{\bf Table 6.} The proportion of the observed intensity of the direct image ($n = 1$) to the total observed intensity in the weak coupling regime. Here, the value of initial scalar field is $\psi_0=0.08$.} \\
	\vspace{2mm}
	\begin{tabular}{c|c|c|c|c}
		\hline
		\diagbox[width=3em, height=2.5em]{$\alpha$}{$\theta$} & $17^\circ$ & $30^\circ$ & $60^\circ$ & $75^\circ$ \\ \hline
		0.16 & 99.8\% & 98.3\% & 89.7\% & 88.6\% \\ \hline
		0.25 & 99.6\% & 96\%   & 81.7\% & 70.7\% \\ \hline
		0.51 & 99.1\% & 88.6\% & 73.5\% & 69\% \\ \hline
		0.7  & 98.9\% & 87.6\% & 70.6\% & 65.8\% \\ \hline
	\end{tabular}
\end{center}
\end{table}
Similarly, irrespective of variations in the observation inclination angle $\theta$ and parameter $\alpha$, no photon ring structure is discernible in the images. Therefore, in Figure \ref{fig11}, we present the relationship between the first derivative of the effective potential $V'_{eff}$ and the radial coordinate $r$ when the parameter $\alpha$ takes different values in the case of weak coupling. Although $V'_{eff}$ exhibits a monotonically increasing behavior with respect to $r$, it does not intersect the horizontal axis (in contrast to the Schwarzschild black hole, which intersects at $r = 3M$). This implies that the equation $V'_{eff}=0$ has no real solution, thereby indicating the absence of photon rings in the spacetime geometry of these boson stars.

\begin{figure}[!h]
\centering 
\subfigure[$\psi_0=0.16,\theta=0.001^{\circ}$]{\includegraphics[scale=0.35]{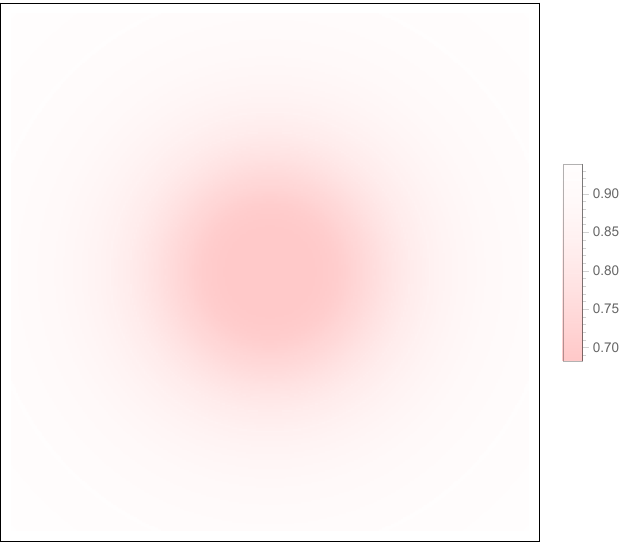}}
\subfigure[$\psi_0=0.16,\theta=30^{\circ}$]{\includegraphics[scale=0.35]{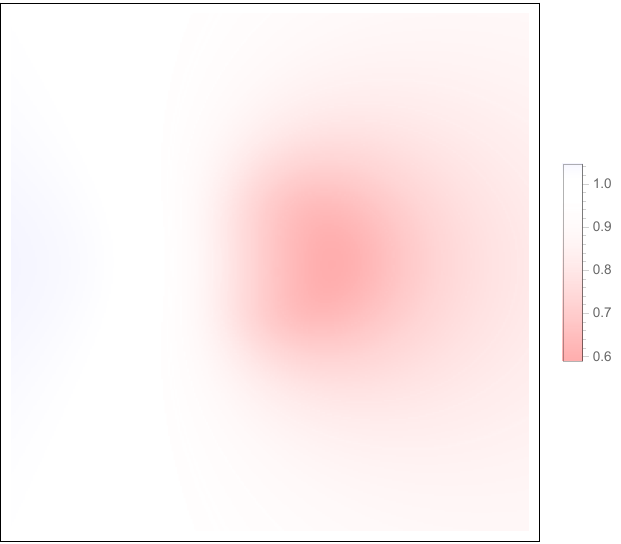}}
\subfigure[$\psi_0=0.16,\theta=60^{\circ}$]{\includegraphics[scale=0.35]{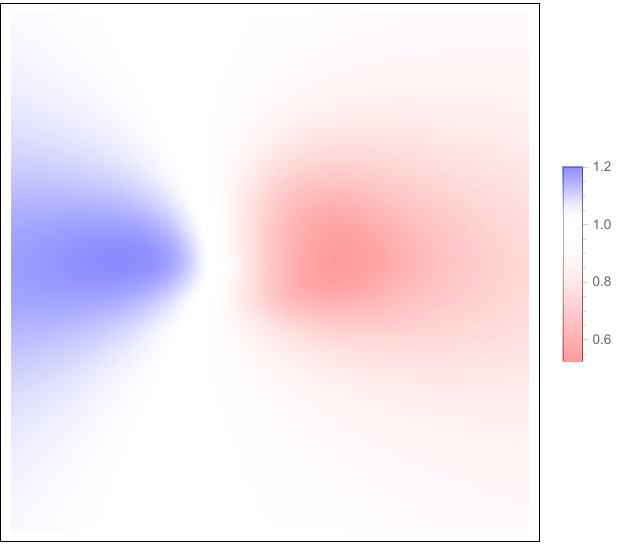}}
\subfigure[$\psi_0=0.16,\theta=75^{\circ}$]{\includegraphics[scale=0.35]{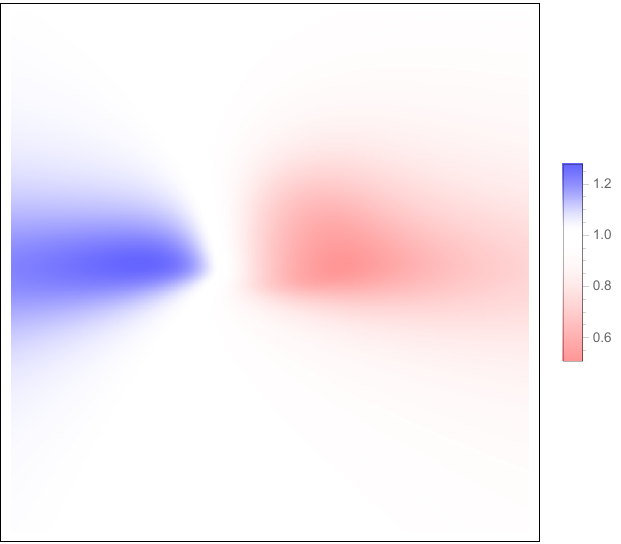}}
\subfigure[$\psi_0=0.25,\theta=0.001^{\circ}$]{\includegraphics[scale=0.35]{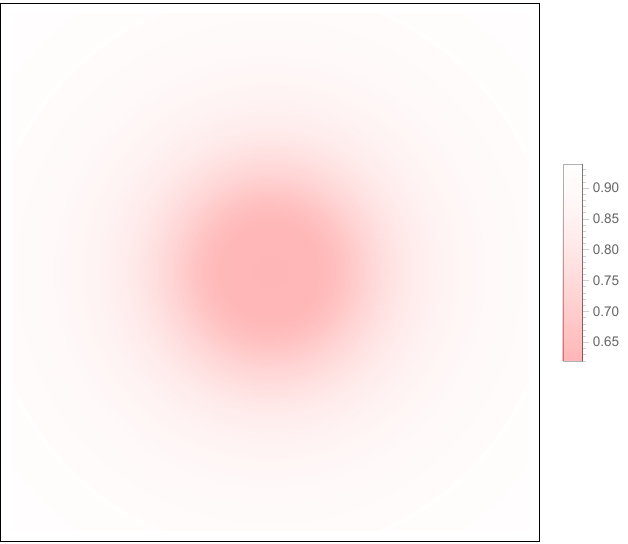}}
\subfigure[$\psi_0=0.25,\theta=30^{\circ}$]{\includegraphics[scale=0.35]{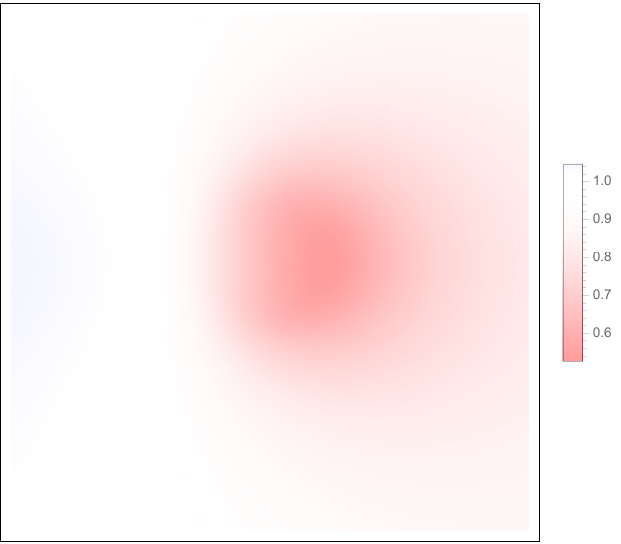}}
\subfigure[$\psi_0=0.25,\theta=60^{\circ}$]{\includegraphics[scale=0.35]{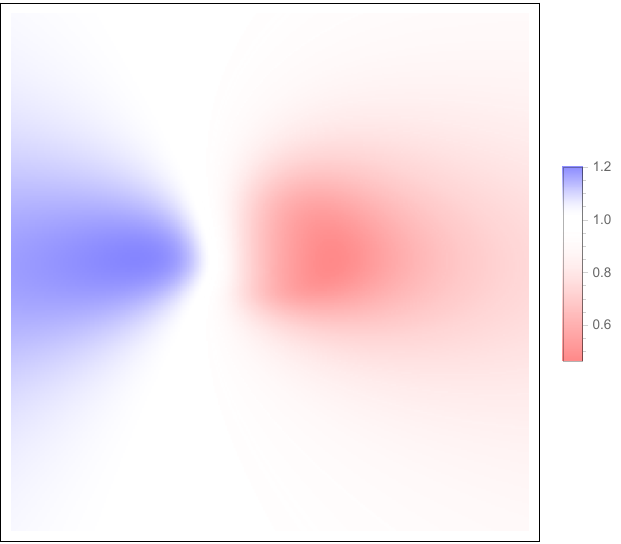}}
\subfigure[$\psi_0=0.25,\theta=75^{\circ}$]{\includegraphics[scale=0.35]{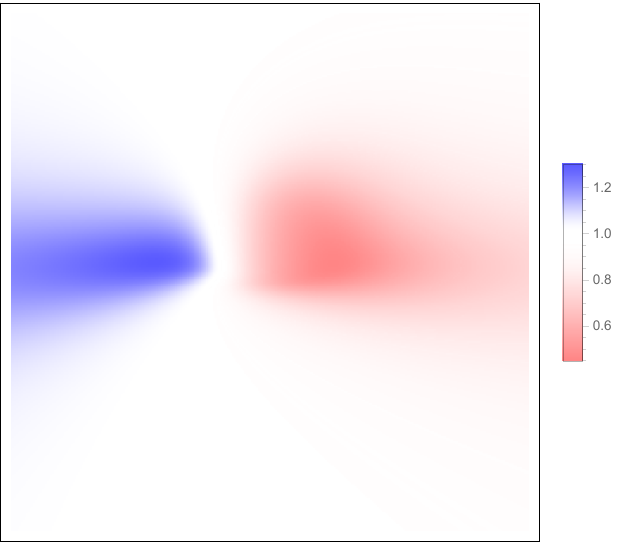}}
\subfigure[$\psi_0=0.51,\theta=0.001^{\circ}$]{\includegraphics[scale=0.35]{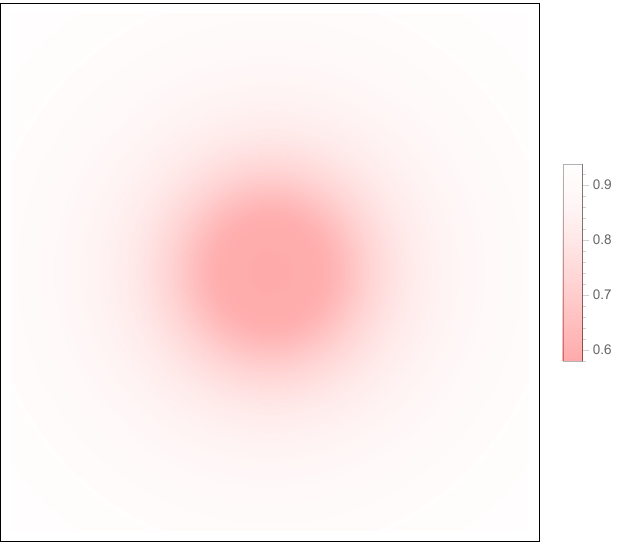}}
\subfigure[$\psi_0=0.51,\theta=30^{\circ}$]{\includegraphics[scale=0.35]{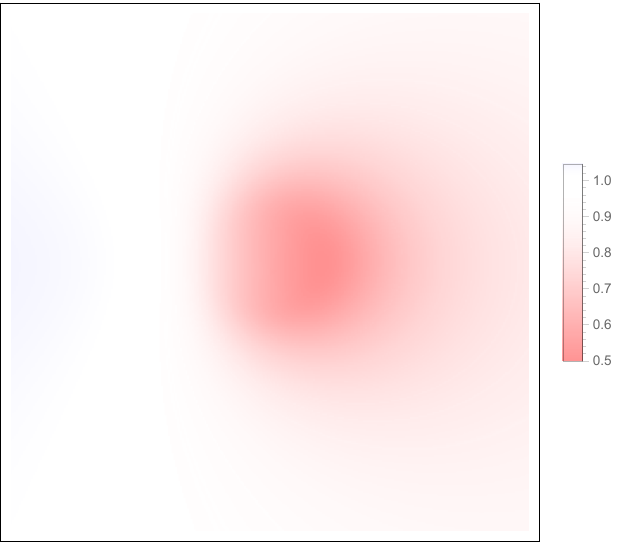}}
\subfigure[$\psi_0=0.51,\theta=60^{\circ}$]{\includegraphics[scale=0.35]{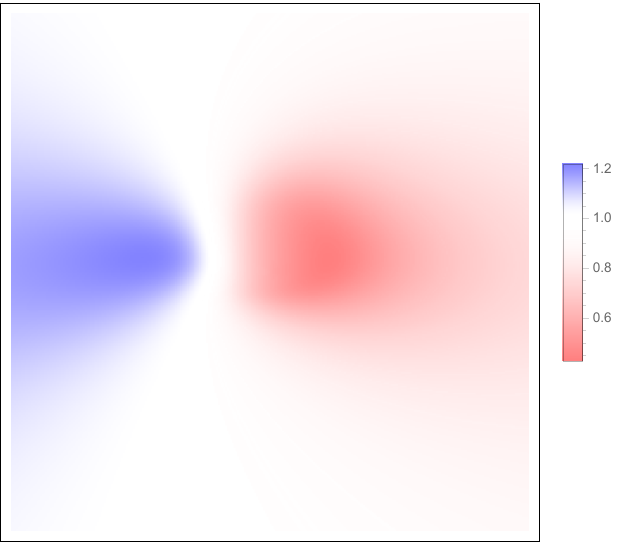}}
\subfigure[$\psi_0=0.51,\theta=75^{\circ}$]{\includegraphics[scale=0.35]{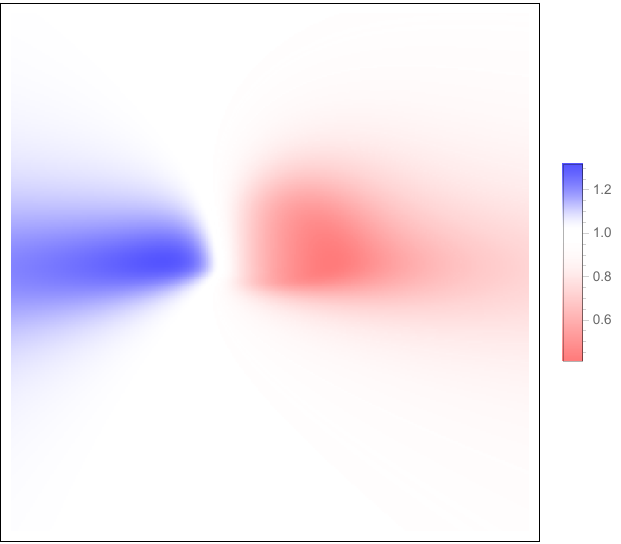}}
\subfigure[$\psi_0=0.7,\theta=0.001^{\circ}$]{\includegraphics[scale=0.35]{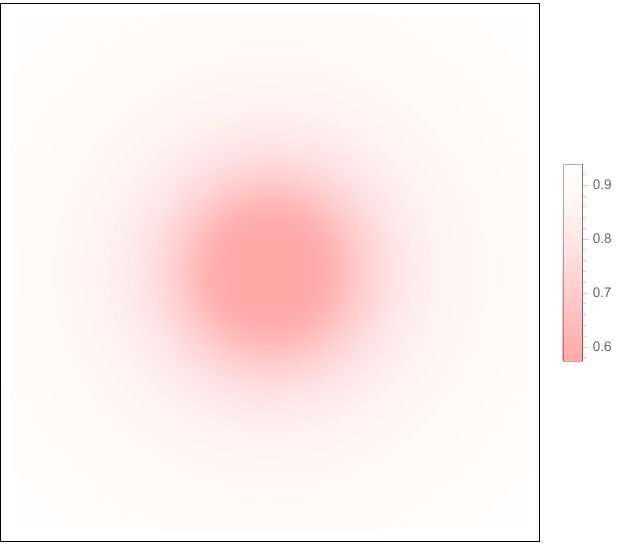}}
\subfigure[$\psi_0=0.7,\theta=30^{\circ}$]{\includegraphics[scale=0.35]{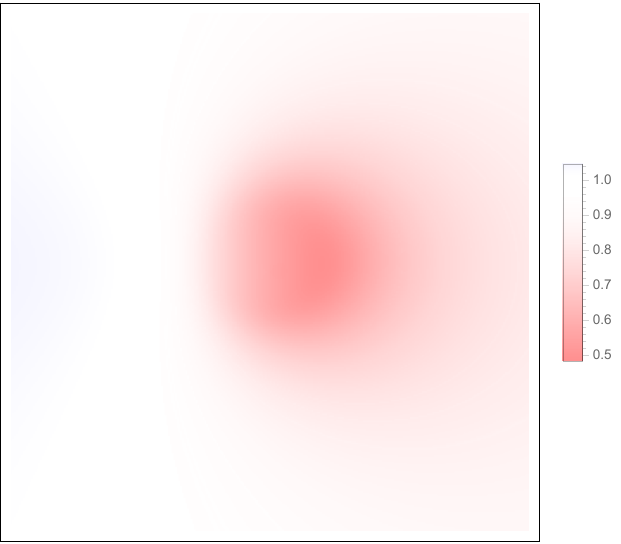}}
\subfigure[$\psi_0=0.7,\theta=60^{\circ}$]{\includegraphics[scale=0.35]{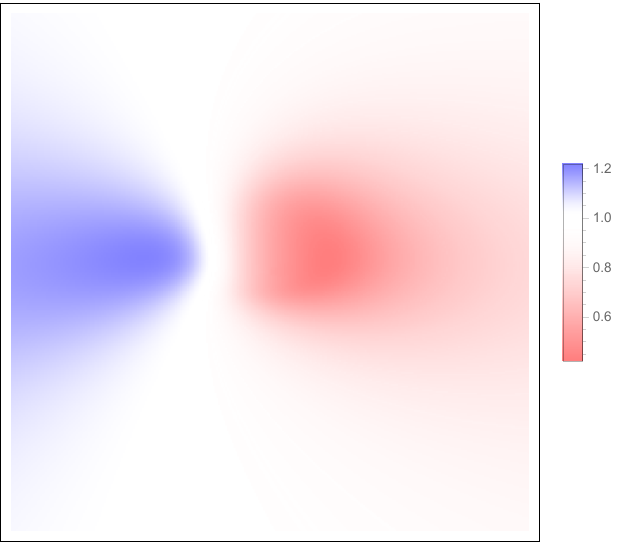}}
\subfigure[$\psi_0=0.7,\theta=75^{\circ}$]{\includegraphics[scale=0.35]{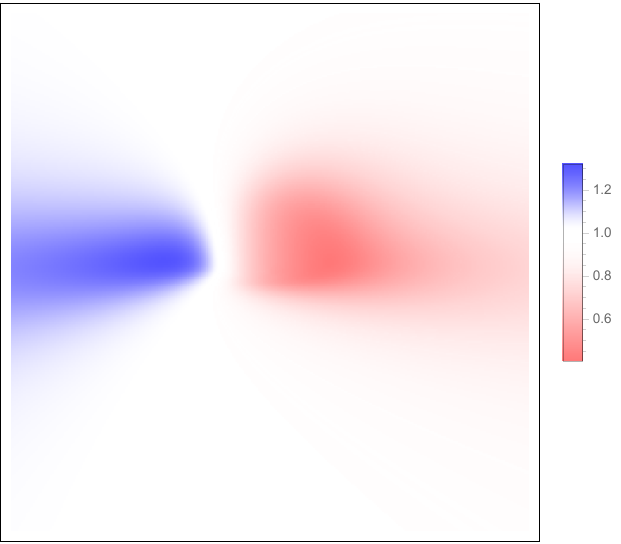}}
\caption{\label{fig14} In the weak coupling regime, the redshift factor distribution corresponds to the direct image under varying coupling parameters $\alpha$. Red and blue colors denote redshift and blueshift, respectively.}
\end{figure}

The corresponding redshift factor distribution of the direct image under weak coupling is illustrated in Figure \ref{fig14}. In Figure \ref{fig14},each row from top to bottom corresponds to coupling parameters $\alpha=0.16, 0.25, 0.51, 0.7$, while each column from left to right corresponds to observer inclination angles $\theta=0^{\circ}, 30^{\circ}, 60^{\circ}, 75^{\circ}$.
By comparing the images in each column, it can be observed that the observer inclination $\theta$ significantly influences the distribution of the redshift factor. When $\theta \to 0^{\circ}$ (the first column), only redshift is present, with no blueshift, and the redshift exhibits a symmetric distribution. As $\theta$ increases, the increasing radial velocity component of the accreting material enhances the blueshift factor, while both the blueshift and redshift distributions become more concentrated. By comparing the images in each row, it can be seen that an increase in $\alpha$ does not affect the distribution of redshift and blueshift but enhances the magnitude of the redshift.

\subsection{Observational characteristics of boson stars in the strong coupling regime}
A reduction in the coupling parameter $\alpha$ leads to an enhancement of nonlinearity within the field equation, consequently increasing the complexity of the numerical solution. Accordingly, we will examine the optical observational features of soliton boson stars under conditions of strong coupling, which correspond to smaller values of $\alpha$.
\begin{table}[h]
\begin{center}
{\footnotesize{\bf Table 7.}
Parameter estimates of $p_i (i=1,\dots,7)$ for the metric component $-g_{tt}$ under strong coupling conditions, with the initial scalar field value $\psi_0 = 0.088$. Here, $m(r)$ represents the mass of the boson star.}
\\
\vspace{2mm}
\begin{tabular}{c|c|c|c|c|c|c|c|c|c}
\hline
Type & $\alpha$ & $m(r)$ & $p_1$ & $p_2$ & $p_3$ & $p_4$ & $p_5$ & $p_6$ & $p_7$ \\ \hline
SBS1 & 0.077 &1.263  &-0.254& 0.05& -0.271& 0.063& -0.011& -0.001& -0.059 \\ \hline
SBS2 & 0.076 & 1.329  &-0.251& 0.048& -0.266& 0.061& -0.01& -0.001& -0.061\\ \hline
SBS3 & 0.075 &1.371  &-0.25& 0.048& -0.268& 0.06& -0.01& -0.001& -0.067\\ \hline
SBS4 & 0.0733 & 1.41  &-0.246& 0.046& -0.26& 0.057& -0.01& -0.001& -0.066\\ \hline
\end{tabular}
\end{center}
\end{table}

\begin{table}	
\begin{center}
{\footnotesize{\bf Table 8.} Parameter estimates of $q_i (i=1,\dots,7)$ for the metric component $g_{rr}$ under strong coupling conditions, with the initial scalar field value $\psi_0 = 0.088$. Here, $m(r)$ represents the mass of the boson star. }
\\
\vspace{2mm}
\begin{tabular}{c|c|c|c|c|c|c|c|c|c}
\hline
Type & $\alpha$ & $m(r)$ & $q_1$ & $q_2$ & $q_3$ & $q_4$ & $q_5$ & $q_6$ & $q_7$ \\ \hline
SBS1 & 0.077 & 1.263 &-6.732& -1.05& 213.839& -27.778& -3.462& 1.699& 1.844 \\ \hline
SBS2 & 0.076 & 1.329  &-4.516& -1.298& 113.743& 25.694& -14.011& 2.474& 2.169\\ \hline
SBS3 & 0.075 & 1.371 &-4.518& -0.989& 77.762& 12.166& -7.204& 1.433& 1.732\\ \hline
SBS4 & 0.0733 & 1.41  &-13.833& -0.429& 145.126& -23.899& 1.953& 0.615& 1.054\\ \hline
\end{tabular}
\end{center}
\end{table}
For different values of $\alpha$, the scalar field $\psi$ rapidly decays and asymptotically approaches zero. We compute the numerical solutions of the metric components $-g_{tt}$ and $g_{rr}$ and fit them using Eqs.(\ref{fit1}) and (\ref{fit2}), ensuring that the fitting functions satisfy the asymptotic flatness condition. The parameter estimates are presented in Table 7 and Table 8, and the result reveal that as $\alpha$ increases, the boson star mass $m(r)$ exhibits a decreasing trend.

\begin{figure}[!h]
\centering 
\subfigure[$\alpha=0.077,\theta=45^{\circ}$]{\includegraphics[scale=0.32]{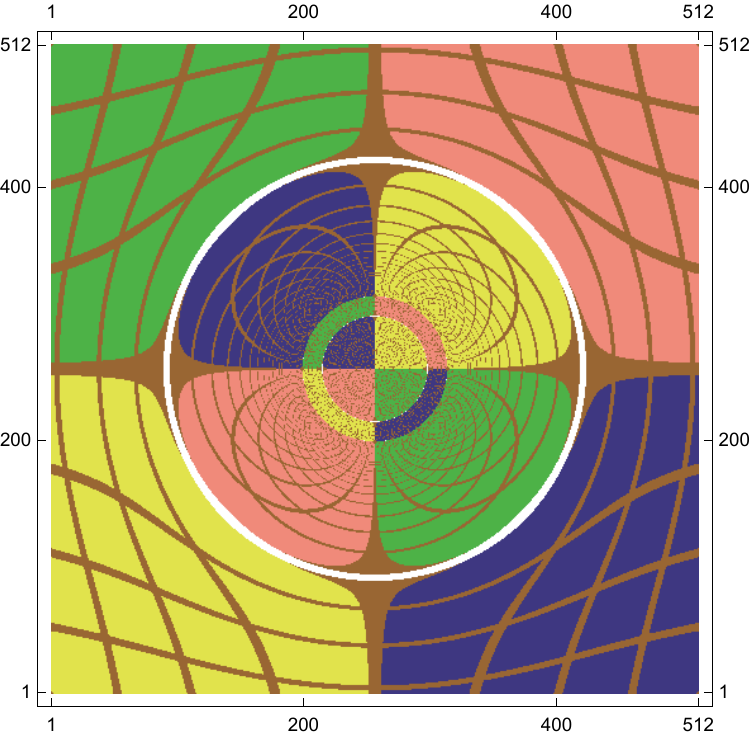}}
\subfigure[$\alpha=0.076,\theta=45^{\circ}$]{\includegraphics[scale=0.32]{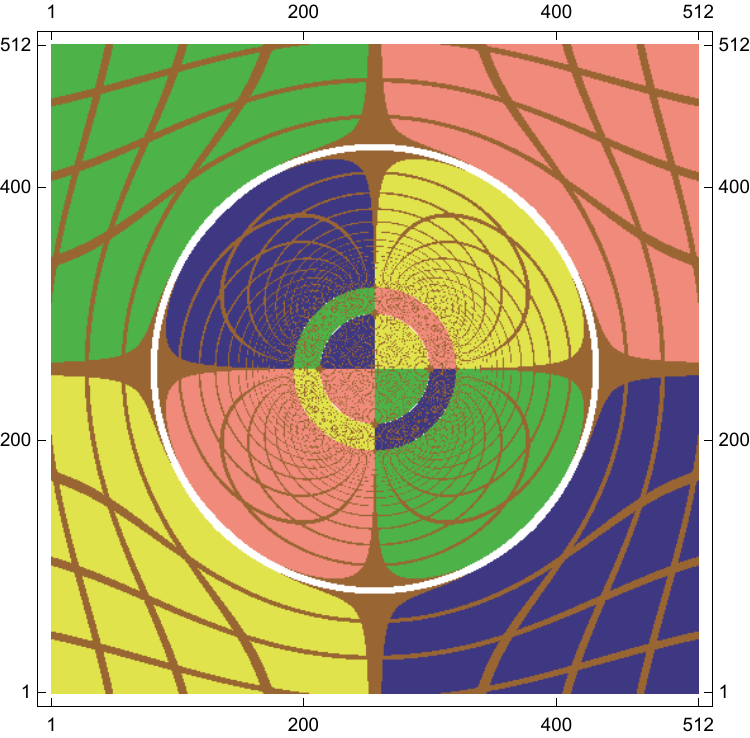}}
\subfigure[$\alpha=0.075,\theta=45^{\circ}$]{\includegraphics[scale=0.32]{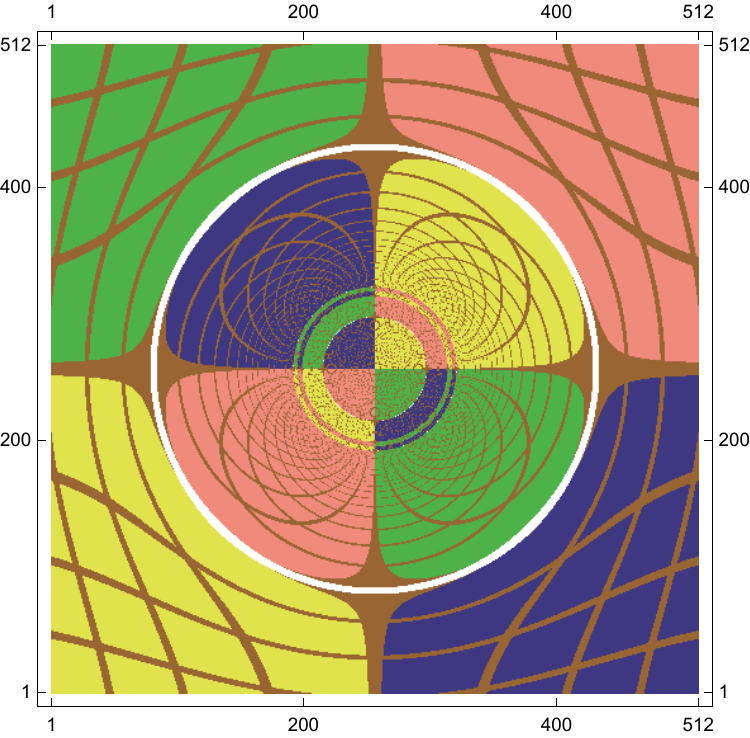}}
\subfigure[$\alpha=0.0733,\theta=45^{\circ}$]{\includegraphics[scale=0.32]{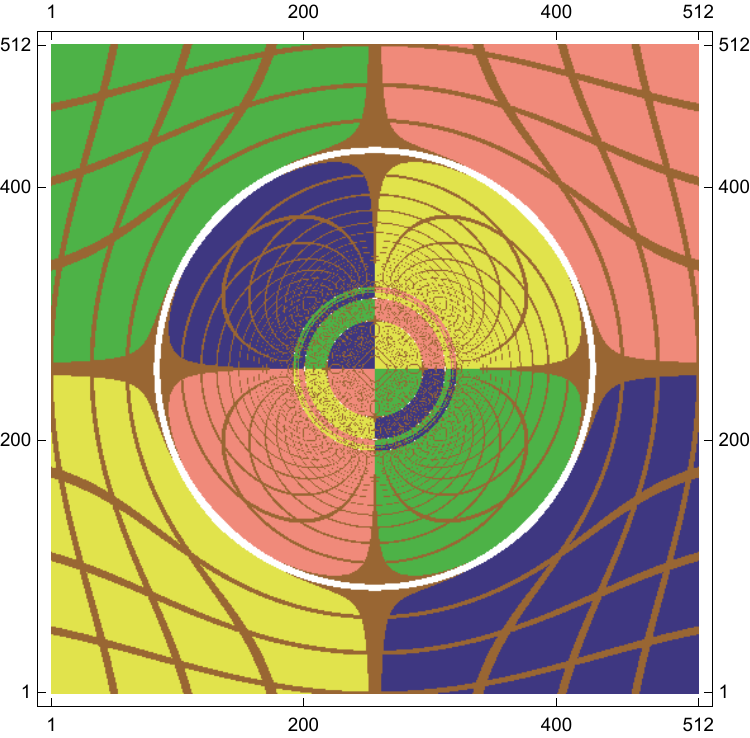}}
\caption{\label{fig15} Optical images of boson stars illuminated by a celestial sphere light source under strong coupling, with the initial scalar field $\psi_0=0.088$, the angle of field of view $\beta_{\mathrm{fov}}=45^{\circ}$, and the observer at $r_{\mathrm{obs}}=50m(r)$, where $m(r)$ represents the mass of the boson star.}
\end{figure}

The optical images of boson stars illuminated by a celestial sphere light source under strong coupling conditions are presented in Figure \ref{fig15}.
The relevant parameter values are  $\psi_0=0.088$, $\beta_{\mathrm{fov}}=45^{\circ}$ and $r_{\mathrm{obs}}=50m(r)$. From Figure \ref{fig15}, it can be observed that the boundary between the blue and yellow quadrants is perpendicular to the horizontal axis. This characteristic is a direct consequence of the boson star's non-rotating nature, indicating the absence of a dragging effect. It is noteworthy that a sub-ring structure has also been observed within the Einstein ring, which presents a significantly different outcome compared to the weak coupling scenario. Indeed, analogous sub-ring structures have been observed in spacetime scenarios where a black hole serves as the gravitational background.
According to \cite{giribet2024sub}, if a bubble made of non-emitting matter of zero optical depth exists around a static black hole, this configuration is stable under a physically reasonable set of parameters and can generate observable distortions in the rings pattern of the image \cite{broderick2022photon}, which correspond to gravitationally lensed secondary images created by the photon ring. This phenomenon generates the sub-annular structure within the photon ring and does not depend on exotic matter or strange spacetime contortions. In the case of strong coupling, as the value of $\alpha$ decreases, the radius of the sub-rings increases slightly, and the number of sub-rings also increases.

\begin{figure}[!h]
\centering 
\subfigure[$\alpha=0.077,\theta=0^{\circ}$]{\includegraphics[scale=0.35]{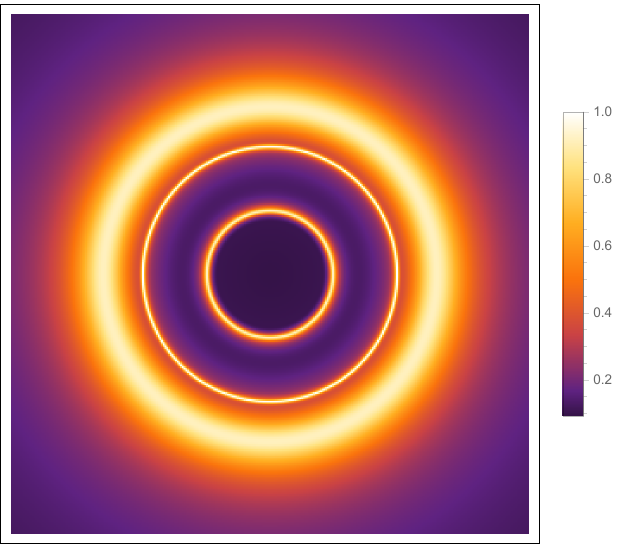}}
\subfigure[$\alpha=0.077,\theta=17^{\circ}$]{\includegraphics[scale=0.35]{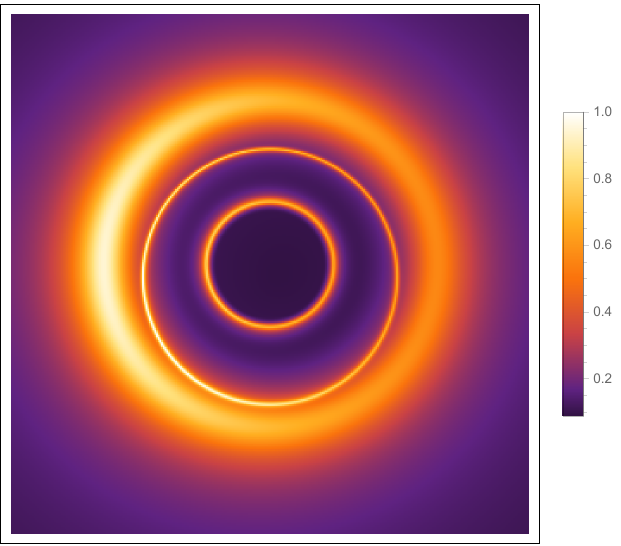}}
\subfigure[$\alpha=0.077,\theta=45^{\circ}$]{\includegraphics[scale=0.35]{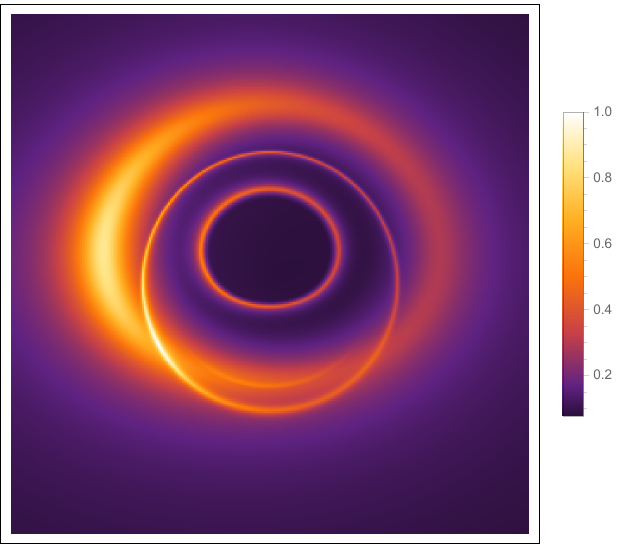}}
\subfigure[$\alpha=0.077,\theta=70^{\circ}$]{\includegraphics[scale=0.35]{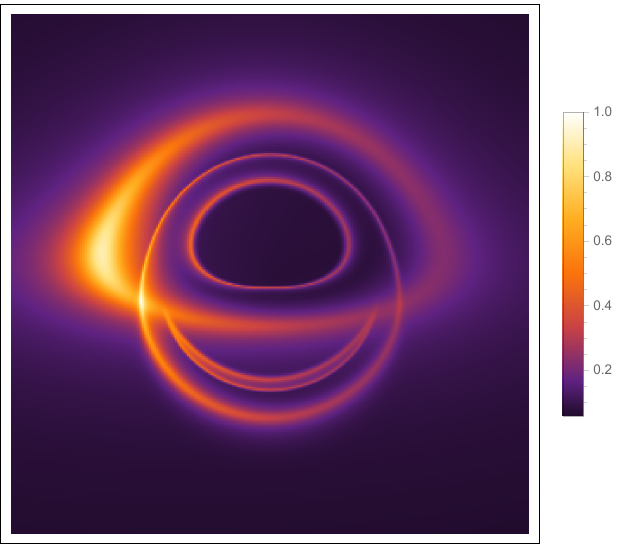}}
\subfigure[$\alpha=0.076,\theta=0^{\circ}$]{\includegraphics[scale=0.35]{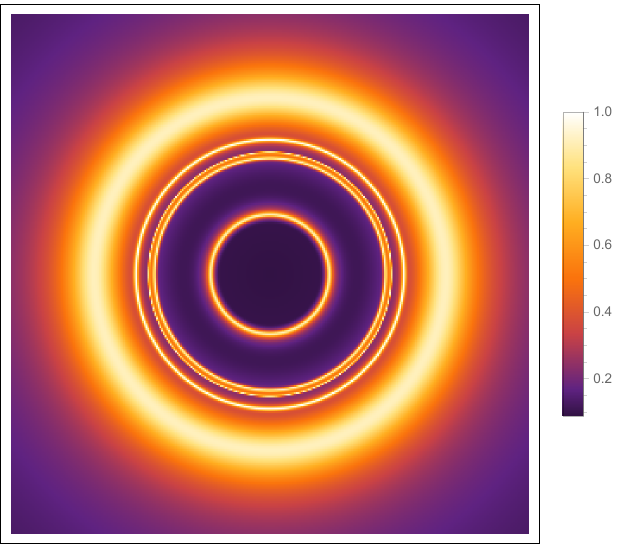}}
\subfigure[$\alpha=0.076,\theta=17^{\circ}$]{\includegraphics[scale=0.35]{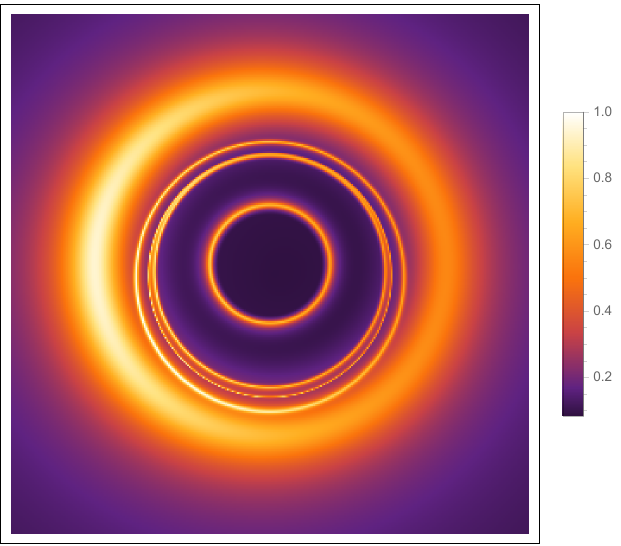}}
\subfigure[$\alpha=0.076,\theta=45^{\circ}$]{\includegraphics[scale=0.35]{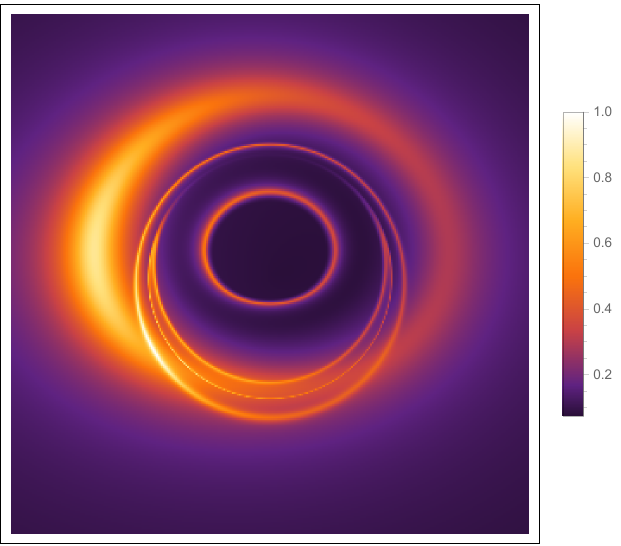}}
\subfigure[$\alpha=0.076,\theta=70^{\circ}$]{\includegraphics[scale=0.35]{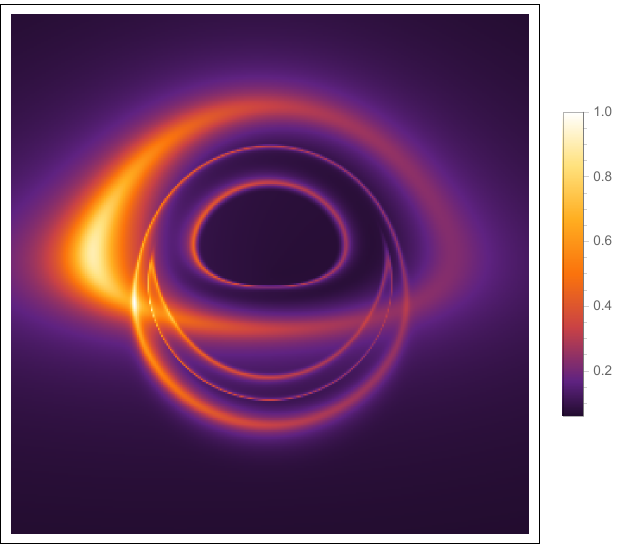}}
\subfigure[$\alpha=0.075,\theta=0^{\circ}$]{\includegraphics[scale=0.35]{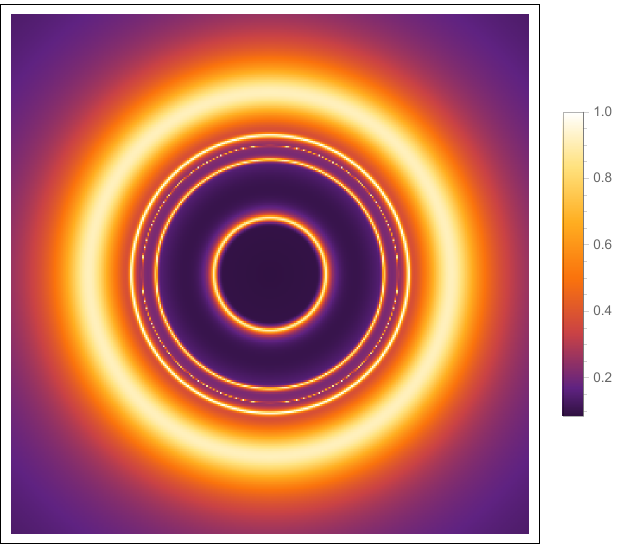}}
\subfigure[$\alpha=0.075,\theta=17^{\circ}$]{\includegraphics[scale=0.35]{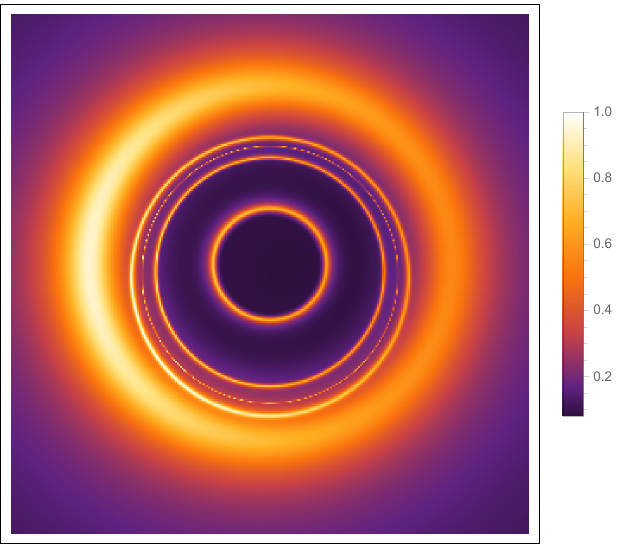}}
\subfigure[$\alpha=0.075,\theta=45^{\circ}$]{\includegraphics[scale=0.35]{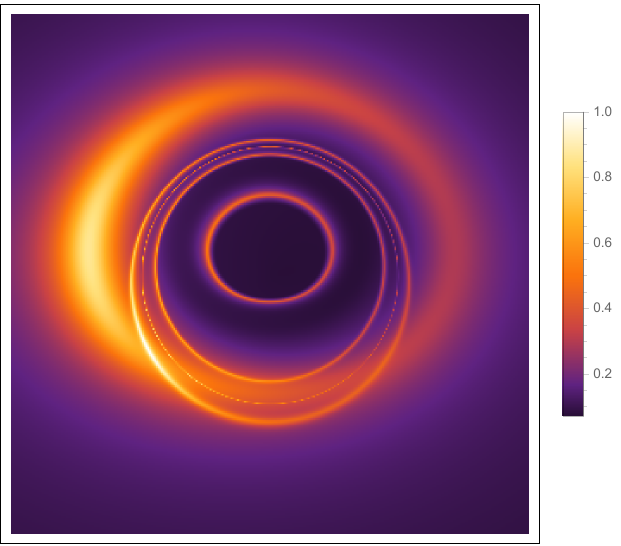}}
\subfigure[$\alpha=0.075,\theta=70^{\circ}$]{\includegraphics[scale=0.35]{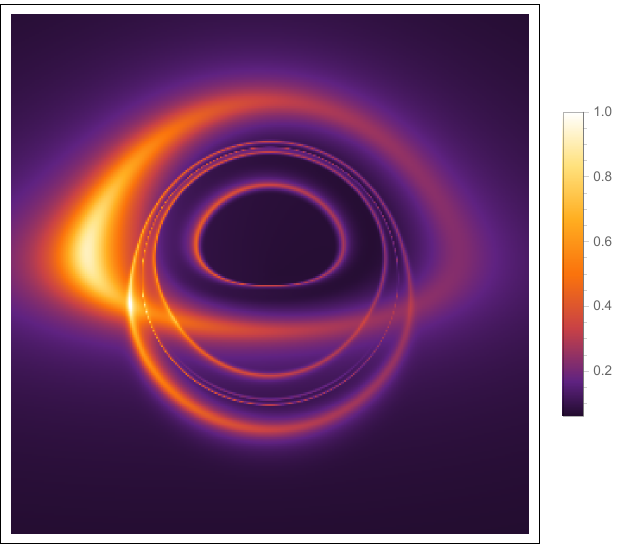}}
\subfigure[$\alpha=0.0733,\theta=0^{\circ}$]{\includegraphics[scale=0.35]{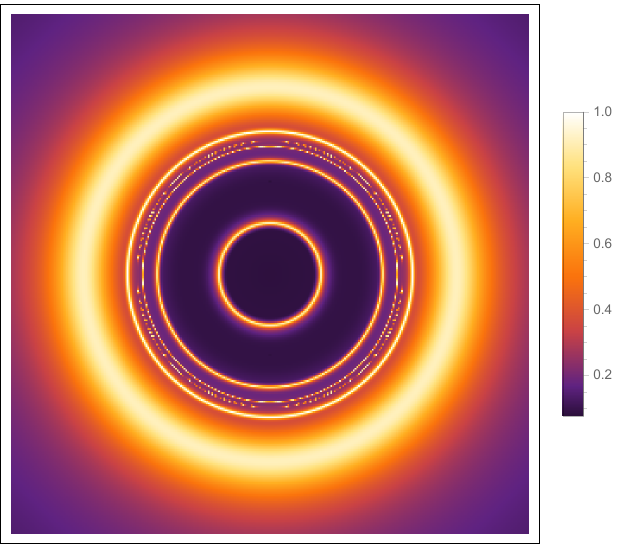}}
\subfigure[$\alpha=0.0733,\theta=17^{\circ}$]{\includegraphics[scale=0.35]{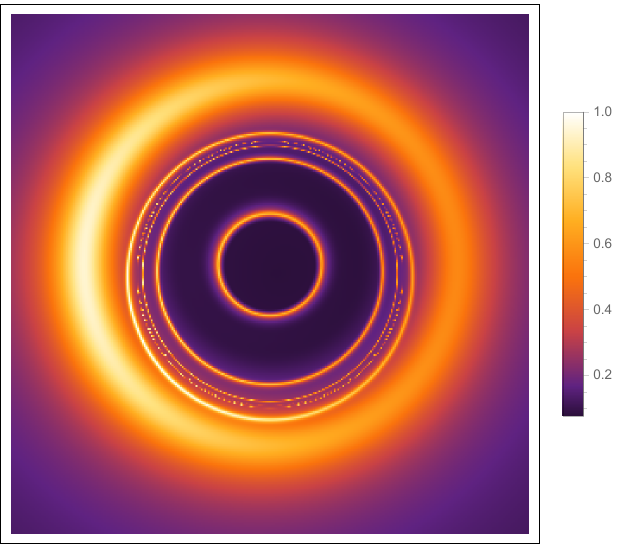}}
\subfigure[$\alpha=0.0733,\theta=45^{\circ}$]{\includegraphics[scale=0.35]{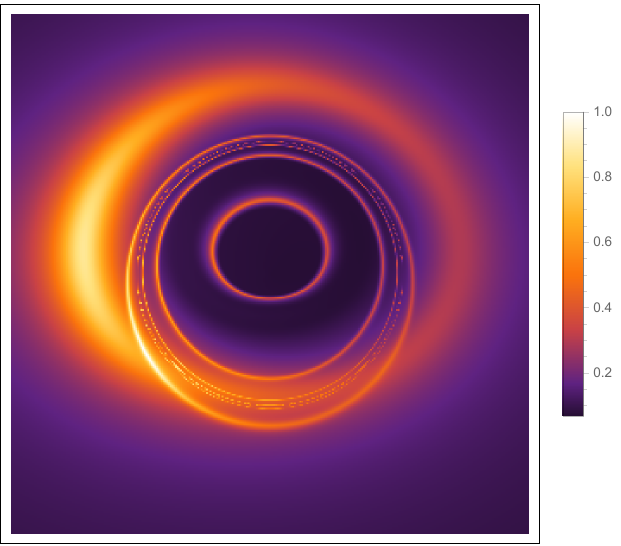}}
\subfigure[$\alpha=0.0733,\theta=70^{\circ}$]{\includegraphics[scale=0.35]{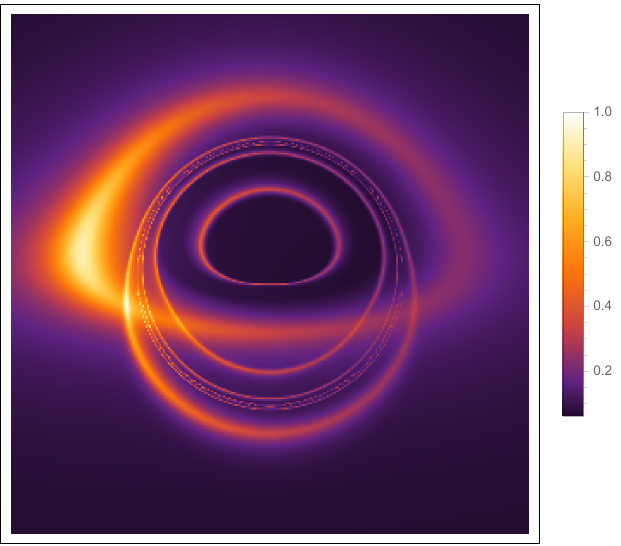}}
\caption{\label{fig16}  Optical images of boson stars illuminated by a thin accretion disk under strong coupling, with the angle of field of view $\beta_{\mathrm{fov}}=8^{\circ}$, the initial scalar field $\psi_0=0.088$, and the observer at $r_{\mathrm{obs}}=200$. Each row from top to bottom corresponds to coupling parameters $\alpha=0.077, 0.076, 0.075, 0.0733$, while each column from left to right corresponds to observer inclination angles $\theta=0.001^{\circ}, 17^{\circ}, 45^{\circ}, 70^{\circ}$.}
\end{figure}

The optical images of boson stars illuminated by a thin accretion disk under strong coupling are shown in Figure \ref{fig16}. When the observer inclination angle $\theta= 0^{\circ}$ (the first column), the outermost bright ring in the optical image corresponds to the direct image of the accretion disk. In addition to the direct image, the image also contains two smaller, brighter rings that are narrower and of reduced size, which correspond to lensed images.  Notably, as parameter $\alpha$ continues to decrease ($\alpha=0.075,0.0733$), a photon ring ($n \geq 3$) emerges in the central region between the two lensed images and becomes distinctly visible. Due to the confinement of the photon ring to an extremely narrow region, its observable intensity is significantly lower compared to that of the direct image. Additionally, these bright rings are consistently arranged in a concentric circular pattern within the image. As the observation inclination angle gradually increases ($\theta=17^{\circ}, 45^{\circ}$), both the direct image and the lensed image experience a certain degree of deformation and no longer exhibit a concentric circular pattern. In particular, when $\theta=70^{\circ}$ (the fourth column), the lensed image transitions into a crescent shape, the direct image transforms into a cap-like structure, and the photon ring remains predominantly circular in form.

\begin{figure}[!h]
	\centering
	\subfigure[$\alpha=0.075, \theta=17^\circ$]{\includegraphics[scale=0.45]{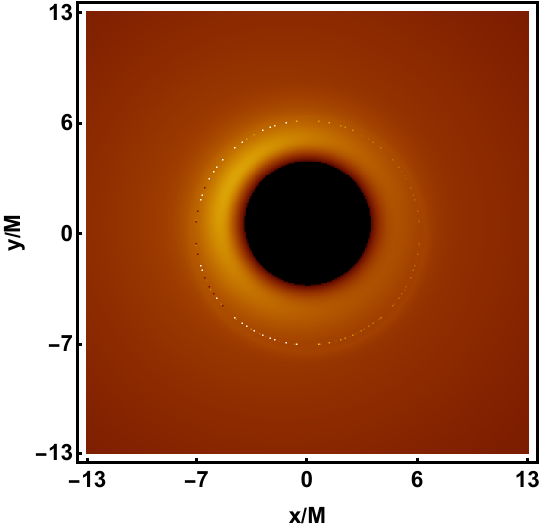}}
	\subfigure[$\alpha=0.075, \theta=17^\circ$]{\includegraphics[scale=0.45]{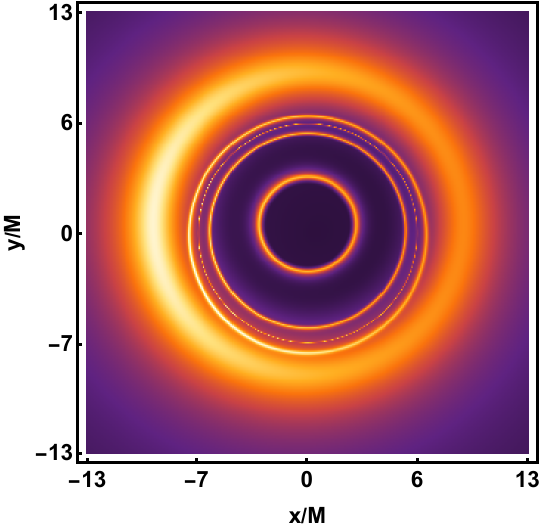}}
	\subfigure[$\alpha=0.076, \theta=17^\circ$]{\includegraphics[scale=0.45]{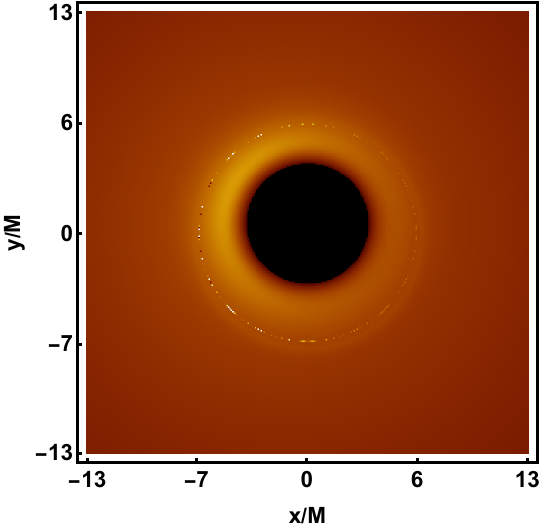}}
	\subfigure[$\alpha=0.076, \theta=17^\circ$]{\includegraphics[scale=0.45]{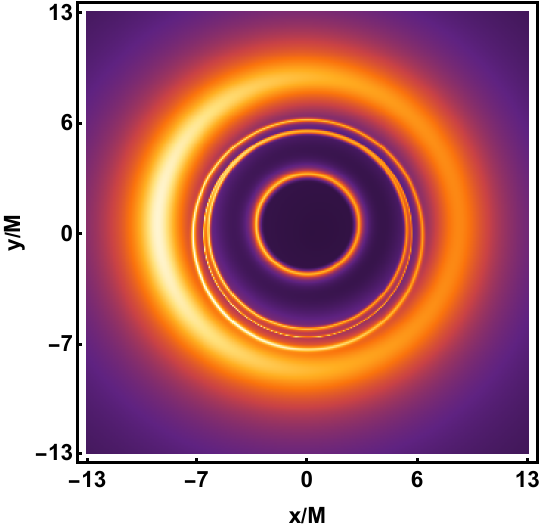}}
\caption{A comparison between the observed images of a boson star and a Schwarzschild black hole, both possessing the same field of view angle $\beta_{\mathrm{fov}}$ and an identical mass of $M = 1.371$. Panels (a) and (c) depict the Schwarzschild black hole, whereas panels (b) and (d) illustrate the boson star with coupling parameters $\alpha = 0.075$. }
	\label{fig99}
\end{figure}

In Table 9, the proportion of the observed intensity of direct images ($n = 1$) with respect to the total observed flux is presented. In the case of strong coupling, although the direct image still constitutes the majority of the observed flux at lower viewing angles, its proportion is significantly reduced compared to that under weak coupling. When $\alpha = 0.0733$ and $\theta =17^\circ$, the observed flux of the direct image decreases to 78.8$\%$, which is approximately 20 percentage points lower than that observed under weak coupling conditions. In other words, under conditions of strong coupling, the total observed intensity arises not only from the direct image, but also from the lensed image and the photon ring. The contributions to the observed intensity from the lensed image and the photon ring cannot be considered negligible.
\begin{table}[h]
\begin{center}
	{\footnotesize{\bf Table 9.} The proportion of the observed intensity of the direct image ($n = 1$) to the total observed intensity in the strong coupling regime. Here, the value of initial scalar field is $\psi_0=0.088$.} \\
	\vspace{2mm}
	\begin{tabular}{c|c|c|c|c}
		\hline
		\diagbox[width=3em, height=2.5em]{$\alpha$}{$\theta$} & $17^\circ$ & $30^\circ$ & $60^\circ$ & $75^\circ$ \\ \hline
		0.077  & 81.3\% & 72.2\% & 63.4\% & 62.1\% \\ \hline
		0.076  & 80.9\% & 71.9\% & 63.1\% &61.7\% \\ \hline
		0.075  & 80.6\% & 71.2\% & 63\%   & 61.5\% \\ \hline
		0.0733 & 78.8\% & 63.7\% & 62.7\% & 60.3\% \\ \hline
	\end{tabular}
\end{center}
\end{table}
Interestingly,  the innermost region of the gravitational lensing imaging in Figure \ref{fig16}, a distinct area with reduced central brightness emerged. The observed intensity of this region was significantly lower than that in the weak coupling scenario and closely resembled the inner shadow of a black hole \cite{rosa2022shadows}. This phenomenon suggests that, under specific configurations of state parameters, boson stars may effectively mimic black holes, as their optical observational features closely resemble the shadow structures observed in black holes.
Therefore, by taking $\alpha = 0.075$ and $\alpha = 0.076$ as illustrative examples and maintaining the mass of the boson star equivalent to that of the Schwarzschild black hole, we conducted a comparative analysis of their optical observational images under identical field view angles $\beta_{\mathrm{fov}}$ and observation inclination angles $\theta$, as presented in Figure \ref{fig99}. It can be observed that, under conditions where the masses of the central compact objects are equivalent, the diameter of the bright ring in the spacetime of a boson star is significantly larger than that in the  Schwarzschild black hole spacetime. Additionally, although boson stars also display a comparatively dark central region, the extent of darkness remains significantly lower than that observed within the inner shadow region of a black hole. Moreover, under the same mass conditions, the dark region of a Schwarzschild black hole is also slightly larger than that of a boson star. In fact, Olivares et al., after performing GRMHD simulations and GRRT calculations within the spacetime of boson stars and reconstructing the corresponding images, found that although boson stars exhibit qualitative similarities with black holes, the observable differences between the two are sufficiently distinct to allow for clear differentiation. These differences stem from dynamical effects directly related to the absence of an event horizoncite\cite{Olivares:2018abq}.

\begin{figure}[!h]
\centering 
\subfigure[$\alpha=0.77,\theta=0^{\circ}$]{\includegraphics[scale=0.35]{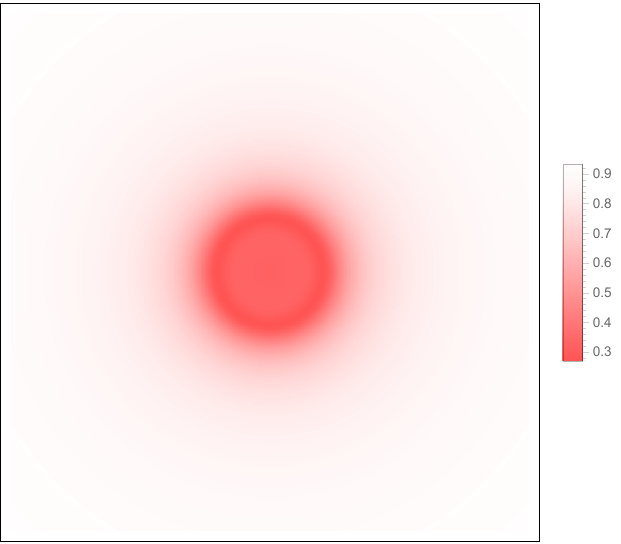}}
\subfigure[$\alpha=0.77,\theta=17^{\circ}$]{\includegraphics[scale=0.35]{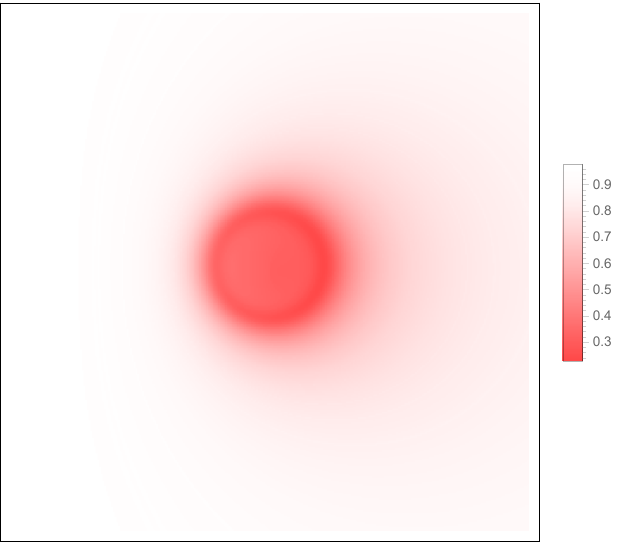}}
\subfigure[$\alpha=0.77,\theta=45^{\circ}$]{\includegraphics[scale=0.35]{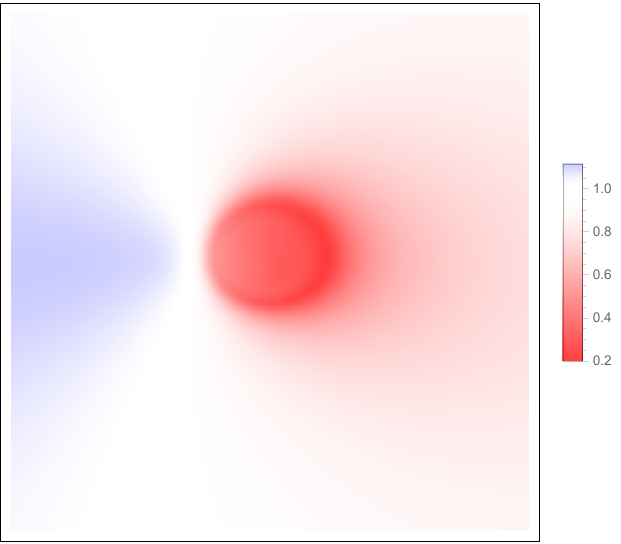}}
\subfigure[$\alpha=0.77,\theta=70^{\circ}$]{\includegraphics[scale=0.35]{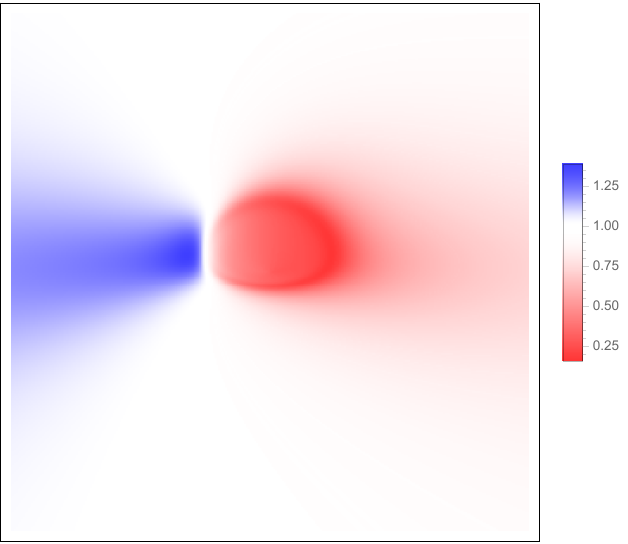}}
\subfigure[$\alpha=0.76,\theta=0^{\circ}$]{\includegraphics[scale=0.35]{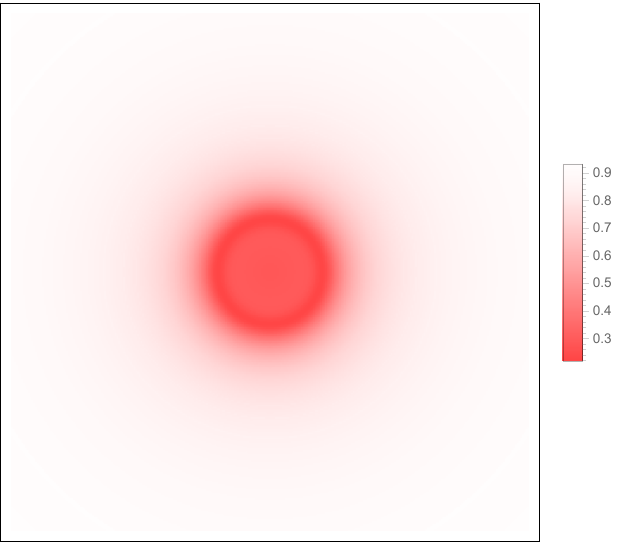}}
\subfigure[$\alpha=0.76,\theta=17^{\circ}$]{\includegraphics[scale=0.35]{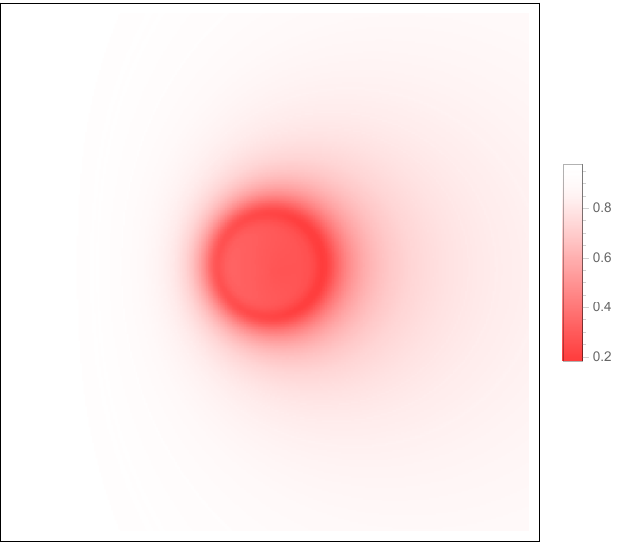}}
\subfigure[$\alpha=0.76,\theta=45^{\circ}$]{\includegraphics[scale=0.35]{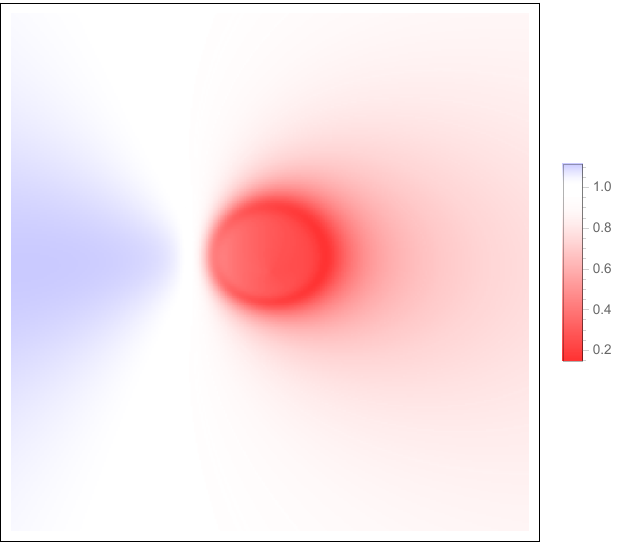}}
\subfigure[$\alpha=0.76,\theta=70^{\circ}$]{\includegraphics[scale=0.35]{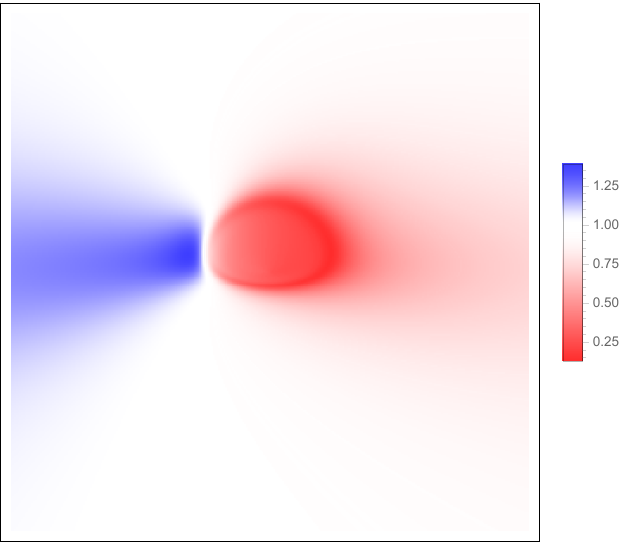}}
\subfigure[$\alpha=0.75,\theta=0^{\circ}$]{\includegraphics[scale=0.35]{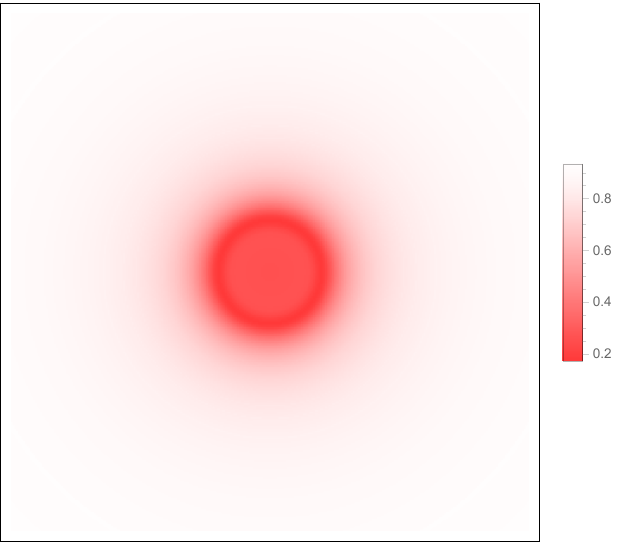}}
\subfigure[$\alpha=0.75,\theta=17^{\circ}$]{\includegraphics[scale=0.35]{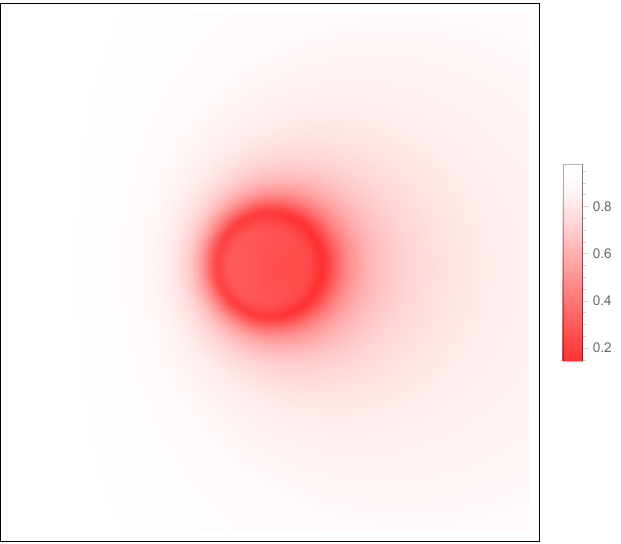}}
\subfigure[$\alpha=0.75,\theta=45^{\circ}$]{\includegraphics[scale=0.35]{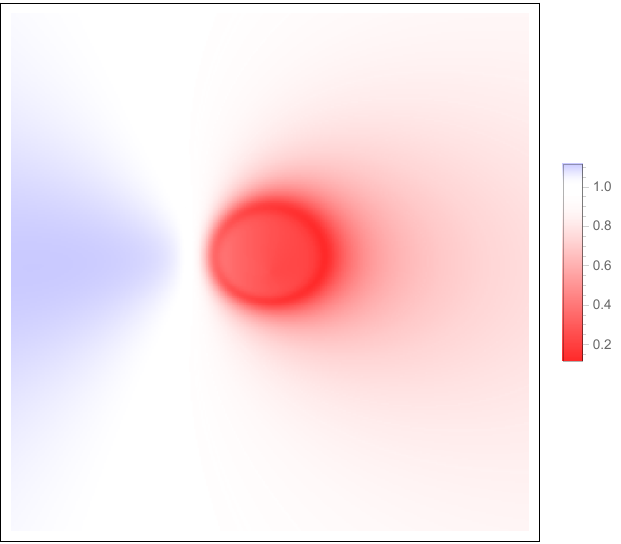}}
\subfigure[$\alpha=0.75,\theta=70^{\circ}$]{\includegraphics[scale=0.35]{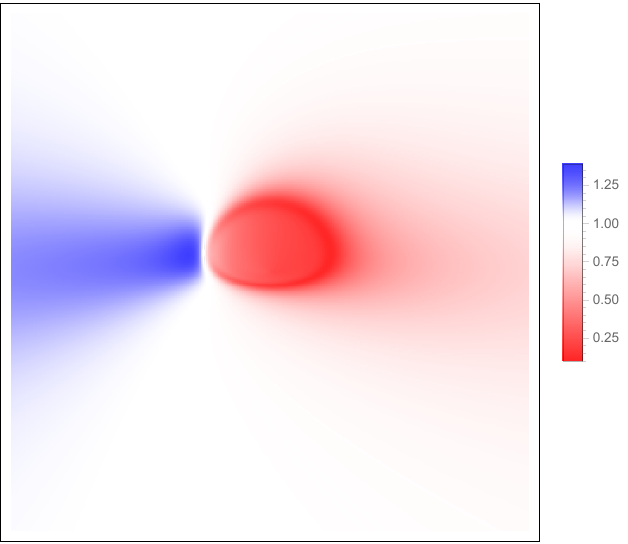}}
\subfigure[$\alpha=0.733,\theta=0^{\circ}$]{\includegraphics[scale=0.35]{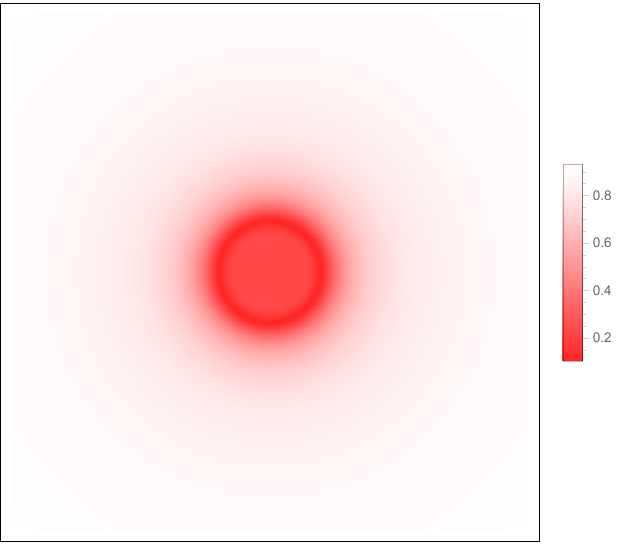}}
\subfigure[$\alpha=0.733,\theta=17^{\circ}$]{\includegraphics[scale=0.35]{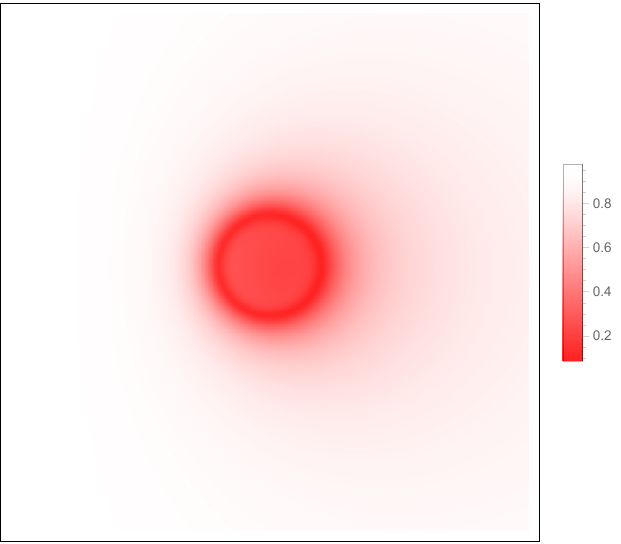}}
\subfigure[$\alpha=0.733,\theta=45^{\circ}$]{\includegraphics[scale=0.35]{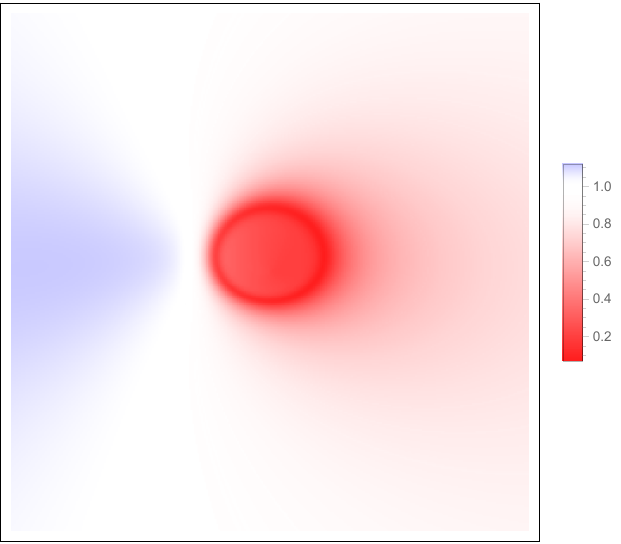}}
\subfigure[$\alpha=0.733,\theta=70^{\circ}$]{\includegraphics[scale=0.35]{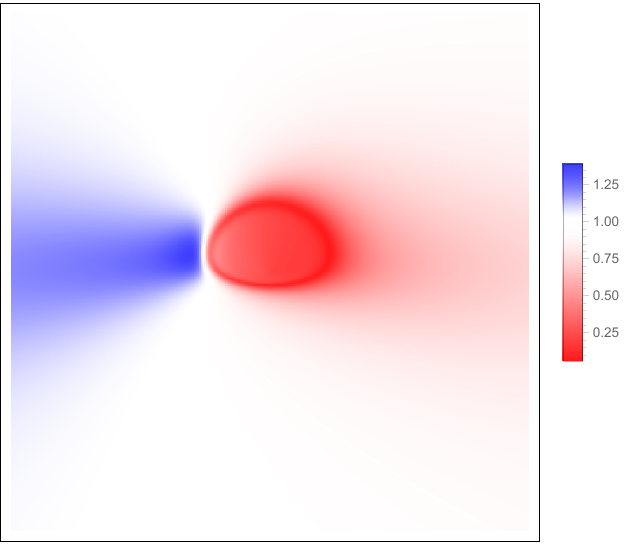}}
\caption{\label{fig17} In the  strong coupling regime, the redshift factor distribution corresponds to the direct image under varying coupling parameters $\alpha$. Red and blue colors denote redshift and blueshift, respectively.}
\end{figure}

In the case of strong coupling, the redshift factor distribution corresponding to the direct image is shown in Figure \ref{fig17}. Each row from top to bottom corresponds to coupling parameters $\alpha=0.077, 0.076, 0.075, 0.0733$, while each column from left to right corresponds to observer inclination angles $\theta=0^{\circ}, 17^{\circ}, 45^{\circ}, 70^{\circ}$. Similar to the weak coupling case, an increase in $\theta$ significantly affects the distribution of the redshift factor and enhances the degree of blueshift, while a decrease in $\alpha$ slightly enhances the redshift. However, in contrast to the weak coupling scenario (as depicted in Figure \ref{fig14}), the redshift factor distribution under strong coupling conditions exhibits a higher degree of concentration toward the center of the image. This results in the formation of a disk-like region when $\theta = 0^{\circ}$ (the first column) or a hat-shaped region when $\theta = 70^{\circ}$ (the fourth column). It corresponds to the central brightness depression observed in Figure \ref{fig16}. In contrast to the redshift factor near the inner shadow of a black hole, the redshift factor within a boson star is located inside the star itself, as boson stars do not possess an event horizon.

\section{Conclusions and discussions}
\label{conclusion}
In terms of observation, the optical characteristics of other compact celestial bodies may exhibit similarities to those of black holes. In this study, we investigated the optical properties of soliton boson stars, and examined the key distinctions between their characteristics and those of black holes. By employing backward ray-tracing techniques, it is possible to systematically analyze the influences of the initial scalar field $\psi_0$, the coupling parameter $\alpha$, and the observer inclination angle $\theta$ on the photon trajectories and optical images of solitonic boson stars. Two types of light sources are considered, that is, the celestial light source and the thin accretion disk model.

Firstly, we investigated the impact of the initial scalar field $\psi_0$  on the observational characteristics of the boson star, as well as the larger coupling parameter $\alpha$ (indicating weak coupling). The non-rotating characteristic of a boson star indicates the absence of the frame-dragging effect when illuminated by a celestial sphere light source, thereby resembling a static black hole. The impact of $\psi_0$ on the optical image of the boson star under the celestial sphere light source is relatively negligible, while variations in $\alpha$ substantially modify the photon trajectories within the Einstein ring. In the context of optical images formed under a thin accretion disk, the direct image plays a dominant role. The parameters $\psi_0$ and $\alpha$ primarily determine the size of the optical image, whereas the observer's inclination angle $\theta$ has a predominant influence on its shape. When $\theta$ is small, the optical image manifests as a single bright ring featuring a central brightness depression, consistent with observations made by the EHT, where gravitational redshift plays a dominant role. As the value of $\theta$ increases, the Doppler effect becomes increasingly pronounced, resulting in a substantial brightness contrast between the left and right sides of the direct image. Moreover, for larger values of $\theta$, a lensed image is formed, with increases in $\psi_0$ and $\alpha$ resulting in a corresponding enlargement of its size. In this scenario, the distribution of the redshift factor exhibits sensitivity to the parameter $\theta$, whereas the intensity of the optical image is predominantly determined by the parameter $\alpha$.

Subsequently, we examined the impact of variations in relevant parameters on the observable characteristics of the boson star under conditions of strong coupling, which is smaller coupling parameter $\alpha$. Interestingly, a smaller value of $\alpha$ leads to the formation of a sub-annular structure within the Einstein ring, while the size of the boson star is inversely proportional to $\alpha$. It is important to highlight that in the optical images under a thin accretion disk scenario, in addition to direct images, a smaller value of $\alpha$ can lead to the formation of lensed images. Among these, the innermost lensed image demonstrates markedly reduced intensity in comparison to other regions, akin to the shadow effect observed near a black hole.  This indicates that, for specific parameter selections, the optical images of a boson star could potentially become indistinguishable from the black hole. As the value of $\alpha$ decreases, the number of lensed images increases, which suggests that the effect of light deflection becomes more pronounced. Meanwhile, an increase in $\theta$ results in substantial distortions of both the direct and lensed images. Since boson stars do not possess an event horizon, under conditions of strong coupling, the redshift factor associated with the direct image is predominantly localized within the boson star itself. Furthermore, an increase in $\theta$ intensifies the blueshift effect.

The variations in the optical images of boson stars across different parameters, along with the distribution of their redshift factors, offer novel avenues for the detection of compact objects. In the subsequent research, we will systematically explore more sophisticated and realistic accretion disk models, including thick accretion disks, as well as modified gravity theories, such as $f(R)$ gravity. These investigations are expected to enhance our comprehension of compact objects and provide critical insights into differentiating boson stars from black holes.

\vspace{10pt}

\noindent {\bf Acknowledgments}

\noindent
This work is supported by the National Natural Science Foundation of China (Grant No. 12375043), and by the Sichuan Science and Technology Program (2024NSFSC1999).



\end{document}